\documentclass[apj]{emulateapj}

\usepackage{natbib}
\usepackage{amsmath}
\usepackage{color}
\newcommand{\herschel}{\textsl{Herschel}}
\newcommand{\planck}{\textsl{Planck}}

\newcommand{\arcm}{$^{\prime}$}
\newcommand{\arcs}{$^{\prime\prime}$}

\slugcomment{Re-submitted to \apj, 2013 June 14}

\shorttitle{Magnetization Imprint in Polarization Morphology}
\shortauthors{Soler et al.}

\begin{document}

%% LaTeX will automatically break titles if they run longer than one line. However, you may use \\ to force a line break if you desire.

\title{An Imprint of Molecular Cloud Magnetization in the Morphology of the Dust Polarized Emission}
%% Use \author, \affil, and the \and command to format
%% author and affiliation information.
%% Note that \email has replaced the old \authoremail command
%% from AASTeX v4.0. You can use \email to mark an email address
%% anywhere in the paper, not just in the front matter.
%% As in the title, use \\ to force line breaks.

%% Notice that each of these authors has alternate affiliations, which
%% are identified by the \altaffilmark after each name.  Specify alternate
%% affiliation information with \altaffiltext, with one command per each
%% affiliation.

\author{J.D.~Soler\altaffilmark{1}}
\author{P.~Hennebelle\altaffilmark{2}}
\author{P.G.~Martin\altaffilmark{3}}
\author{M.-A.~Miville-Desch\^{e}nes\altaffilmark{3,4}}
\author{C.B.~Netterfield\altaffilmark{1,5}}
\author{L.M.~Fissel\altaffilmark{1}}

\altaffiltext{1}{Department of Astronomy \& Astrophysics, University of Toronto,
    Toronto, ON Canada M5S 3H4; soler@astro.utoronto.ca}
\altaffiltext{2}{Laboratoire de Radioastronomie, \'{E}cole Normale Sup\'{e}riure and Observatoire de Paris, UMR CNRS 8112. 24 rue Lhomond 75231 Paris Cedex 05, France}
\altaffiltext{3}{Canadian Institute for Theoretical Astrophysics, University of Toronto, 60 St. George Street, Toronto, ON M5S 3H8, Canada}
\altaffiltext{4}{Institut d'Astrophysique Spatiale CNRS, Universit\'{e} Paris-Sud, b\^{a}timent 121, 91405, Orsay, France}
\altaffiltext{5}{Department of Physics, University of Toronto, 60 St. George Street, Toronto, ON M5S 1A7, Canada}

\begin{abstract}
We describe a morphological imprint of magnetization found when considering the relative orientation of the magnetic field direction with respect to the density structures in simulated turbulent molecular clouds. This imprint was found using the Histogram of Relative Orientations (HRO): a new technique that utilizes the gradient to characterize the directionality of density and column density structures on multiple scales. We present results of the HRO analysis in three models of molecular clouds in which the initial magnetic field strength is varied, but an identical initial turbulent velocity field is introduced, which subsequently decays. The HRO analysis was applied to the simulated data cubes and mock-observations of the simulations produced by integrating the data cube along particular lines of sight. In the 3D analysis we describe the relative orientation of the magnetic field $\mathbf{B}$ with respect to the density structures, showing that: 1.The magnetic field shows a preferential orientation parallel to most of the density structures in the three simulated cubes. 2.The relative orientation changes from parallel to perpendicular in regions with density over a critical density $n_{T}$ in the highest magnetization case. 3.The change of relative orientation is largest for the highest magnetization and decreases in lower magnetization cases. This change in the relative orientation is also present in the projected maps. In conjunction with simulations HROs can be used to establish a link between the observed morphology in polarization maps and the physics included in simulations of molecular clouds.
\end{abstract}

%% Keywords should appear after the \end{abstract} command. The uncommented example has been keyed in ApJ style. See the instructions to authors for the journal to which you are submitting your paper to determine what keyword punctuation is appropriate.


%\tableofcontents
%% Mark off your abstract in the ``abstract'' environment. In the manuscript style, abstract will output a Received/Accepted line after the title and affiliation information. No date will appear since the author does not have this information. The dates will be filled in by the editorial office after submission.

\section{Introduction}\label{intro}
The study of the magnetic field in molecular clouds is crucial to understand the physical processes relevant in molecular cloud evolution and star formation \citep{shu1987, mckee2007, crutcher2012}. Although some theoretical models point to supersonic turbulence as an important possibly dominant mechanism in the formation of structure in molecular clouds \citep{padoan1999, maclow2004}, recent studies indicate the magnetic field is also very important in particular regimes and scales \citep{kudoh2008, li2009, vazquezsemadeni2011}. The formulation of a complete model of the process that transforms gas into stars requires the comparison between the observations of the magnetic field and predictions from simulations.

The intensity of the magnetic field along the line of sight can be estimated using the Zeeman effect \citep{crutcher1983, crutcher1999}, the Faraday rotation of linearly polarized radiation from radio sources \citep{simard1980,han2006}, and the synchrotron unpolarized emission \citep{beuermann1985, miville2008, jaffe2010}. The morphology of the magnetic field projected onto the plate of the sky is observable by measuring polarized radiation \citep{heiles1993, crutcher2012}.

A longstanding technique used to infer the morphology of the magnetic field projected on the plane of the sky is the measurement of polarization of visible and near-infrared light background stars \citep{hiltner1949, davis1951, heiles2012}. This technique assumes that aspherical dust grains are aligned with respect to the interstellar magnetic field and the observed position angle of linear polarization is parallel to the plane-of-the-sky-projection of the field \citep{davis1951, dolginov1976}. A more recent complementary technique is the measurement of the thermal polarized emission of dust in the submillimeter wavelengths \citep{hildebrand1984, novak1997, vaillancourt2007}. Measurable degrees of polarization on the order of a few percent are typical for interstellar clouds in the submillimeter and the far-infrared \citep{benoit2004, vaillancourt2011}. The mechanism which produces the alignment of the grains is still the subject of active research \citep{draine1996, draine1997, weingartner2003, lazarian2007, hoang2009}. However, there is evidence indicating that it results from radiative torques produced by anisotropic radiation flux with respect to the magnetic field \citep{hoang2008}. Theoretical understanding of the grain alignment indicates that even in relatively dense clouds (visual extinction $A_{\rm V}\sim10$), dust polarization traces the underlying magnetic field \citep{bethell2007}.

The strength of the magnetic field has been estimated from observations of the projected component on the plane of the sky using the Chandrasekhar-Fermi method (CF) \citep{chandrasekhar1953, pereyra2007, houde2009, novak2009}. This method is based on the dispersion of the orientation angles of the polarization pseudovectors. Theoretical modeling of CF shows that the magnetic field strength estimates are improved by separating the turbulent and uniform components of the magnetic field \citep{falceta2008, hildebrand2009}. Maps of local magnetic field strength around pre-stellar cores have been obtained by comparing the polarization orientation angles and the intensity gradient, assuming ideal magneto-hydrodynamics (MHD) equations and spherical symmetry for the dust distribution in the core \citep{koch2012}.

Significant studies have been made to relate the magnetic field morphology inferred from polarization to the density structure in both observed \citep{li2009, chapman2011} and simulated clouds \citep{ostriker2001, nakamura2008}. In the case of pre-stellar cores, a number of observations reveal an hourglass morphology in the magnetic field \citep{girart2006, attard2009}. In the case of larger structures, the morphological analysis relies on visual inspection and statistics of the orientation angles \citep{goldsmith2008}. The results of observations are inconclusive in establishing a global trend of relative orientation between the magnetic field and the structure \citep{goodman1990}. Theory and simulations try to identify extreme cases which illustrate the scenarios where the formation of the structure in the cloud is dominated by either the magnetic field or the turbulence. Some of the predictions of the models are helical fields around filaments \citep{FiegePudritz2000ApJ, nakamura2008}, elongated structures along the magnetic field lines when the medium is strongly compressed supersonically \citep{ostriker2001, heitsch2001, falceta2008}, and linear correlation of the magnetic field strength with column density for supersonic turbulence \citep{burkhart2009}.

Recent observations by the ESA \herschel\ Mission reveal ubiquitous filamentary structures in molecular clouds \citep{peretto2012, andre2010, molinari2010}. These observations together with starlight polarization measurements have been used to investigate the physical conditions of gravitational collapse and star formation within the filaments. Seminal examples in which the inferred magnetic field direction in adjacent lower density material is perpendicular to the densest filaments are in Taurus \citep{chapman2011, palmeirim2013}, Serpens \citep{sugitani2011}, the Musca Dark Cloud \citep{pereyra2004}, and the Pipe Nebula \citep{alves2008}. These studies provide insight into the physical processes involved in the formation of filamentary structure but they do not probe the complex relationship of column density and magnetic field morphologies in the dense regions of molecular clouds, where stars predominantly form and which are fundamental to establishing a global picture of molecular cloud formation \citep{lada2010}.

The advent of new instruments for measuring the thermal dust polarized emission such as ALMA \citep{peck2008}, BLASTPol \citep{pascale2012} and \planck\ \citep{Planck2011} will produce an unprecedented volume of polarization maps in scales ranging from pre-stellar cores to entire molecular cloud complexes, including regions too extincted for starlight polarization observations. This large data set motivates a common scheme for the morphological analysis of polarization maps. It is in this context that we introduce the Histogram of Relative Orientations HRO, a novel statistical tool for characterizing column density morphologies on multiple scales. This technique, developed for pattern recognition in machine vision, uses the gradient to characterize the directionality of the density structure in a 2D map and here in a 3D simulated cube. It provides a robust characterization of the density field that can be used with polarization observations to investigate the relative orientation of density structure with respect to the magnetic field.

This article is organized as follows: Section \ref{hro:hro} introduces the Histogram of Relative Orientations. Section \ref{hro:sims} introduces the simulations of molecular clouds used to characterize our statistical tool. Section \ref{hro:HOG3D} presents the HRO analysis of 3D simulated clouds and Section \ref{hro:HOG2D} shows how the proposed diagnostic obtained in 3D can also be found in the projected 2D maps. Finally Section \ref{hro:conclusions} summarizes the results and discusses the application of HROs to observations of real molecular clouds.

% -------------------------------------------------------------------------------------------
\section{The Histogram of Relative Orientations}\label{hro:hro}

The Histogram of Relative Orientations (HRO) is inspired by a family of detection algorithms usually called Histograms of Oriented Gradients (HOG) \citep{dalal2006}. HOGs are used in pattern recognition and are developed in the context of machine vision. These algorithms use the gradient to describe the orientation of the edges in an image. The gradient is based on the difference between pixels in a neighborhood and quantifies the edges: the magnitude carries the information on the difference (edge strength) and the orientation defines the direction perpendicular to a contour line (edge direction). A histogram of the orientation angles of the gradient rotated by 90\degr\ therefore summarizes statistically the orientation of the features in an image. Further refinement of this method includes weighting the contribution of each pixel (or voxel in 3D) to the histogram by the magnitude of the gradient at each point, allowing the characterization of the image using only the stronger edges.

In the HRO, the gradient of the density (column density) characterizes the directionality of the structures in a vector field which is directly compared to the magnetic field on each voxel (pixel) using the scalar product of vectors. The result of this operation is a relative orientation angle between both vectors in each pixel which is characterized by a histogram. In 3D, the HRO is the histogram of relative orientation angles between the magnetic field vector $\mathbf{B}$ and the density gradient $\mathbf{\nabla} n$. In 2D, the HRO is the histogram of relative orientation angles between the projected magnetic field pseudovector $\mathbf{B}$ and the column density gradient $\mathbf{\nabla} \Sigma$. The calculation of the gradient and the estimation of the angle between vectors are described in the following subsections.

\subsection{Calculation of the Gradient}
The orientation of the isodensity contours is characterized by the gradient of density:%
\begin{equation}
\mathbf{\nabla}n = \left(\frac{\partial n}{\partial x}\right)^{(l)}\hat{x} + \left(\frac{\partial n}{\partial y}\right)^{(l)}\hat{y} + \left(\frac{\partial n}{\partial z}\right)^{(l)}\hat{z}.
\end{equation}\label{DensityGradient}
The subindex $l$ is related to the size of the area on which the gradient is calculated or as subsequently shown, the size of the Gaussian derivative kernel. To zeroth order, the components of the gradient can be calculated using forward differences of adjacent pixels:%
\begin{eqnarray}
\frac{\partial n}{\partial x}  & \sim & n(x+1,y)-n(x,y).
\end{eqnarray}

Derivatives are linear and shift-invariant, and so the gradient calculation can be done by convolving the image with a particular kernel. The result of the convolution of a slice of the simulation cube n(x,y) with a $l\times l$ derivative kernel $K^{(\partial/\partial x)}$ is:
\begin{eqnarray}
\nonumber \left(\frac{\partial n}{\partial x}\right)^{(l)} & = & n(x,y) \star  K^{(\partial/\partial x)}, \\
\left(\frac{\partial n}{\partial x}\right)^{(l)}_{ij} & = & \sum^{l-1}_{t,u} n_{i+t-l/2,j+u-l/2} K^{(\partial/\partial x)}_{t,u}.
\end{eqnarray}

Forward differences can be calculated using the $2\times2$ Roberts kernels, but such a small kernel is too sensitive to noise. The calculation of the gradient can be improved by using the central differences:%
\begin{eqnarray}
\nonumber \frac{\partial n}{\partial x} & = & \frac{n(x+1,y)-n(x-1,y)}{2},\\
\frac{\partial n}{\partial y} & = & \frac{n(x,y+1)-n(x,y-1)}{2}.
\end{eqnarray}
Applying this operation to all the pixels in an image is equivalent to convolving the image with $3\times3$ Prewitt kernels:%
\begin{eqnarray}
K^{(\partial/\partial x)}  =
\frac{1}{6} \left(\begin{array}{ccc} -1 & 0 & 1  \\
-1 & 0 & 1 \\
-1 & 0 & 1
\end{array} \right) & , &
K^{(\partial/\partial y)} =
\frac{1}{6}\left(\begin{array}{ccc} -1 &  -1 & -1\\
0 & 0 & 0 \\
1 & 1 & 1
\end{array} \right).
\end{eqnarray}%

Noise reduction is achieved by averaging over the vicinity of each pixel. This is implemented to first order using the $3\times3$ Sobel kernels. Convolving with these kernels is equivalent to smoothing the image over a $3\times3$ region and then calculating first derivatives, which is a particular case of the Gaussian Derivatives method:%
\begin{eqnarray}
K^{(\partial/\partial x)} =
 \left(\begin{array}{ccc} -1 & 0 & 1  \\
-2 & 0 & 2 \\
-1 & 0 & 1
\end{array}     \right) & , &
K^{(\partial/\partial y)} =
\left(\begin{array}{ccc} -1 &  -2 & -1\\
0 & 0 & 0 \\
1 & 2 & 1
\end{array}     \right).
\end{eqnarray}%

\subsubsection{Gaussian Derivatives}\label{GaussianDerivatives}
In HROs, the computation of the gradient is performed using Gaussian derivative kernels. The size of the Gaussian determines the area of the vicinity over which the gradient will be calculated. Varying the size of the Gaussian kernel enables the sampling of different scales and reduces the effect of noise in the pixels.

Convolution and differentiation are commutative and associative; therefore the smoothing and derivative operators can be written as:%
\begin{eqnarray}\label{GaussianDerivativeEq}
\frac{\partial}{\partial x}(I \star G)  & = & I \star \frac{\partial}{\partial x}G
\end{eqnarray}%
G is the 2D-Gaussian kernel and I is a 2D image. This equation implies that filtering an image and subsequently calculating the gradient is equivalent to the convolution of the image with a kernel that is the first derivative of a gaussian kernel. This operation is called a Gaussian derivative \citep{young1986}.

For the present study, we use the Gaussian derivatives method and following Equation \ref{GaussianDerivativeEq} we obtained each component of the gradient by convolving slices of the density cube $n(x,y)$ and $n(y,z)$ with a kernel formed by the derivative of a $l\times l$ map of a two-dimensional Gaussian $G^{(l)}$:%
\begin{eqnarray}
\nonumber\left(\frac{\partial n}{\partial x}\right)^{(l)} & = & n(x,y) \star \frac{\partial}{\partial x}G^{(l)}(x,y), \\
\left(\frac{\partial n}{\partial y}\right)^{(l)} & = & n(x,y) \star \frac{\partial}{\partial y}G^{(l)}(x,y), \\
\nonumber\left(\frac{\partial n}{\partial z}\right)^{(l)} & = & n(y,z) \star \frac{\partial}{\partial z}G^{(l)}(y,z).
\end{eqnarray}%
In the 2D projected maps%
\begin{eqnarray}
\left(\frac{\partial \Sigma}{\partial x}\right)^{(l)} & = & \Sigma \star \frac{\partial}{\partial x}G^{(l)} \\
\nonumber\left(\frac{\partial \Sigma}{\partial y}\right)^{(l)} & = & \Sigma \star \frac{\partial}{\partial y}G^{(l)}.
\end{eqnarray}

The orientation angle of the iso-$\Sigma$ contours in the 2D project maps is:%
\begin{eqnarray}
\psi & \equiv & \arctan\left(\frac{\partial \Sigma/\partial x}{\partial \Sigma/\partial y}\right),
\end{eqnarray}
which according to the convention for position angles of polarization is measured counterclockwise from the top of the map.

\begin{figure}
\centering
\includegraphics[angle=0,width=0.47\linewidth]{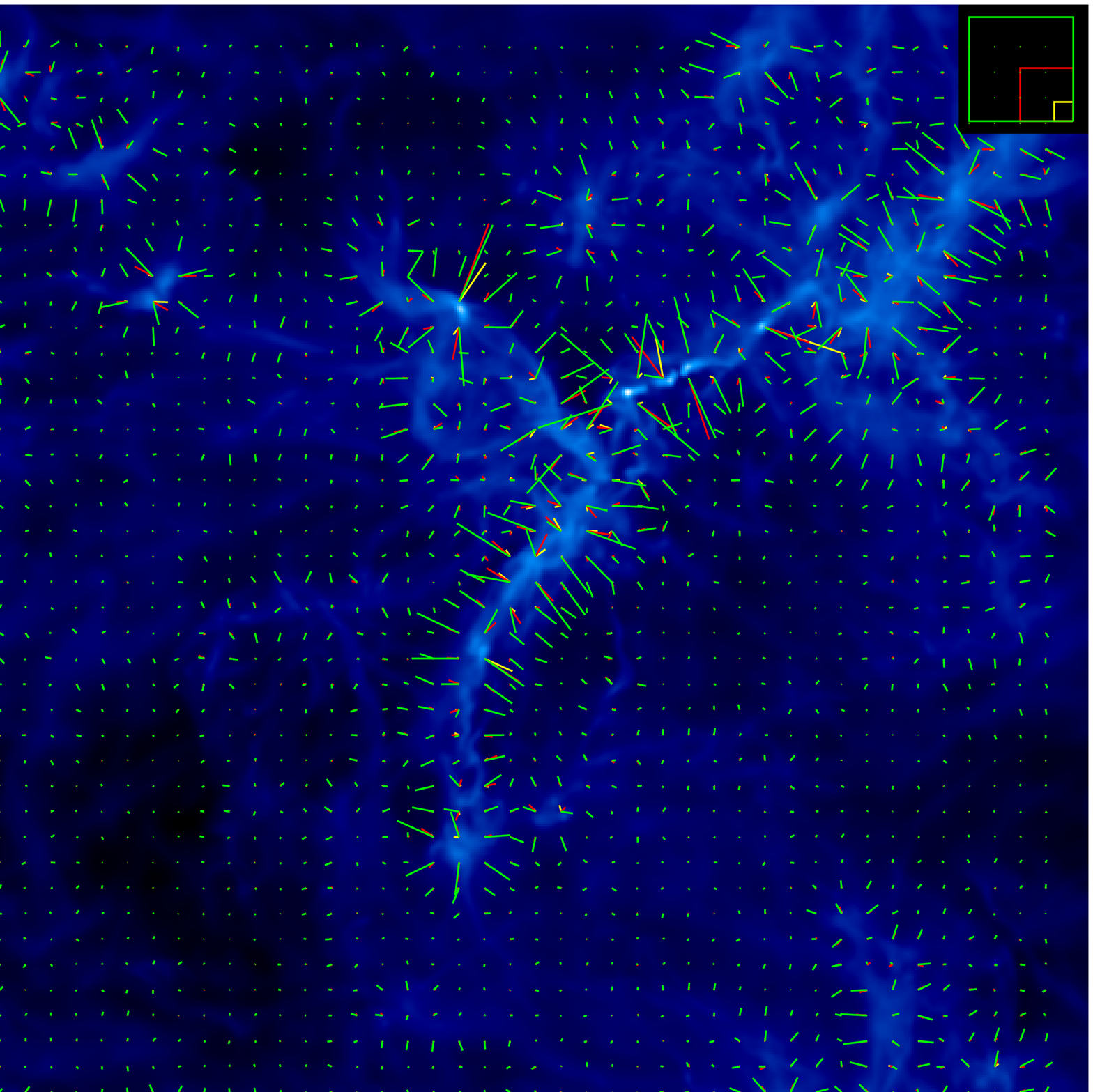}
\includegraphics[angle=0,width=0.49\linewidth]{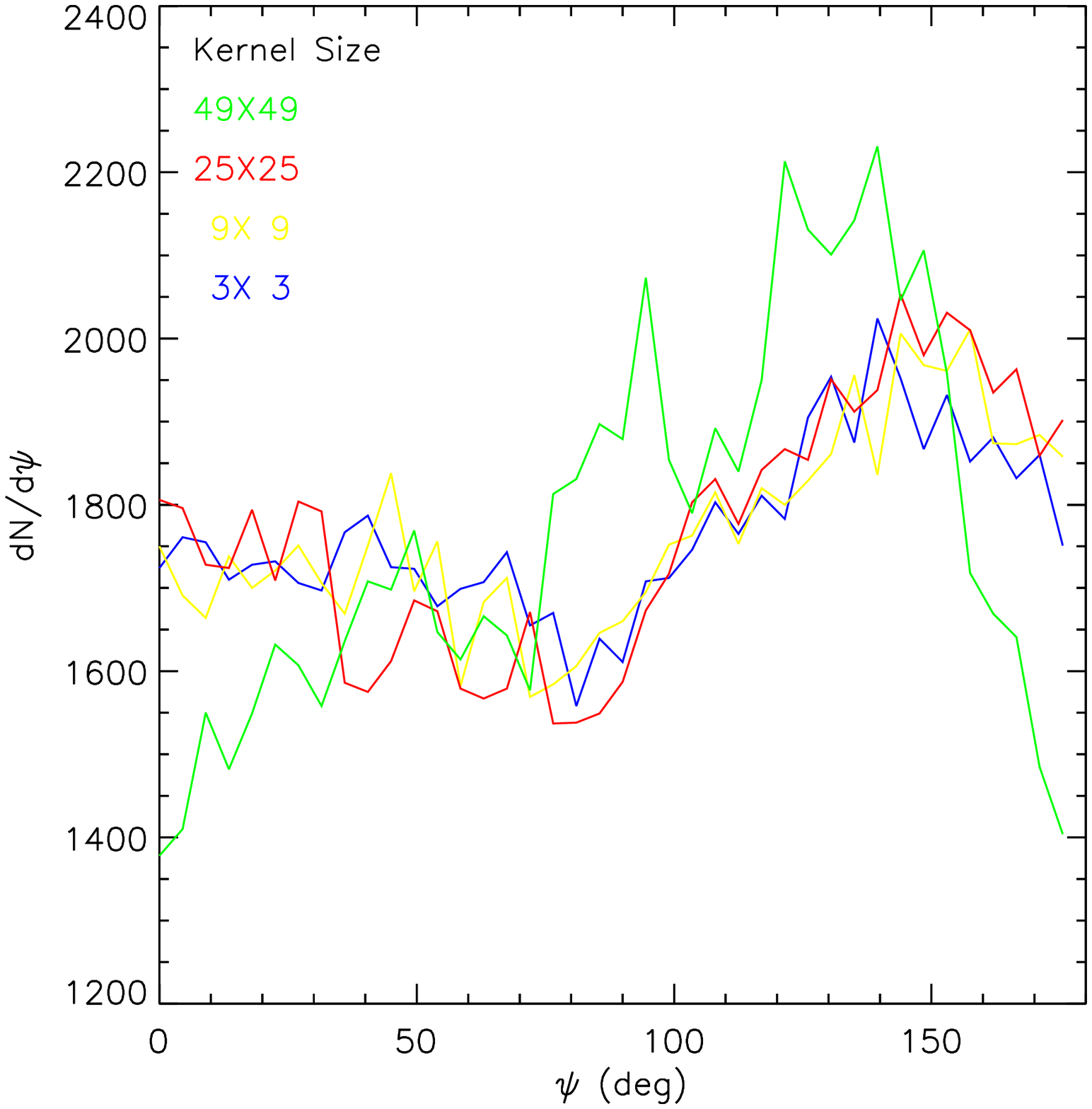}
\caption[Map Characterization Using Gradients]{Left: Filamentary structure in the simulated column density. Overlaid are the gradient vectors obtained with derivative kernels with 9$\times$9 (yellow), 25$\times$25 (red), and 49$\times$49 (green) pixels. The squares on the upper right corner of the image show the sizes of these kernels. Right: Histogram of orientation angles of the iso-$\Sigma$ contours ($\psi$) calculated with each derivative kernel. The histograms show that the structure is predominantly oriented at $\psi$ from 135\degr\ to 150\degr\ but also shows a secondary structure at 45\degr.}\label{GradientExample}
\end{figure}

Figure \ref{GradientExample} illustrates the characterization of a simulated image using the gradient of the column density $\mathbf{\nabla} \Sigma$. The vectors plotted over the image show the direction of the gradient obtained using Gaussian kernels of different sizes. The histogram shows the orientation angles of the structures in the image, measured counterclockwise with respect to the y-axis of the image (convention for polarization position angles). Each peak in the histogram corresponds to a dominant orientation of the features in the image. A histogram of a completely random map would be flat. The histogram of a straight rod would be a delta function centered on its orientation angle. The histogram of a blob would have a peak corresponding to the orientation of its semi-major axis.

The HROs use the gradient of the density in 3D simulated cubes to generate a vector field which characterizes isodensity contours and can be directly compared to the magnetic field vectors. In 2D, the gradient of the column density ($\mathbf{\nabla} \Sigma$) is compared to the orientation of the linear polarization field, which is a pseudovector since it is invariant to a $\pm 180$\degr\ rotation.

\subsection{Calculation of the Angle}

The angle $\phi$ between the gradient and the magnetic field at each voxel is calculated using a combination of the scalar and vector product of vectors:%
\begin{equation}\label{AngleCalculation}
\phi \equiv \arctan\left( \frac{\mathbf{B} \times \mathbf{\nabla}n}{\mathbf{B} \cdot \mathbf{\nabla}n} \right)
\end{equation}%
The result of this calculation is a cube (a map in 2D) with values of $\phi$ at each voxel (pixel in 2D). As explained in Section \ref{hro:HOG3D}, in the study of the alignment in 3D it is more natural to construct the HROs as a function of $\cos\phi$ and not directly $\phi$. The angle between the isodensity contours and the magnetic field is $\phi \pm 90$\degr. The histogram of values of $\cos\phi$ ($\phi$ in 2D) weighted by the magnitude of the gradient at each voxel (pixel) is what we call Histogram of Relative Orientations.

\begin{figure}
\centering
\includegraphics[angle=270,width=0.99\linewidth]{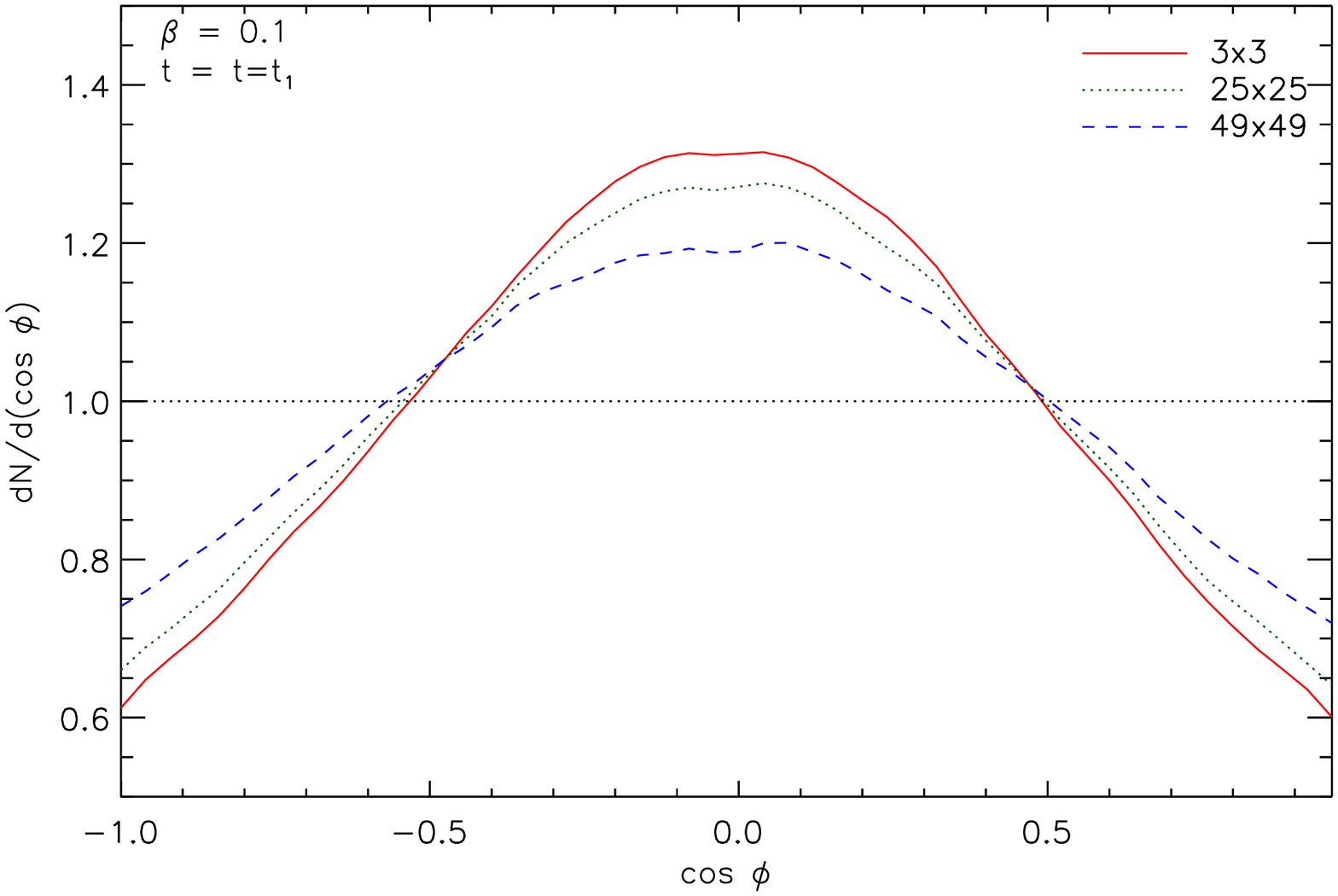}\\
\includegraphics[angle=270,width=0.99\linewidth]{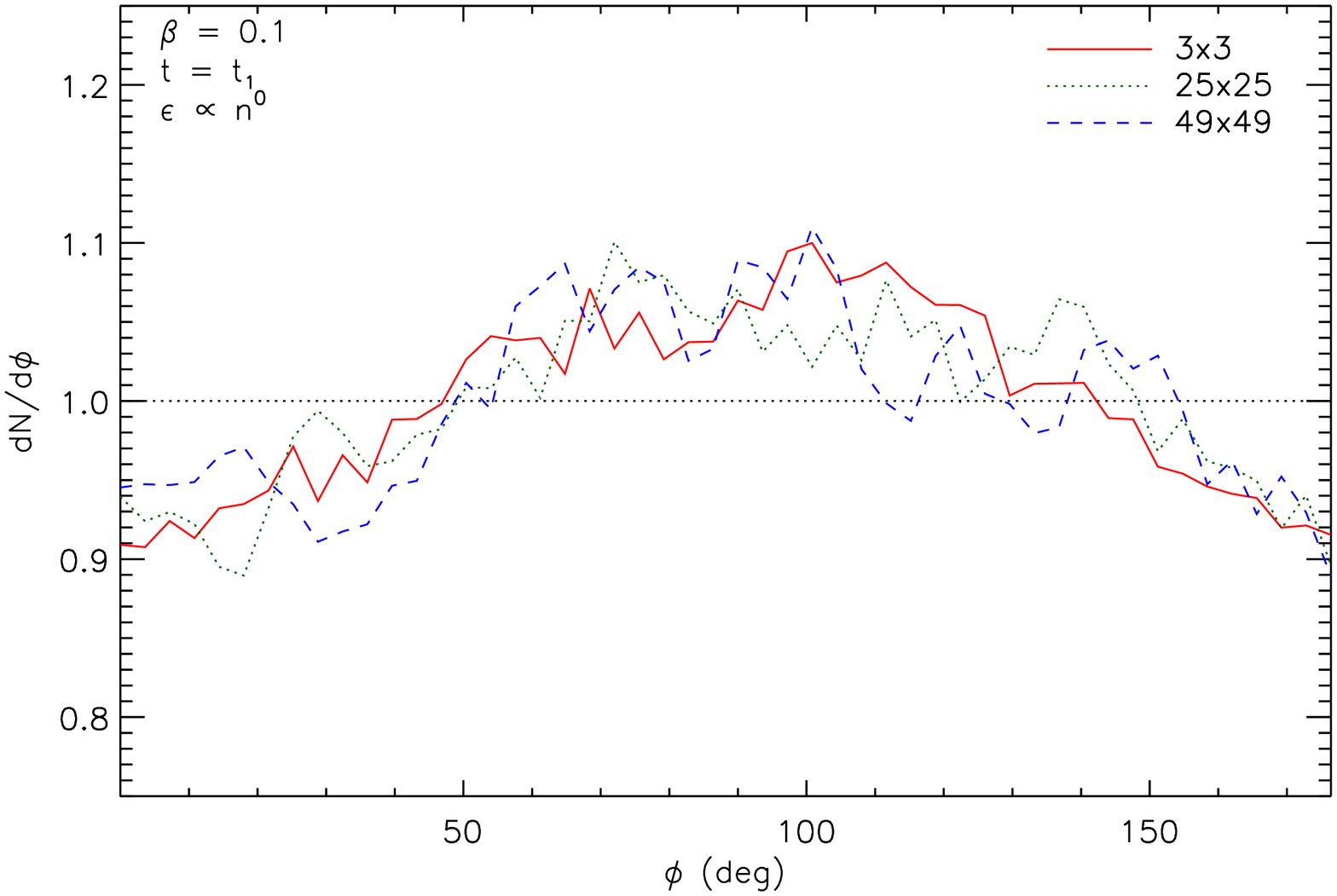}
\caption[HRO with Multiple Derivative Kernels]{HROs obtained by applying three different derivative kernels to the $\beta=0.1$ simulation cube (left) and to a projected map obtained by integrating the cube along the z-axis (right). The similarity of the curves within each plot shows that the predominant relative orientation is present in regions with scales ranging from 3 pixels to 49 voxels (pixels).}\label{HOGmultikernel}
\end{figure}

Figure \ref{HOGmultikernel} shows a histogram of $\cos\phi$ calculated on a simulated cube and $\phi$ calculated on a projection of the cube using three derivative kernels. Using these three kernels we observe the same preferential relative orientation between $\mathbf{B}$ and $\mathbf{\nabla} n$ in the cube and $\mathbf{B}_{POS}$ and $\mathbf{\nabla} \Sigma$ in the projection.

\subsection{Segmentation}\label{Segmentation}

Individual regions of a cube (map) are studied by dividing it into bins of the parameters that define it and subsequently masking each bin. This process is known in computer science as image segmentation. Segmentation is a subject of active research in computer science and its implementation in 3D (2D) HROs is limited to dividing the cube (image) in density (column density) bins with the same number of voxels (pixels).

\begin{figure}
\centering
\includegraphics[angle=270,width=0.49\linewidth]{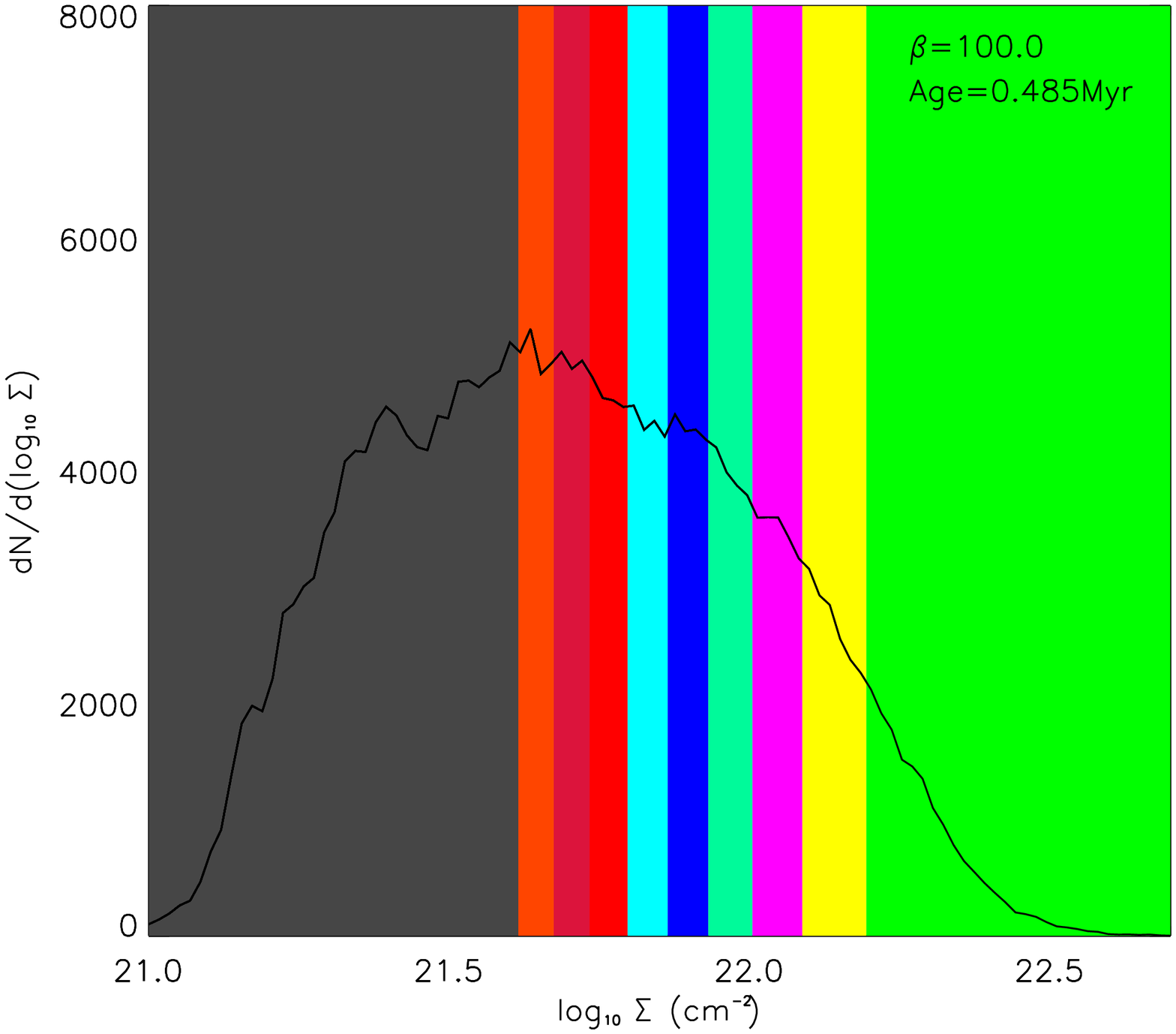}
\includegraphics[angle=270,width=0.47\linewidth]{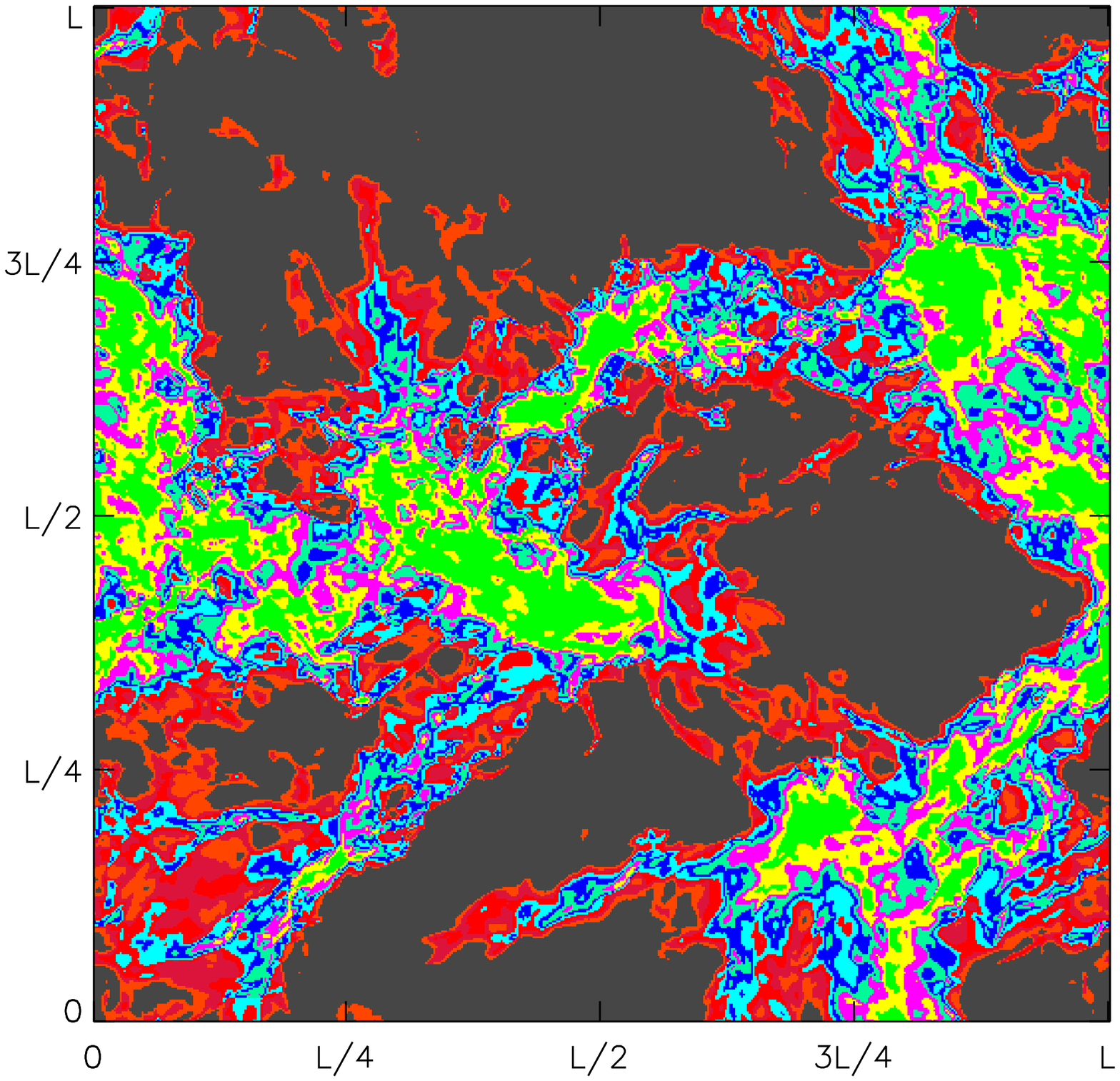}\\\
\includegraphics[angle=270,width=0.49\linewidth]{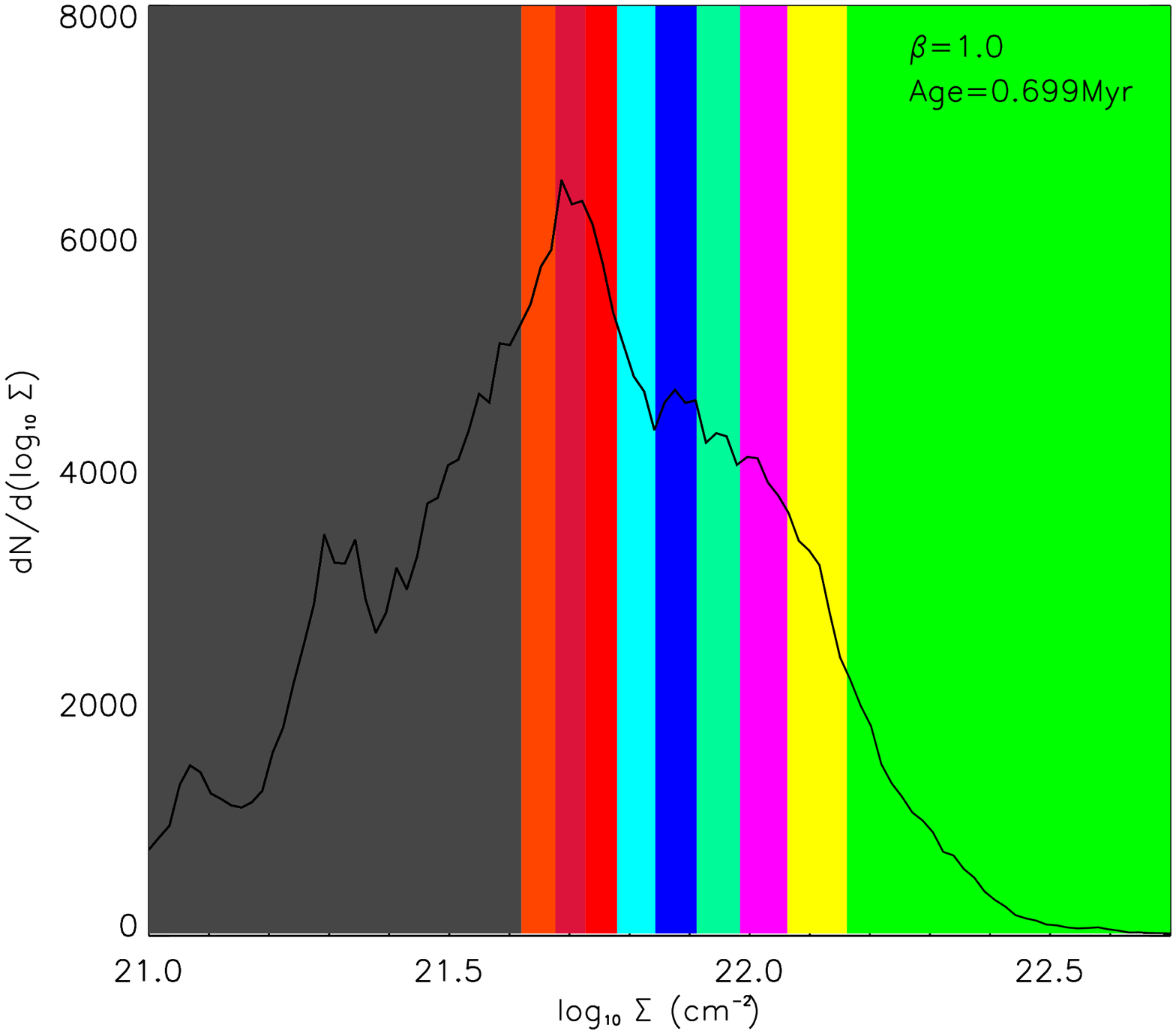}
\includegraphics[angle=270,width=0.47\linewidth]{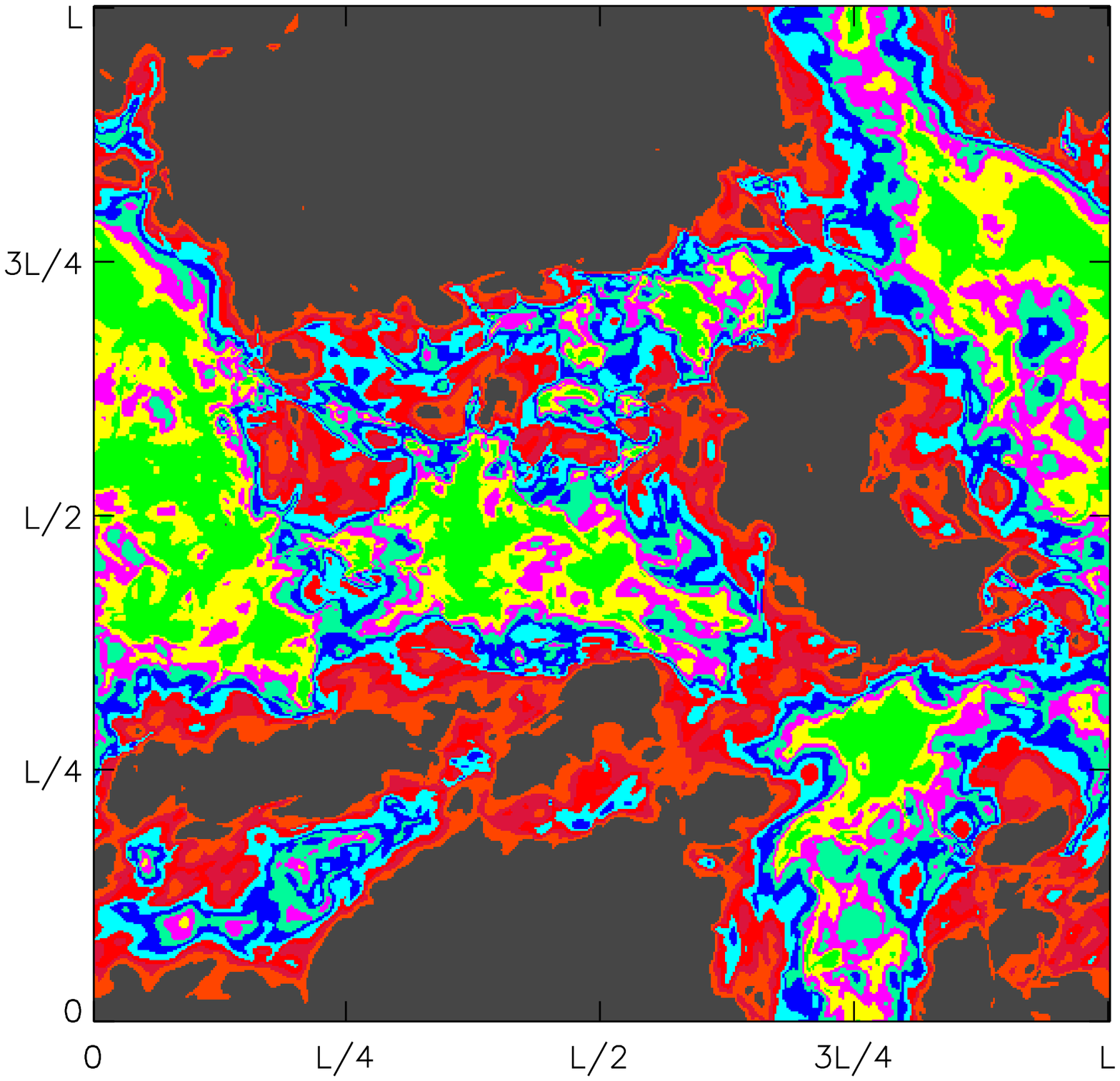}\\
\includegraphics[angle=270,width=0.49\linewidth]{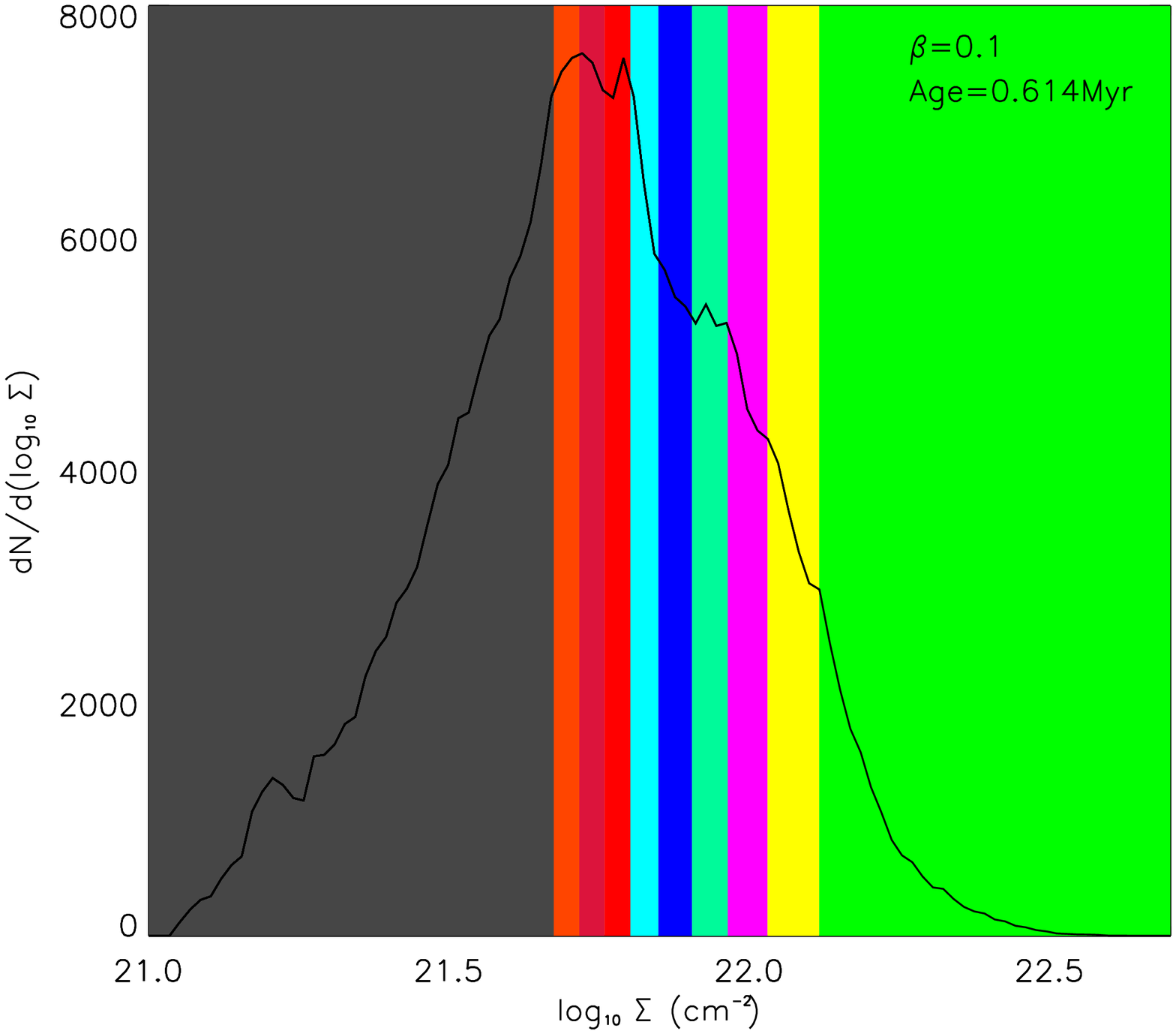}
\includegraphics[angle=270,width=0.47\linewidth]{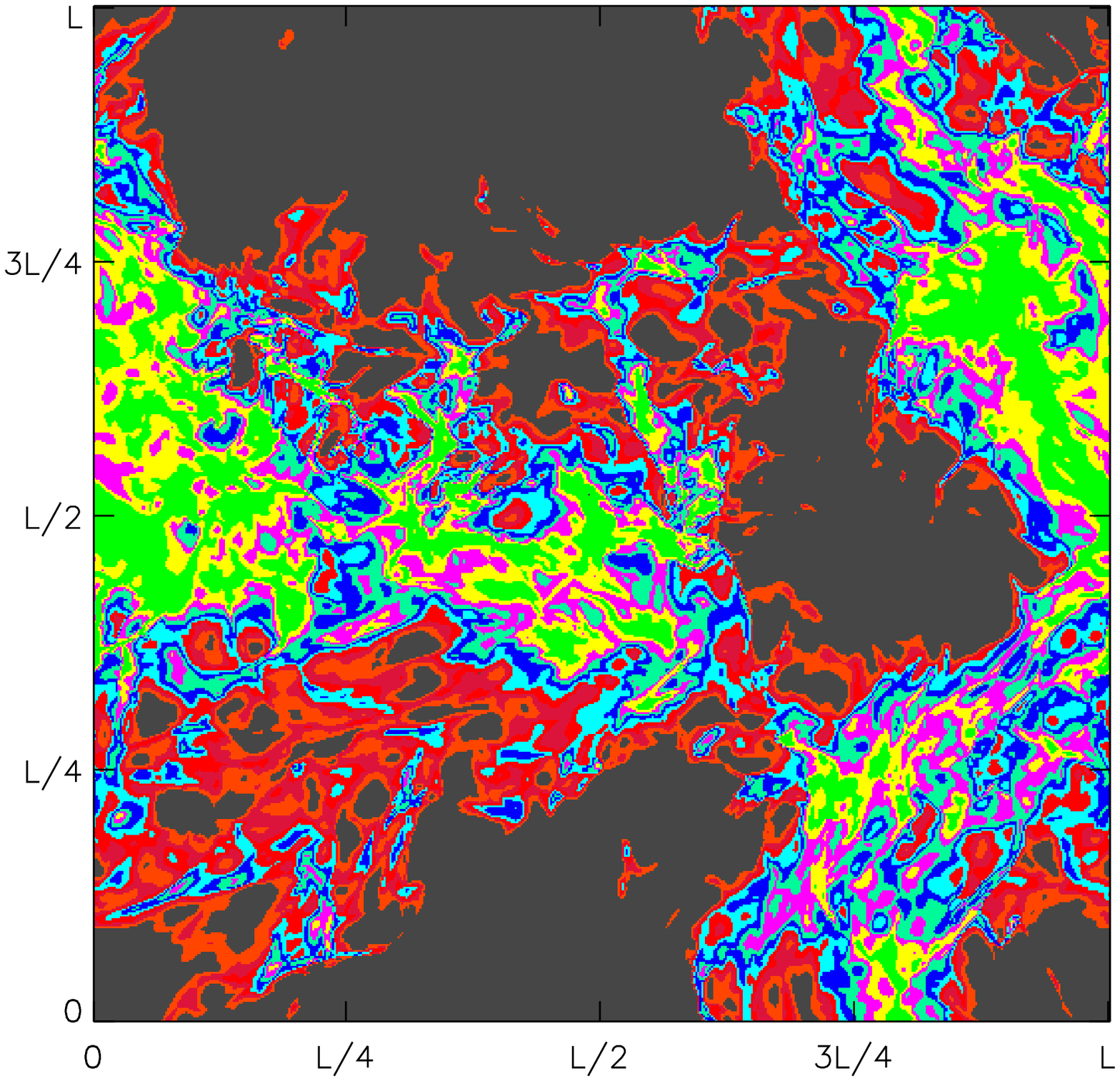}
\caption[Segmentation of Column Density Map]{Example of the map segmentation technique used in the HRO analysis. Left: Column density distribution in projections of three simulations. Overlaid colors correspond to density bins with equal number of pixels in regions with $\Sigma>\bar{\Sigma}$. Gray regions corresponds to pixels with $\Sigma<\bar{\Sigma}$. Right: Column density maps of the projections of three simulations. Overlaid colors correspond to regions with column densities shown in the histograms.}\label{DensHist}
\end{figure}

The objective of the segmentation is the study and comparison of the relative orientation in density regimes where the dominant physical processes and the dust alignment efficiency $\epsilon$ are different. The segmentation by column density of the maps resulting from the projection of the simulated cubes is illustrated in the left-hand side of Figure \ref{DensHist}. Using the column density distribution, which is close to log-normal, we produce density bins with the same number of pixels. An equal number of gradient vectors in each bin guarantees comparable statistics for each density bin.

The segmentation by density or column density is motivated by the change of physical conditions in the densest regions with respect to less dense regions. In the densest regions self-gravity becomes relevant and the kinetic energy density is lower than the magnetic energy density, in contrast with the less dense regions where the balance between magnetic and turbulent energy is different. The segmentation of maps by polarization percentage, line of sight velocity or polarization angle dispersion is a potentially useful way to study physical processes in observations of molecular clouds and will be addressed in future works.

%--------------------------------------------------------------------------------------------------------------
\section{Model Parameters}\label{hro:sims}

The cloud models used to characterize HRO are created by integrating the compressible ideal MHD equations using the RAMSES-MHD code \citep{teyssier2002, fromang2006}. RAMSES-MHD is a N-body and MHD code with 3D Adaptive Mesh Refinement (AMR). The refinement criteria are based on density as described in \citet{dib2010}. RAMSES-MHD uses \emph{constrained transport} to guarantee that $\mathbf{\nabla}\cdot\mathbf{B}=0$ to machine accuracy at all time and uses the MUSCL-Hancock scheme, a finite volume method that combines good accuracy with fast execution. The solutions are obtained in a cubic box of side L with grids of 512$^3$ zones, integrated from the AMR cubes with effective resolution 2048$^3$.

The energy equation used in this simulation set is the barotropic equation of state: these clouds are isothermal with $T = 11.44$~K (sound speed $c_{s} = 0.2$~km~s$^{-1}$) for regions with density less than $10^4$ times the initial density $n_{0}$ and adiabatic for larger densities. In the absence of fully time-dependent radiative transfer this represents a reasonable first approximation for the gas at volume densities higher than the mean (comprising most of the matter) for conditions appropriate to molecular clouds \citep{scalo1998}. The solutions are obtained in a cubic box of side $L$ with initial uniform density $n_{0}=$ $n_{H_{2}}=$ $536.41$~cm$^{-3}$, which is comparable to the density in the Taurus-Auriga Dark Cloud Complex \citep{stahler2005}. We apply periodic boundary conditions in all models, simulating a portion of the interior of a molecular cloud.

The gravitational potential is computed from the density using standard Fourier methods. The periodic boundary conditions exclude the $\mathbf{k}=0$ component of the density. Therefore, the gravitational potential $\phi_{G}$ obeys the modified Poisson equation $\nabla^{2}\phi_{G}=4\pi G(\rho-\bar{\rho})$, where $\bar{\rho}=M/L^{3}$ is the mean density (mass divided by the volume of the box).

The relative importance of gravity and thermal pressure forces is related to $L$ and $n_{0}$. For this analysis we have chosen $L = 4$~pc; the Jeans length $L_{J}\equiv c_{s}(\pi/G\rho_{0})^{1/2}$ is 0.88~pc for $\rho_{0} = \mu n_{0}$ and $\mu=2.4\, m_{p}$. The total mass in the simulated cubes is $2.04\times10^3\, M_{\odot}$ and with $L=4$~pc, it is comparable to small regions within Dark Cloud Complexes such as Lupus I \citep{Cambresy1999}.

Turbulence in the simulation is introduced as an initial isotropic random velocity field $\delta\mathbf{v}$ with a Kolmogorov energy spectrum $E(k) \propto k^{-5/3}$. $\delta\mathbf{v}$ is a Gaussian random perturbation field with a power spectrum $|\delta v_{\mathbf{k}}|^{2} \propto k^{-11/3}$. The Kolmogorov energy spectrum is comparable to the spectrum inferred for large-scale cold interstellar clouds \citep{larson1981} and the spectrum that naturally follows from the evolution of incompressible turbulence. An identical realization of the initial velocity field is used for all of the models, so that the initial states of the simulations differ only in the strength of the mean magnetic field. The velocity field is a mixture of compressible and solenoidal modes. In the current study we focus on the characterization models with different magnetization. To keep this comparison simple, the turbulence decays during the cloud evolution (it is not driven), avoiding the complexity introduced by modeling the energy injection rate and the spectrum of driven turbulence. All simulations are initiated with kinetic energy $E_{K}=100\bar{\rho}L^{3}c^{2}_{s}$ corresponding to an initial Mach number $\mathcal{M}\equiv\sigma_{v}/c_{s}=$10 and include gravity.

Three different initial magnetization cases are considered and parameterized by the ratio of the isothermal sound speed to the initial Alfv\'{e}n speed or equivalently, the ratio of the plasma pressure to the initial magnetic pressure: $\beta\equiv c^{2}_{s}/v^{2}_{A,0}=\bar{\rho}c^{2}_{s}/(B^{2}_{0}/4\pi)$. We consider a ``quasi-hydrodynamical'' model with $\beta = 100.0$, an ``intermediate magnetization'' model with $\beta = 1.0$, and a ``high magnetization'' model with $\beta = 0.1$. The physical value of the magnetic field is given by:
\begin{equation}
B_{0} = \beta^{-1/2} \mu\mbox{G} \left(\frac{T}{10\mbox{ K}}\right)^{1/2}\left(\frac{n_{H_{2}}}{100\mbox{ cm}^{-3}}\right)^{1/2}.
\end{equation}
For the chosen initial density and temperature, the corresponding uniform magnetic field strengths are 0.35, 3.47, and 10.97~$\mu$G. The evolved fields are spatially nonuniform and can differ greatly from the initial values although the volume-averaged magnetic field is a constant $B_{0} \hat{x}$ in time. The value of $\beta$ is proportional to the mass-to-magnetic flux ratio in the simulation and cannot change with time.

The initial Alfv\'{e}n Mach number $\mathcal{M}_{A} \equiv \left<v^{2}/v_{A}^{2}\right>^{1/2}$ $=\mathcal{M}\beta^{1/2}$ is 100, 10, and 3.16 for $\beta$ = 100.0, 1.0, and 0.1 respectively, and therefore all models considered have initially supersonic and super-Alfvenic flow. In this work, ``high magnetization'' is relative to thermal and magnetic pressures only, given that the turbulence is the dominant energy density in all considered models. The initially uniform clouds are unstable to compressions transverse to the mean magnetic field (``magneto-Jeans unstable'') when the magneto-sonic wave crossing time exceeds the characteristic gravitational contraction time, $t_{g}$ \citep{chandrasekhar1953}. All models would be unstable by the magneto-Jeans criterion ($L/L_{J}=4.56<\beta^{-1/2}$). The three simulations are initially supercritical, with the mass $M$ to magnetic flux $\Phi$ ratio over the critical value $(M/\Phi)_{crit}$ $=1/(2\pi G^{1/2})$, and $(M/\Phi)/(M/\Phi)_{crit}\approx$ $(t_{g}v_{A}/(\pi L))^{-1}$ is equal to 142.9, 14.3, and 4.52 respectively.

\begin{deluxetable}{ccccc}
\tabletypesize{\scriptsize}
\tablecaption{Model parameters and times of snapshots\label{tableSim}}
\tablewidth{0pt}
\tablehead{
\colhead{$\beta$} & \colhead{$<B_0>$} &\multicolumn{3}{c}{Snapshot Age}\\
 & \colhead{($\mu$G)} & \colhead{(Myr)} & ($t_{g}$\tablenotemark{a}) & ($t_{f}$\tablenotemark{b})
}
\startdata
100.0 & 0.35 & 0.48 & 0.11  & 0.25\\
100.0 & 0.35 & 1.16 &  0.27 & 0.60 \\
1.0   & 3.47 & 0.69 &  0.16  & 0.36  \\
1.0   & 3.47 & 1.32 & 0.31 & 0.68  \\
0.1   & 10.97 & 0.61 & 0.14  & 0.32   \\
0.1   & 10.97 & 1.15 & 0.27  & 0.59
\enddata
%% Text for table notes should follow after the \enddata but before
%% the \end{deluxetable}. Make sure there is at least one \tablenotemark
%% in the table for each \tablenotetext.
%\tablecomments{Table \ref{tbl-1} is published in its entirety in the electronic edition of the {\it Astrophysical Journal}.  A portion is shown here for guidance regarding its form and content.}
\tablenotetext{a}{Characteristic gravitational time scale $t_{\mbox{g}} = 4.27$~Myr}
\tablenotetext{b}{Flow crossing time scale $t_{\mbox{f}} = 1.95$~Myr}
\end{deluxetable}

In Table \ref{tableSim}, we describe the simulation snapshots under consideration in terms of the sound crossing time $t_{v}\equiv L/c_{s}= 19.5$~Myr, which is fixed owing to the \emph{isothermal} equation of state, the flow crossing time $t_{f}\equiv L/\sigma_{v}$ $= 1.95$~Myr, where we use the Mach number associated with the initial turbulent velocity and the characteristic gravitational time scale, $t_{g}\equiv(\pi/\mbox{G}\bar{\rho})^{1/2}= 4.27$~Myr. Two snapshots in this study are taken: one at $t_{1}\sim 0.03\, t_{v}$ and another at $t_{2}\sim 0.06\, t_{v}$. As in preceding studies \citep{ostriker2001, heitsch2001}, this paper concentrates on structures that form as a consequence of turbulence, and subsequently collapse gravitationally. The snapshots are taken when the gas has started to collapse and form dense structures at few places. At $t_{1}$, shock fronts moving through the gas initiate the formation of filaments and knots. At $t_{2}$ the collapse into very dense structures has taken place in certain regions producing over-densities with $n\sim10^{3}\overline{n}$.

% 3D HOGS =====================================================================

\section{HRO Applied to Simulation Cubes}\label{hro:HOG3D}

The simulated cube contains the density, velocity, and magnetic field values for every voxel. The HROs are used to summarize the relative orientation of the magnetic field vector $\mathbf{B}$ with respect to the gradient of the density $\mathbf{\nabla}n$ in each one of these voxels following the procedure described in Section \ref{hro:hro}. The histogram of the angle $\phi$ between two sets of random vectors in 3D is not uniform, i.e., in 3D, two random vectors are more likely to be perpendicular than parallel. In 3D the uniformly distributed quantity is $\cos\phi$ and therefore we choose that quantity for the histogram.

\begin{figure}
\centering
\includegraphics[angle=270,width=1.00\linewidth]{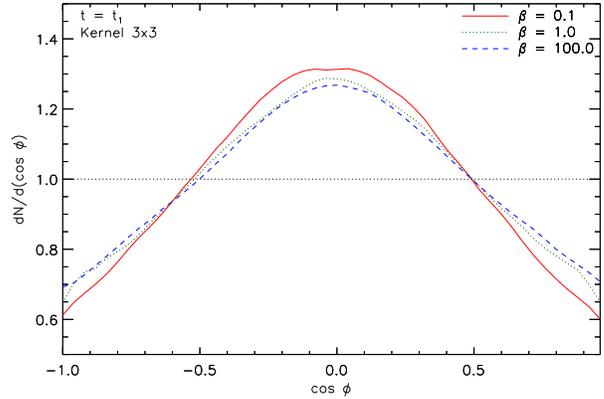}
\caption[Comparison of HROs of Simulations Cubes with Different Magnetization]{Histogram of Relative Orientation showing the cosine of the angle between the magnetic field vector $\mathbf{B}$ and the gradient of the density $\mathbf{\nabla} n$ for the low, intermediate, and high-magnetization simulation cubes ($\beta = 100$, $1.0$, and $0.1$) in a snapshot taken at $t\sim 0.06\, t_{v}$. The histogram is normalized such that a random distribution of $\mathbf{B}$ and $\mathbf{\nabla} n$ would equal unity in each bin (black dotted line). The histograms calculated from the simulated cubes show a peak at $\cos\phi\sim 0$ which corresponds to the magnetic fields predominantly tracing the isodensity contours.}\label{HOG3Dmulti}
\end{figure}

Figure \ref{HOG3Dmulti} shows the HRO corresponding to all of the voxels in each of the three simulations for a snapshot at $t_{2}\sim 0.3\, t_{g}$. The peak at $\cos\phi=0$ shows that $\mathbf{\nabla}n$ is predominantly oriented perpendicular to $\mathbf{B}$ in the three simulations. This corresponds to $\mathbf{B}$ being mostly parallel to the isodensity contours in the simulated volume.

The two snapshots considered here correspond to $t<t_{g}$ and $t<t_{v}$. At these time scales, the dynamics of the gas are dominated by shocks and the over-densities are not formed by self-gravitation. The relative orientation revealed by the HRO is the result of the magnetic field becoming strongly bent and stretched. The imprint of the initial magnetization was found by studying and comparing the relative orientation in the highest densities regions. In these regions, the relative orientation is the consequence of the locally strong self-gravity and the interaction of the gas and the magnetic field.

\begin{figure*}
\includegraphics[angle=270,width=0.46\linewidth]{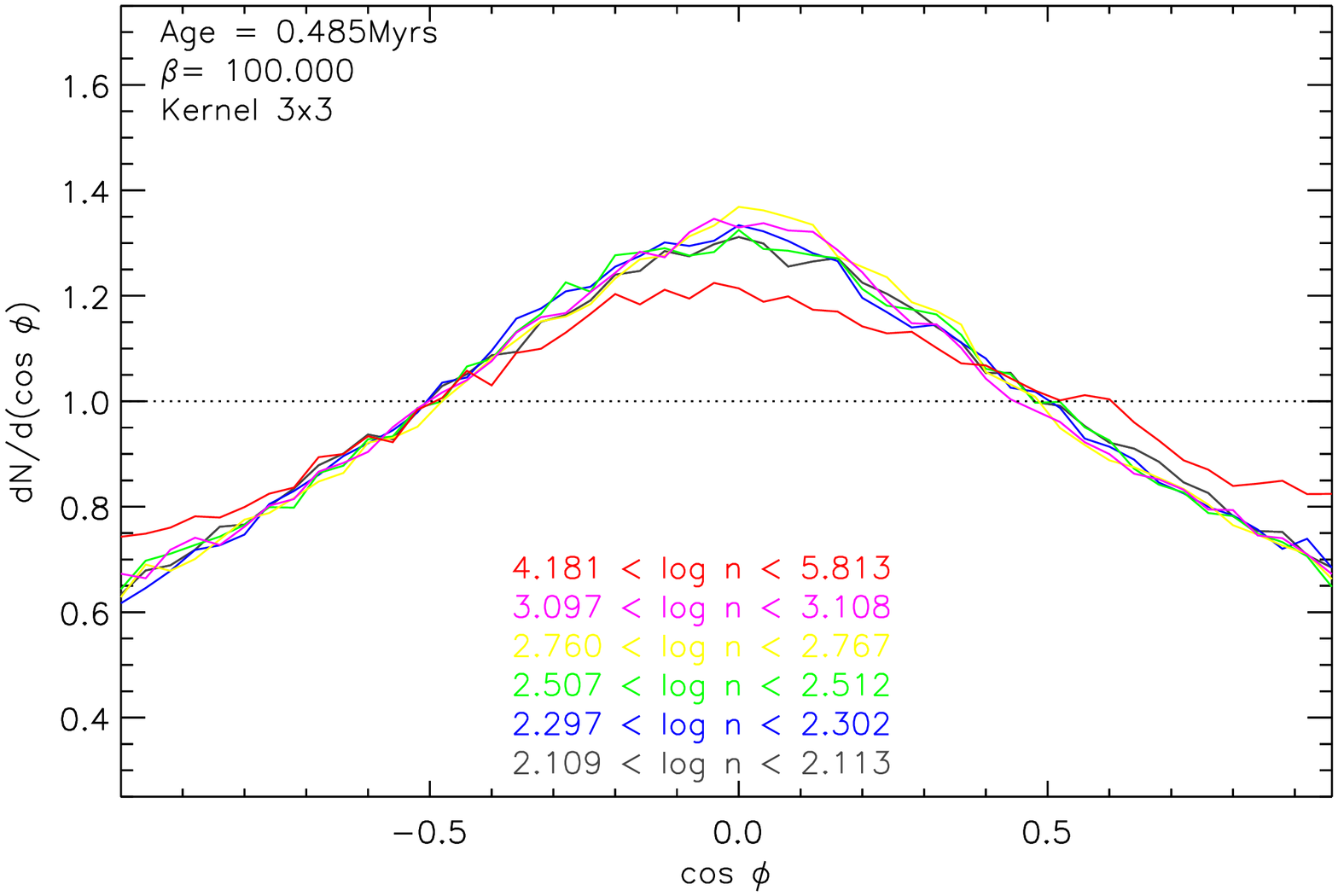}
\includegraphics[angle=270,width=0.46\linewidth]{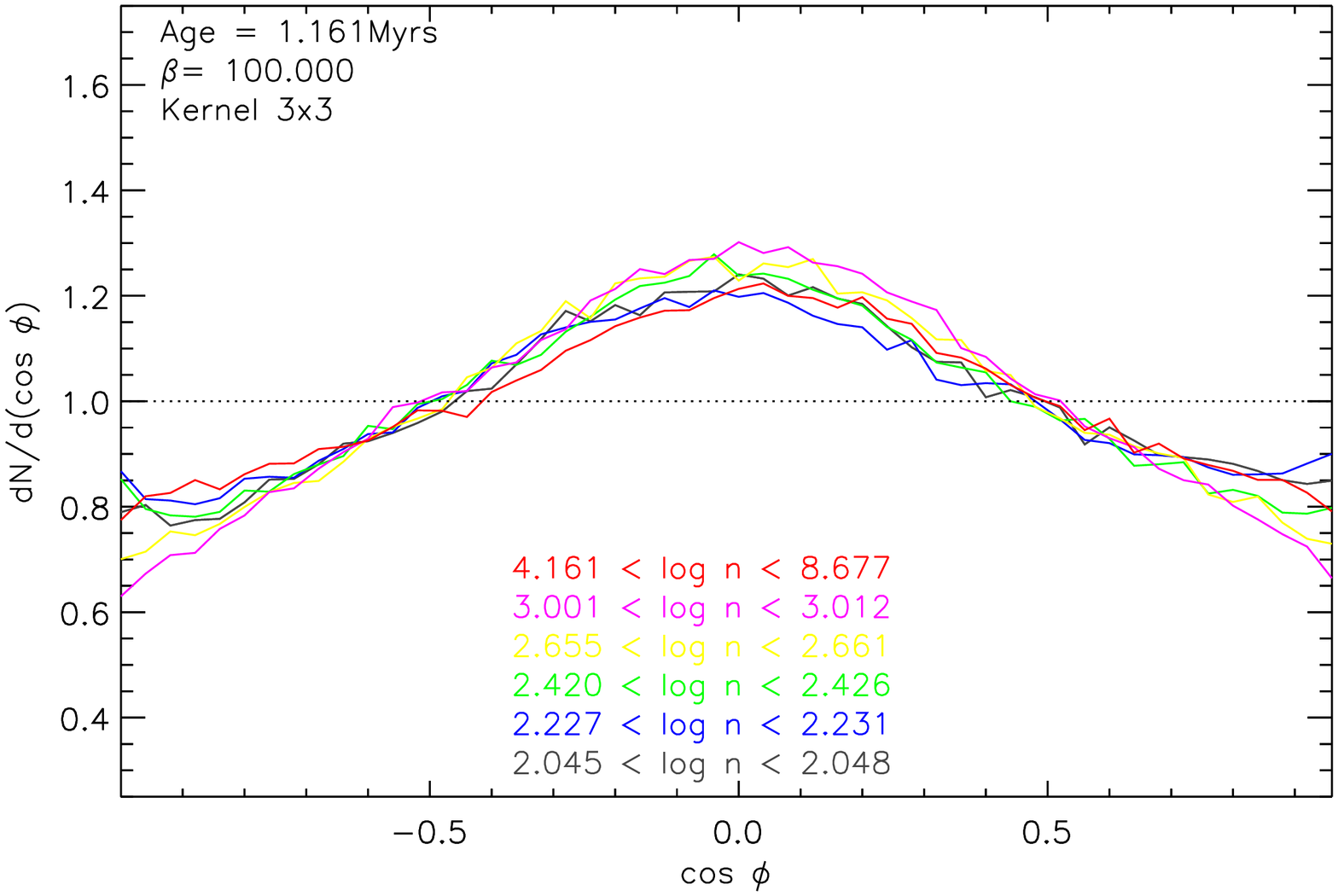}\\
\includegraphics[angle=270,width=0.46\linewidth]{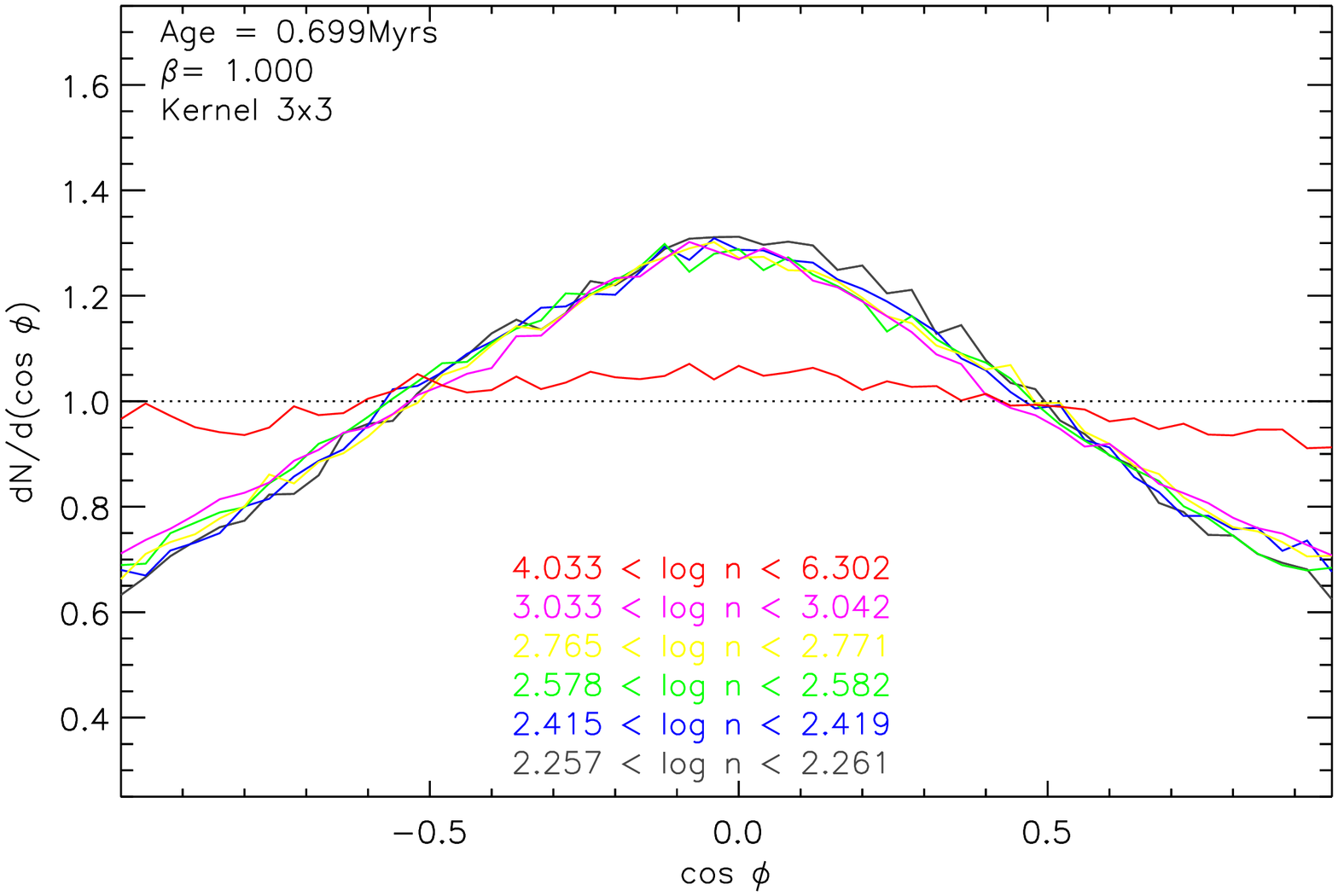}
\includegraphics[angle=270,width=0.46\linewidth]{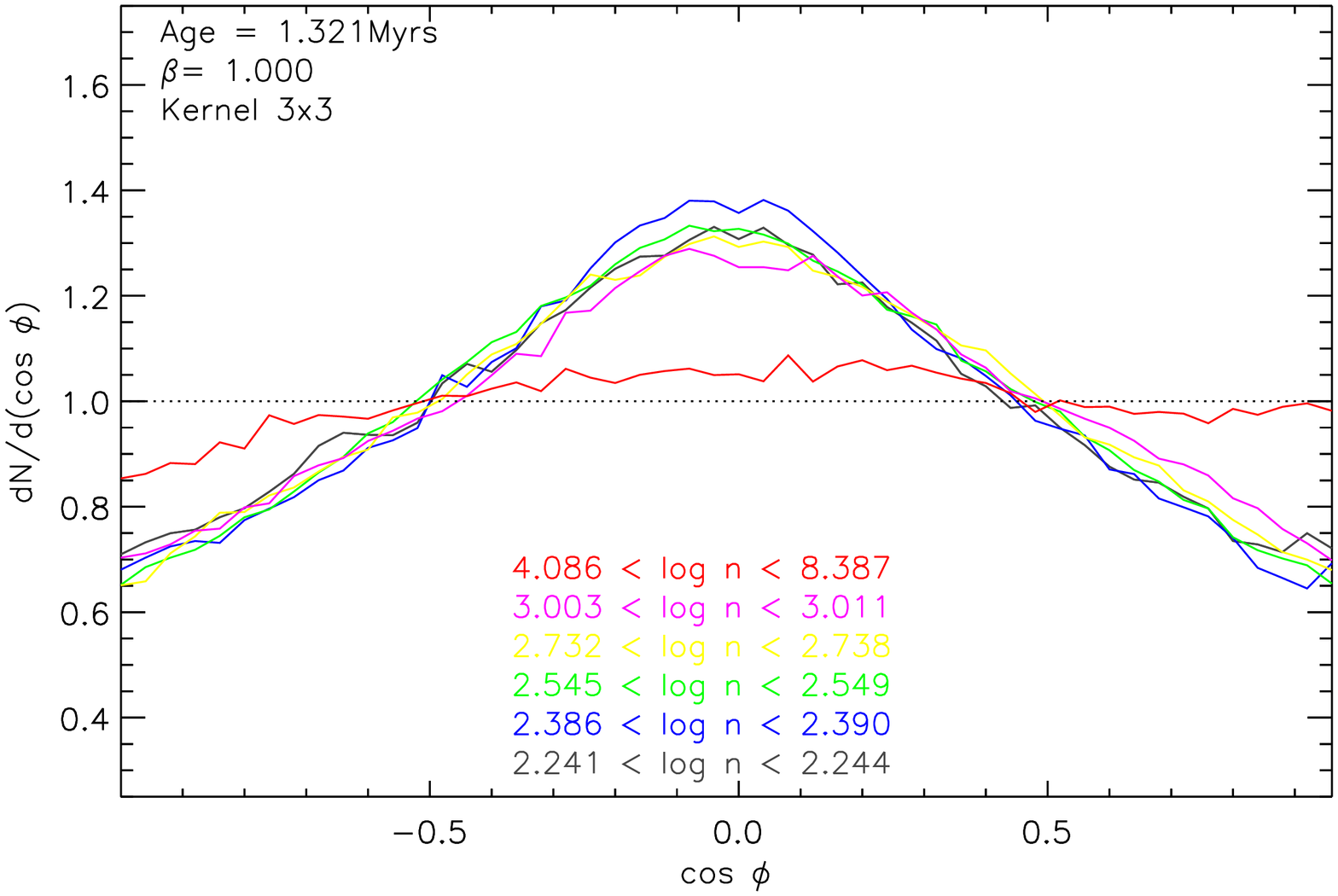}\\
\includegraphics[angle=270,width=0.46\linewidth]{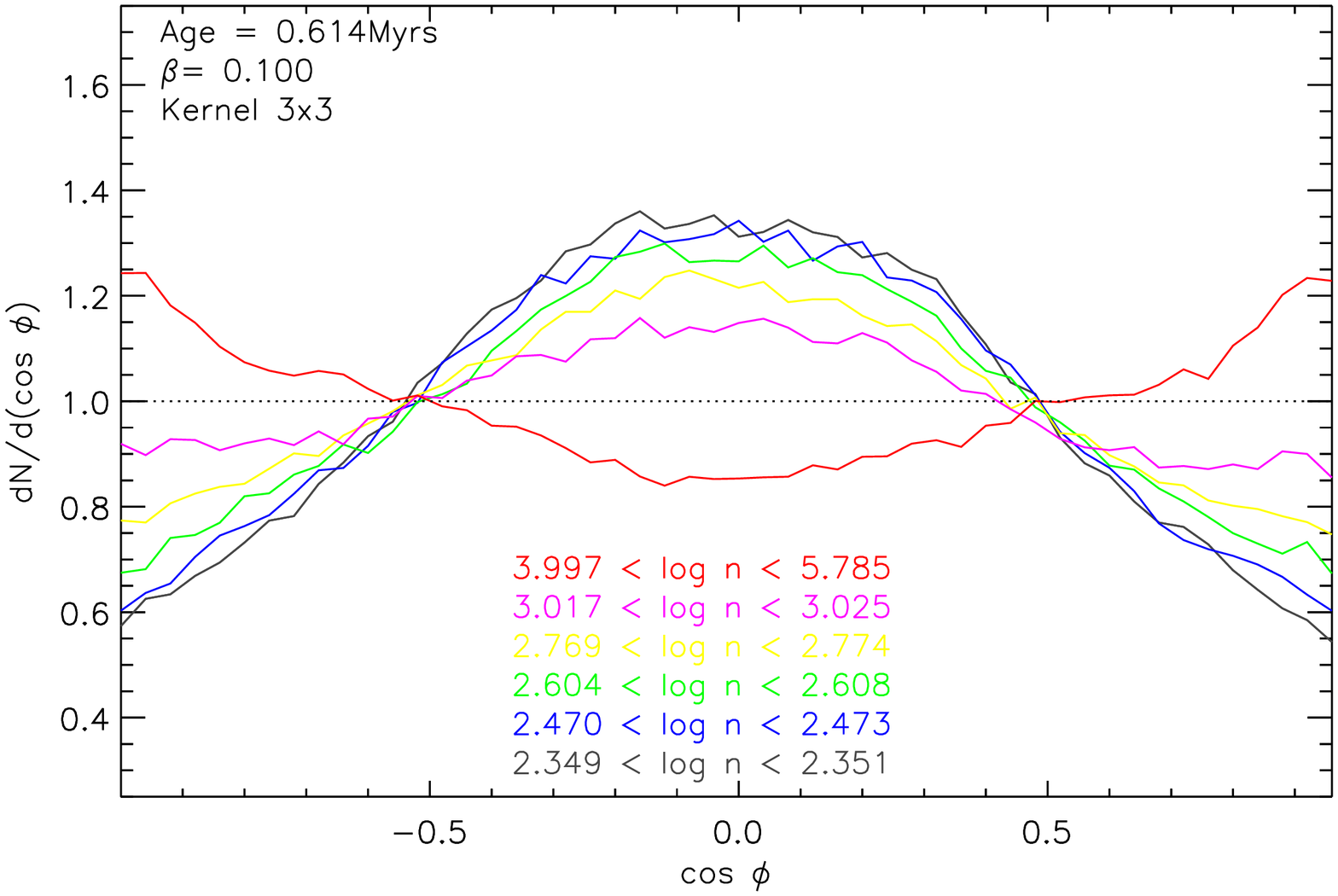}
\includegraphics[angle=270,width=0.46\linewidth]{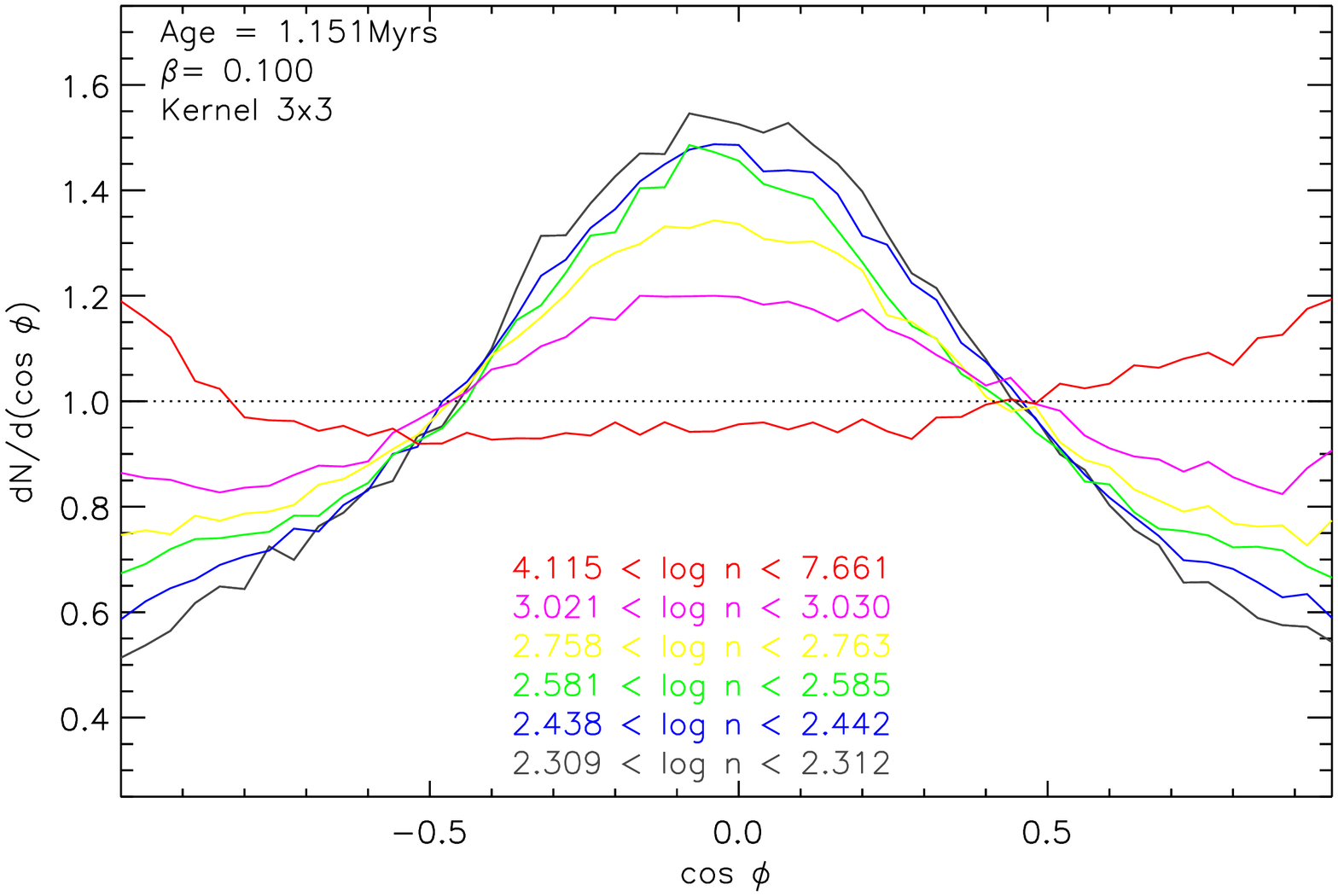}
\caption[HROs of Simulation Cubes]{HROs corresponding to simulated cubes with $\beta$=100.0 (top), 1.0 (middle), and 0.1 (bottom) in snapshots taken at $t\sim0.03\, t_{v}$ (left) and $t\sim0.06\, t_{v}$ (right). The colored curves within each plot correspond to voxels in the density ranges indicated in the figure. The histograms in the low magnetization case (top) peak at $\cos\phi\sim 0$ in regions with densities $n \geq\bar{n}$ which corresponds to the magnetic field ($\mathbf{B}$) predominantly tracing the isodensity contours even at the greatest densities. The histograms from the intermediate (middle) and high magnetization (bottom) cases also peak at $\cos\phi\sim 0$ in regions with densities $n \sim\bar{n}$. However, the histograms flatten in higher density regions. At the highest densities the histograms peak at $\cos\phi\sim\pm 1$. This corresponds to $\mathbf{B}$ tracing the isodensity contours in regions with $n \sim\bar{n}$, then showing no particular relative orientation in intermediate density regions, and being oriented perpendicular to the isodensity contours in the highest density regions.}\label{HOGsegments3D}
\end{figure*}

The colored curves in Figure \ref{HOGsegments3D} correspond to the HRO of regions in a particular density bin. The red and magenta curves correspond to regions with the highest densities, the gray and blue curves to regions with densities close to the mean density of the cube ($\log \bar{n} = 2.73$) and all others to intermediate densities. The segmentation of the cube in these density bins is described in Section \ref{Segmentation}.

Figure \ref{HOGsegments3D} illustrates the difference in the relative orientation of $\mathbf{\nabla}n$ and $\mathbf{B}$ in low and high density regions in simulations with low and high magnetization. In the low and intermediate magnetization cases ($\beta=100$ and $\beta=1$ ), $\mathbf{\nabla}n$ and $\mathbf{B}$ are preferentially perpendicular to each other, which corresponds to $\mathbf{B}$ being parallel to the isodensity contours. In the case of greater magnetization ($\beta=0.1$), $\mathbf{\nabla}n$ is mostly perpendicular to $\mathbf{B}$ at densities close to the mean but the relative orientation progressively changes when considering regions with greater densities until $\mathbf{\nabla}n$ and $\mathbf{B}$ are predominantly parallel to each other.

With this progressive change in relative orientation, the HRO curve changes from convex to concave. To quantify this behavior we define the histogram shape parameter:
\begin{equation}
 \zeta \equiv A_{\mbox{c}} - A_{\mbox{e}}.
\end{equation}\label{zeta}
$A_{\mbox{c}}$ is the area under the central region of the HRO curve ($-0.25 < \cos\phi < 0.25$) and $A_{\mbox{e}}$ is the area in the extremes of the HRO curve ($-1.00 < \cos\phi < -0.75 $ and $0.75 < \cos\phi < 1.00 $). This parameter characterizes a curve peaking at $\cos\phi\sim0.0$ (convex) as $\zeta>0.0$ whereas a curve peaking at $\cos\phi\sim\pm1.0$ (concave) corresponds to $\zeta<0.0$ and a flat distribution corresponds to $\zeta\sim 0$. The uncertainty in the determination of $\zeta$ is related to the standard deviation around the calculated area in each region, i.e.,
\begin{equation}
\sigma_{\zeta} \equiv \sqrt{ \sigma^{2}_{A_{c}} + \sigma^{2}_{A_{e}} }.
\end{equation}
With these definitions, a noisy HRO will either produce $\zeta\sim 0$ which is consistent with a random relative orientation or will increase $\sigma_{\zeta}$ to a level in which it describes no preferential orientation.

\begin{figure*}
\begin{center}
    \includegraphics[angle=270,width=0.90\linewidth]{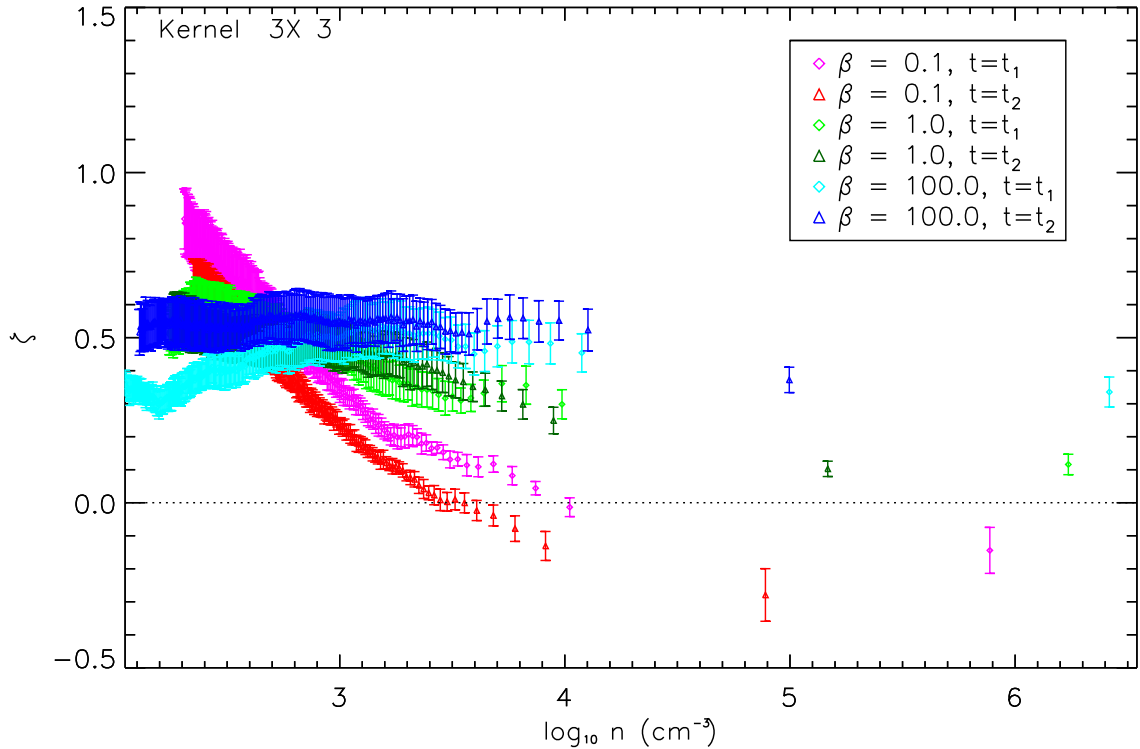}
    \caption[HRO Shape Parameter as a Function of Density]{HRO shape parameter $\zeta$ which parameterizes the relative orientation of the magnetic field ($\mathbf{B}$) and the gradient of the density $\mathbf{\nabla} n$. $\zeta > 0.0$ corresponds to a HRO showing $\mathbf{B}$ predominantly perpendicular to $\mathbf{\nabla} n$ ($\mathbf{B}$ parallel to the isodensity contours). $\zeta\sim0.0$ corresponds to a flat HRO showing no predominant relative orientation between $\mathbf{B}$ and $\mathbf{\nabla} n$. $\zeta<0.0$ corresponds to a HRO showing $\mathbf{B}$ predominantly parallel to $\mathbf{\nabla} n$ ($\mathbf{B}$ perpendicular to the isodensity contours). The HROs of the low magnetization case show $\mathbf{B}$ predominantly parallel to the isodensity contours with $\zeta>0.0$ even in the higher density regions. In contrast, the HROs of the high magnetization cases show $\mathbf{B}$ parallel to the isodensity contours with $\zeta>0.0$ in the low density regions, changing to $\mathbf{B}$ perpendicular to the isodensity contours, with $\zeta<0.0$ in the highest density regions. The shape parameter $\zeta$ as a function of density decreases faster in higher magnetization simulations.}\label{HOG3Dturnover}
\end{center}
\end{figure*}

Figure \ref{HOG3Dturnover} shows the value of $\zeta$ for the different column density segments in all six simulations. The red curve illustrates the progressive change of the relative orientation of $\mathbf{\nabla}n$ and $\mathbf{B}$ with density in the simulations with the greatest magnetization: it changes from $\mathbf{B}$ preferentially parallel to the isodensity contours ($\zeta>0$) for densities close to the mean ($\log n$~[cm$^{-3}$] $\sim 2.7$) to $\mathbf{B}$ perpendicular to the isodensity contours ($\zeta<0$) in the regions with densities higher than about $50\bar{n}$. The curves corresponding to low and intermediate magnetization also show a similar change in the relative orientation of $\mathbf{\nabla}n$ and $\mathbf{B}$ with density although it is less pronounced. Thus we find that the slope of $\zeta$ as a function of the mean density in the density bin is a parameter which characterizes the initial magnetization state of the simulated cloud.

Additionally, we find that $\zeta$ becomes negative ($\mathbf{B}$ is preferentially perpendicular to the isodensity contours) at the highest densities in the highest magnetization case. The value of the threshold density, $n_{T}$, above which $\mathbf{\nabla}n$ and $\mathbf{B}$ are predominantly parallel ($\zeta$ becomes negative) is also a parameter which characterizes the magnetization state of the simulated cloud: higher magnetization corresponds to lower values of $n_{T}$.

% 2D HOGS =========================================================================================================
\section{HRO Applied To Observations of Simulation Cubes}\label{hro:HOG2D}

The HRO analysis is applied to the maps of the column density $\Sigma$ and polarization obtained by integrating the six simulated data cubes along three different lines of sight (x, y, or z-axis). The projection of the Stokes parameters into the plane of the sky incorporates a simple grain alignment efficiency model. This model uses the density in each voxel as a proxy for what might actually be a column density dependence in grain alignment mechanism and intends to illustrate the results of HROs in projections obtained with different dust grain alignment efficiencies $\epsilon$. A detailed treatment of grain alignment by anisotropic radiation flux with respect to the magnetic field, such as in the radiative torques (RATs) mechanism \citep{hoang2008}, requires ray tracing studies which are beyond the scope of this work.

The observed intensity of the polarized dust emission results from the combined effect of $\epsilon$ and integration along the line of sight. The maps of the Stokes parameters I, Q, and U are produced by integrating the density $n$ and the magnetic field $\mathbf{B}$ in the cube along the k-direction (either x, y or z) according to:%
\begin{eqnarray}\label{ProjectionEquation}
\nonumber I_{ij} &\propto& \Sigma_{ij} = \sum_{k}n_{ijk}\\
Q_{ij} &\propto& \sum_{k} \epsilon_{ijk}n_{ijk}(B^{(i)}_{ijk}B^{(i)}_{ijk}-B^{(j)}_{ijk}B^{(j)}_{ijk})\\
\nonumber U_{ij} &\propto& \sum_{k} 2\epsilon_{ijk}n_{ijk}B^{(i)}_{ijk}B^{(j)}_{ijk}
\end{eqnarray}
This model is based on the integration of the Stokes parameters used in previous studies \citep{martin1974, lee1985}. The projected magnetic field $\mathbf{B}_{POS}$ is calculated from the projected Stokes parameters using:
\begin{eqnarray}\label{BProjectionEquation}
|\mathbf{B}_{POS}| &=& \sqrt{Q^{2} + U^{2}}, \\
\psi_{B} &=& \left(\frac{1}{2}\right)\arctan(U/Q).
\end{eqnarray}

For each cell of the simulated data cube we evaluate:%
\begin{equation}\label{AlignmentEfficiency}
\epsilon_{ijk} = \left\{ \begin{array}{ll} (n_{ijk})^{p}  & \mbox{if } n_{ijk} \geq n_{0}\\
 		1.0 & \mbox{if } n_{ijk} < n_{0}
\end{array}\right.
\end{equation}%
In this toy-model we have chosen $n_{0}=500$~cm$^{-3}$, corresponding to a relatively diffuse region within a molecular cloud. The case with $p=0$ is used to test the line of sight integration independently of the environment dependence of $\epsilon$. The case with $p<0$ simulates depolarization effects in high column density regions \citep{matthews2001II, houde2002}. It also accounts to zeroth-order for the more efficient grain alignment in regions of relatively low extinction as suggested by RATs. Similar models assuming polarizability in each volume element proportional to the local density have been previously studied \citep{ostriker2001, padoan2001}.

We compare results for uniform alignment efficiency ($\epsilon = 1$) with the results of a model with $\epsilon = 1$ at $n < n_0$ but decreasing $\epsilon$ with increasing $n$ in regions with $n > n_0$. This intends to contrast the results of line of sight integration with a particular environmental dependence of the alignment mechanism. The study of the projection effects produced by a particular alignment mechanism requires detailed analysis of the radiative environment which are beyond the scope of this work.

The result of the integration of the data cubes is a map of column density $\Sigma$ and a weighted projection of the magnetic field, a pseudovector that we call $\mathbf{B}_{POS}$. The HROs of the simulated maps are calculated using a process analogous to that used in the 3D data cubes. The histogram of the angle $\phi$ between two sets of random vectors in 2D is uniform, and therefore we choose $\phi$ as the variable.

\begin{figure}
\begin{center}
\includegraphics[angle=270,width=0.99\linewidth]{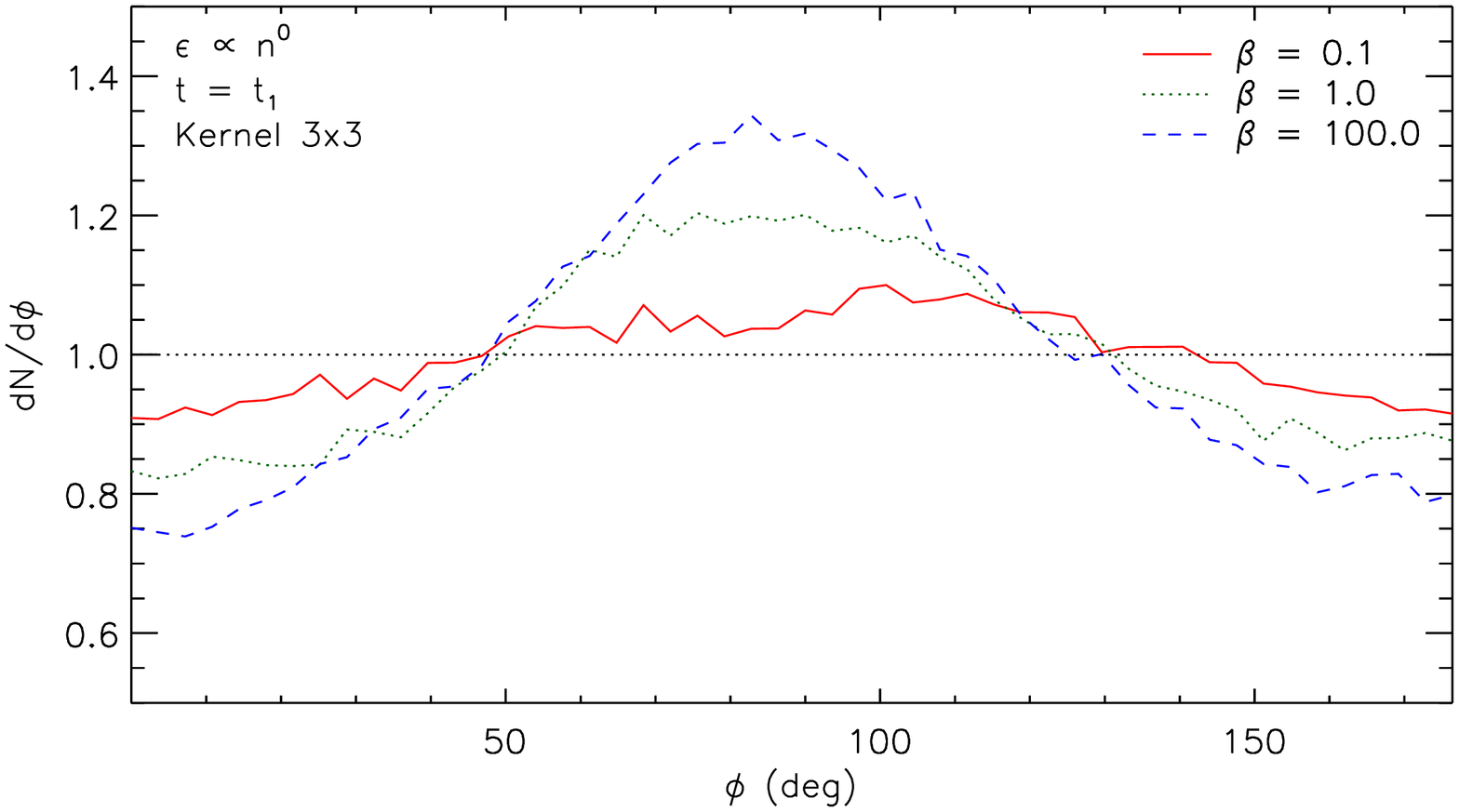}\\
\includegraphics[angle=270,width=0.99\linewidth]{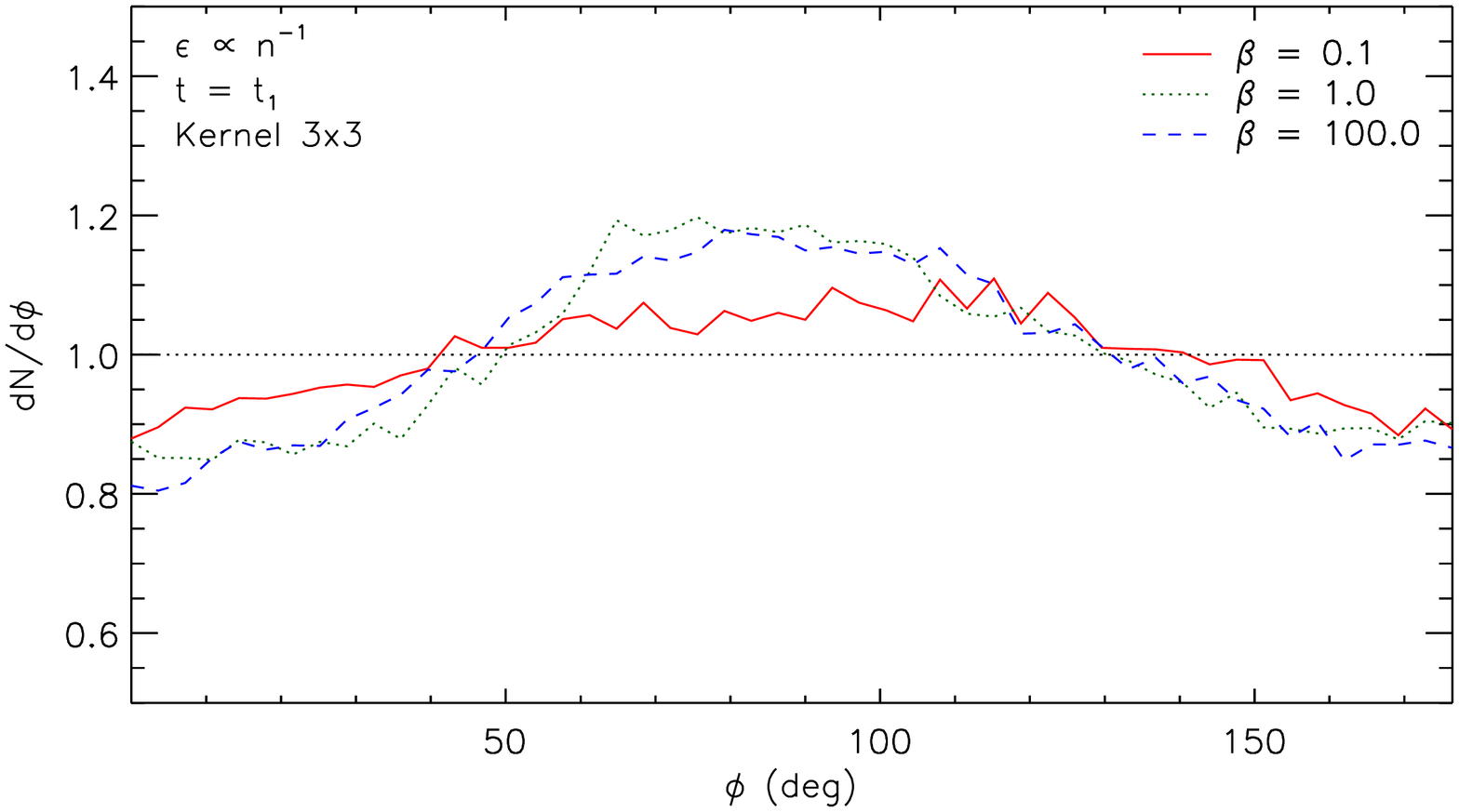}
\caption[HRO of Projected Maps]{HROs showing the angle between the projected magnetic field pseudovector $\mathbf{B}_{POS}$ and the gradient of the column density $\mathbf{\nabla} \Sigma$ in projections of the low, intermediate, and high-magnetization simulation cubes ($\beta = 100$, $1.0$, and $0.1$) in a snapshot taken at $t\sim0.06\, t_{v}$ and using grain alignment efficiency from Equation \ref{AlignmentEfficiency} with $p=$ (top) and $p=-1$ (bottom). The projections are obtained following Equation \ref{ProjectionEquation}. The histogram is normalized such that a random distribution of $\mathbf{B}_{POS}$ and $\mathbf{\nabla} \Sigma$ would equal unity in each bin (black dotted line). The histograms calculated from the simulated cubes show a peak at $\phi\sim 90$\degr\ which corresponds to the magnetic fields predominantly tracing iso-$\Sigma$ contours.}\label{HOG2Dmulti}
\end{center}
\end{figure}

\begin{figure*}
\begin{center}
    \includegraphics[width=0.42\linewidth,angle=0]{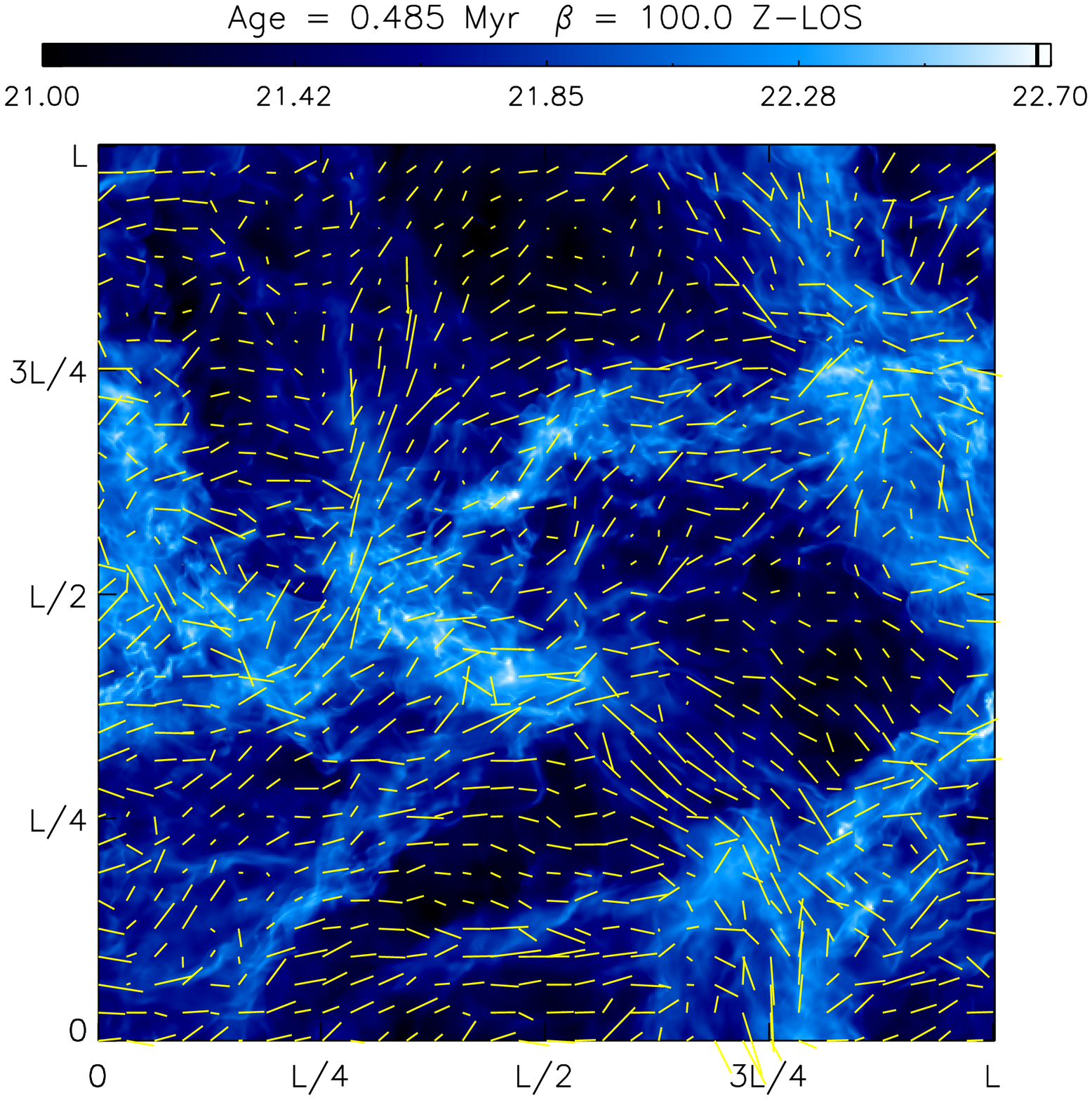}
    \includegraphics[width=0.42\linewidth,angle=0]{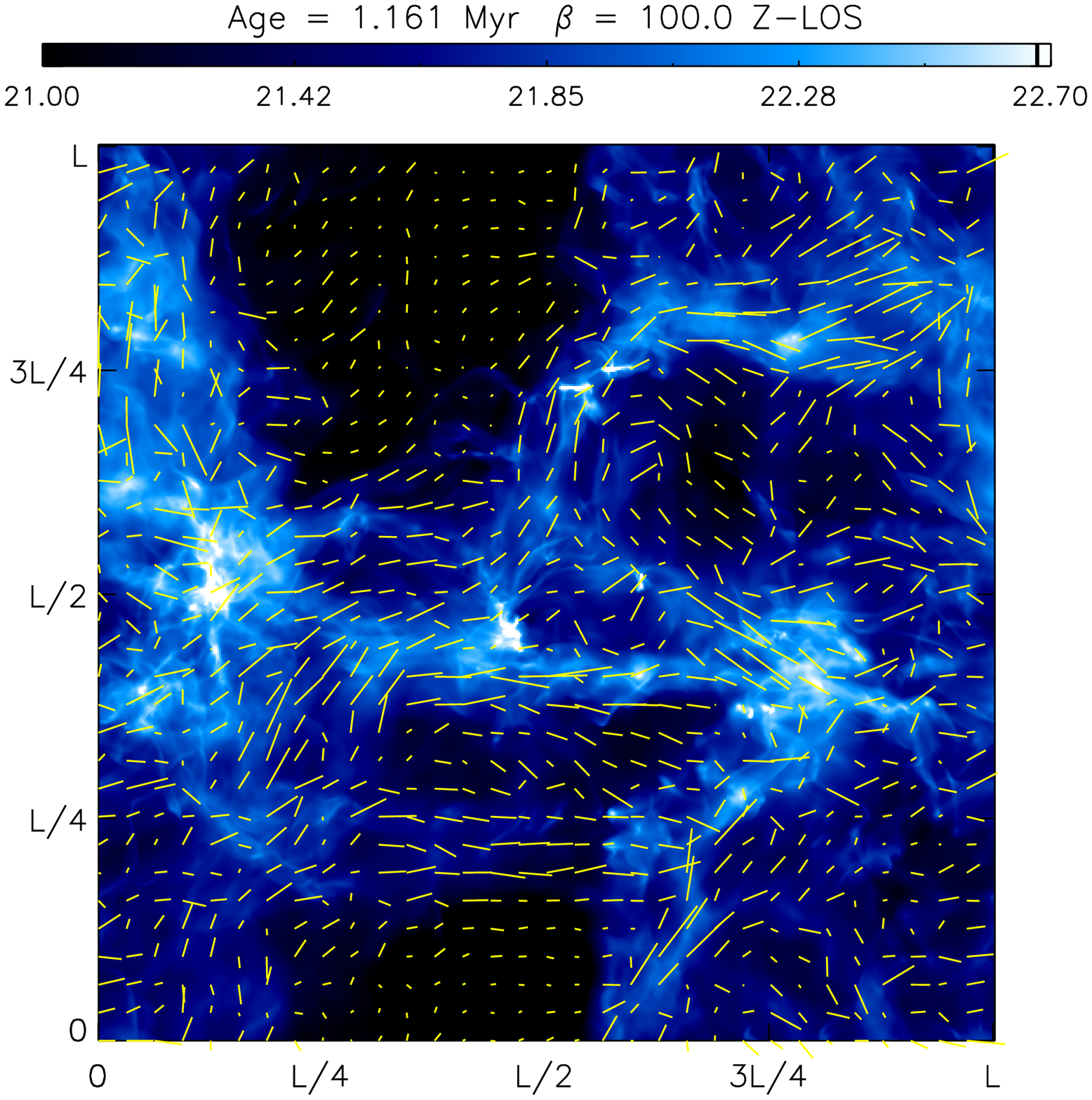}\\
    \includegraphics[width=0.42\linewidth,angle=0]{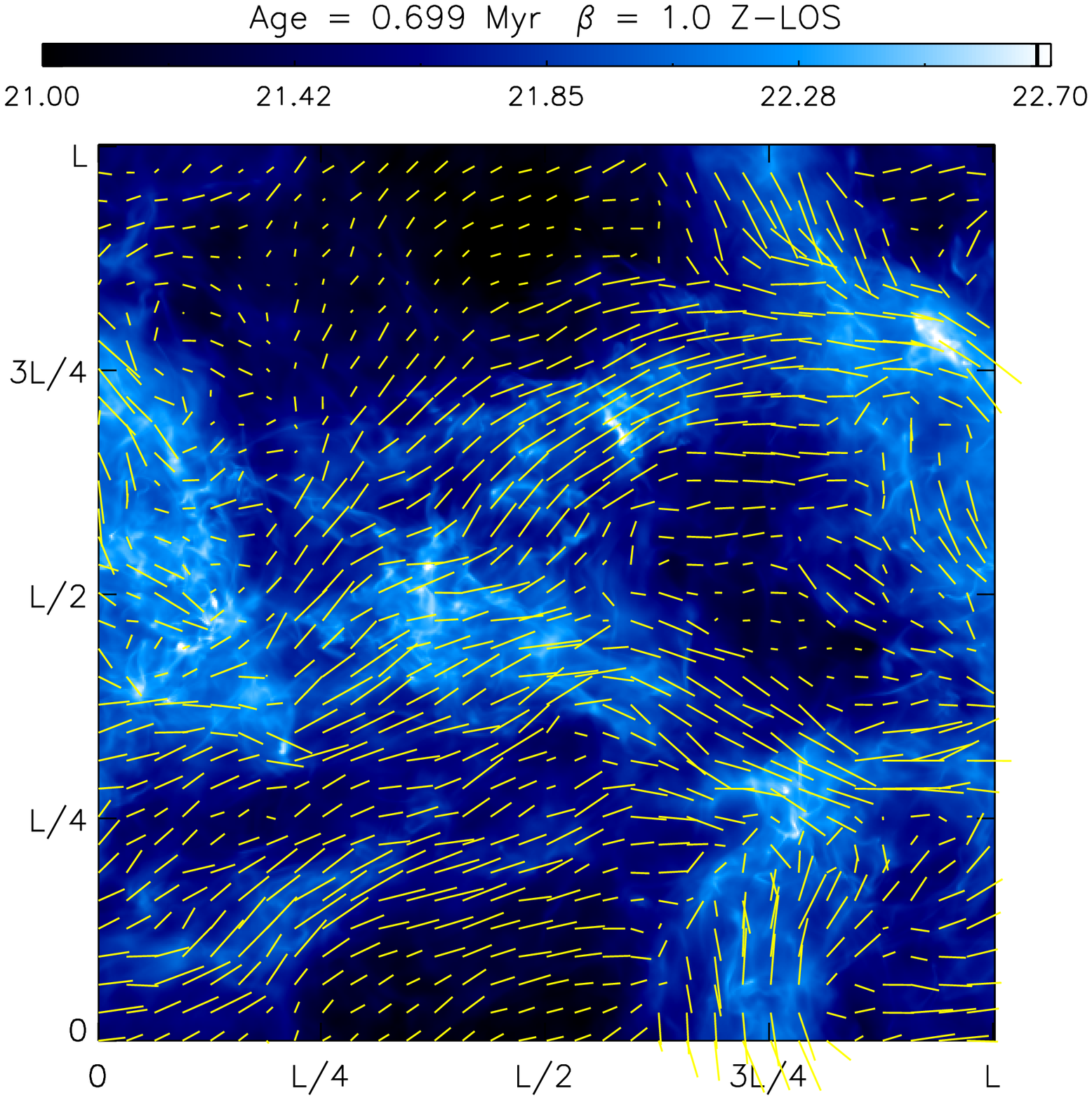}
    \includegraphics[width=0.42\linewidth,angle=0]{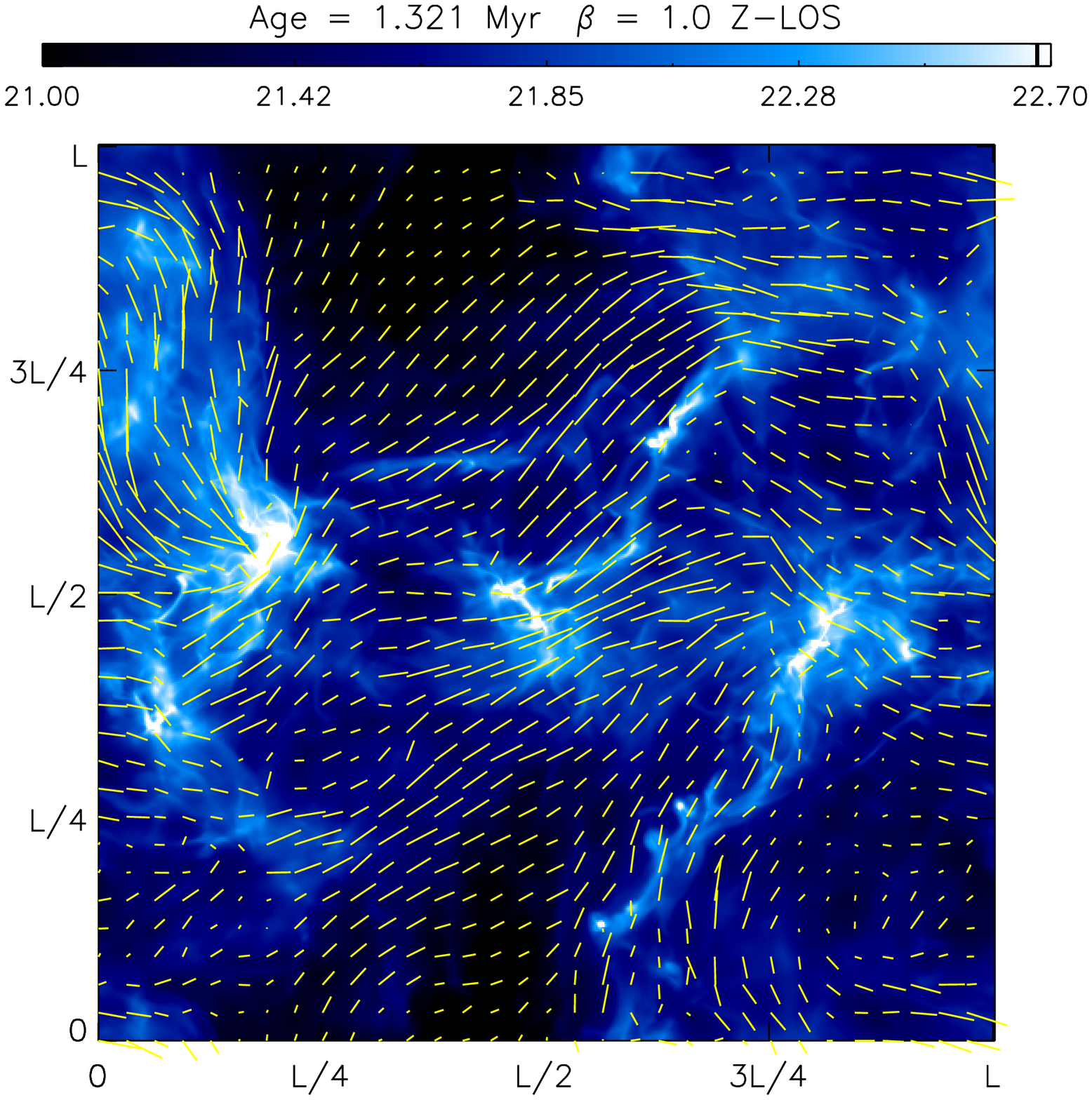}\\
    \includegraphics[width=0.42\linewidth,angle=0]{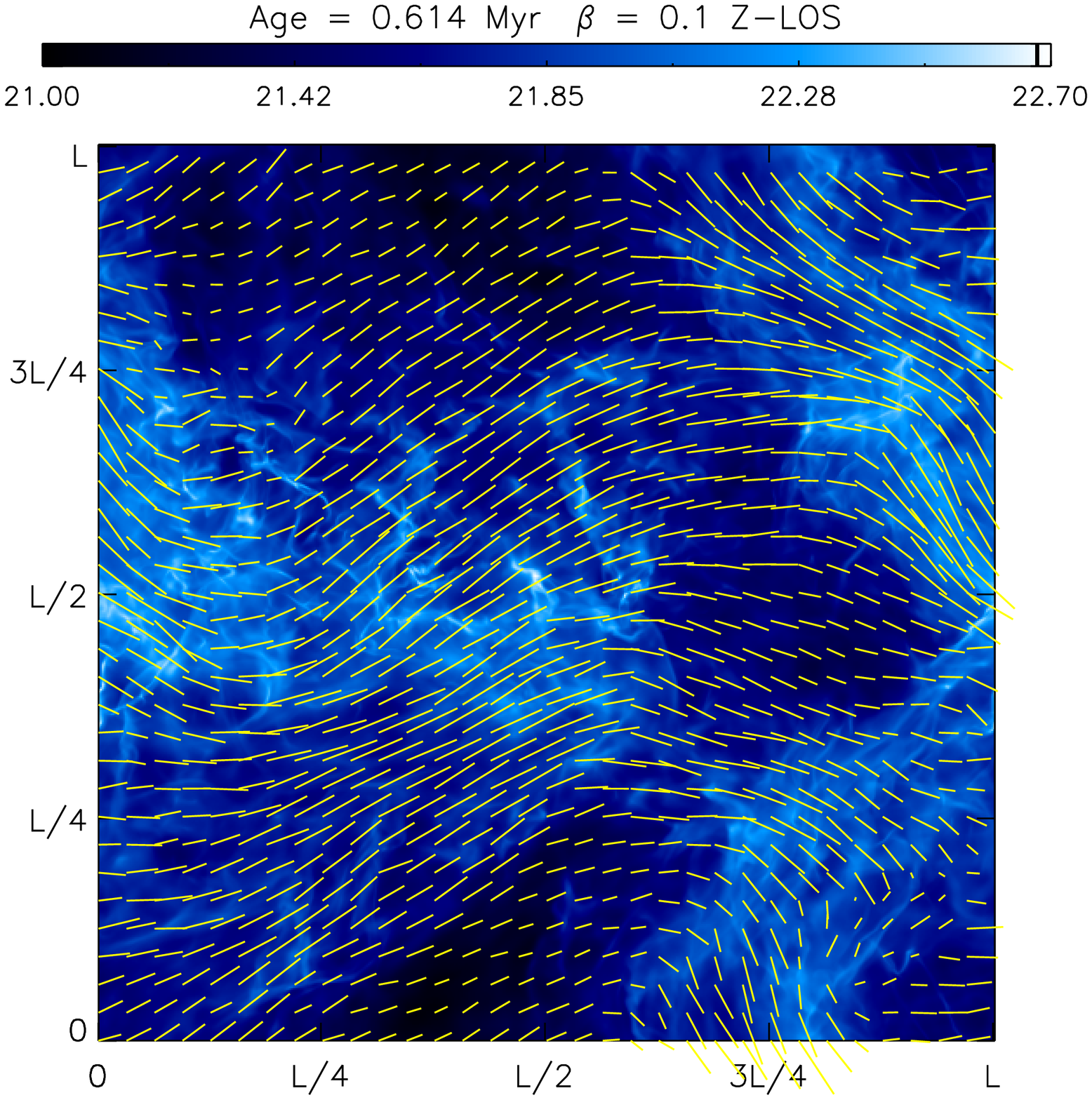}
    \includegraphics[width=0.42\linewidth,angle=0]{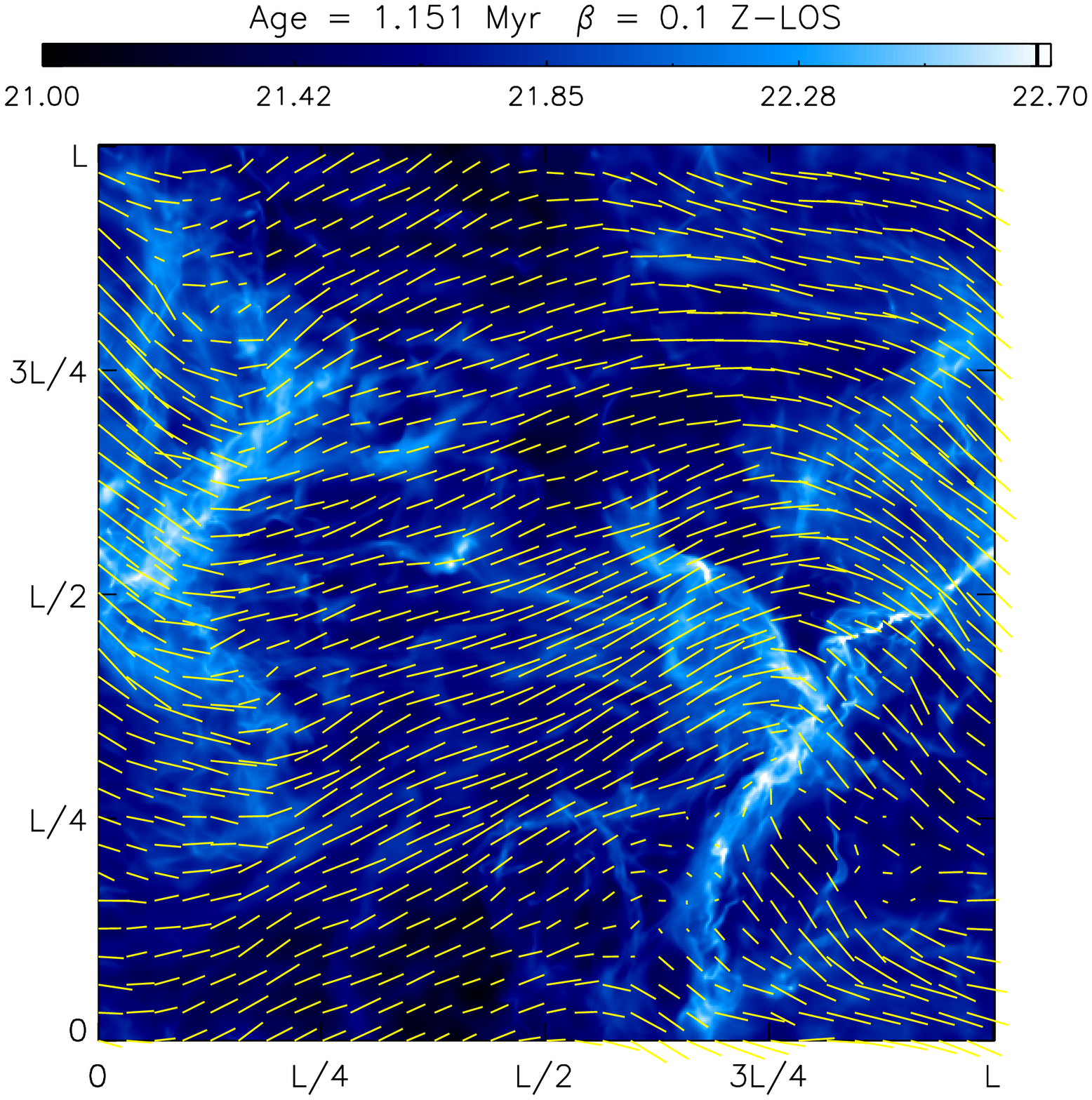}
  \caption[Projected Magnetic Field and Column Density Maps (Z-LOS)]{Maps of the logarithm of the column density ($\log \Sigma$) and overlaid magnetic field pseudovectors as determined by Equations \ref{ProjectionEquation} and \ref{AlignmentEfficiency} with $p=-1$ and $n_{0}=500$~cm$^{-3}$. These maps correspond to projections along the Z-axis of simulations with $\beta=$ 100 (top), 1.0 (middle), and 0.1 (bottom) in snapshots taken at $t\sim0.03\, t_{v}$ (left) and $t\sim0.06\, t_{v}$ (right).}\label{BImap}
\end{center}
\end{figure*}

Figure \ref{HOG2Dmulti} shows the HROs corresponding to the angle $\phi$ between the projected magnetic field pseudovector $\mathbf{B}_{POS}$ and the gradient of the column density $\mathbf{\nabla} \Sigma$ in the projected maps. The peak at 90\degr\ reveals that the relative orientation between $\mathbf{\nabla} n$ and $\mathbf{B}$ observed in the 3D (Figure \ref{HOG3Dmulti}) is also present in the polarization and column density maps and $\mathbf{B}_{POS}$ predominantly follows the iso-$\Sigma$ contours. This result can be confirmed by visual inspection of the $\mathbf{B}_{POS}$ and $\Sigma$ maps as shown in Figure \ref{BImap}.

Figure \ref{BImap} shows maps of the logarithm of the column density ($\log \Sigma$) and overlaid magnetic field pseudovectors. The orientation of the pseudovectors shows that the magnetic field is less affected by turbulence in the high magnetization case where the direction projected field is more coherent than in the quasi-hydrodynamic case. The amplitude of the projected pseudovectors is also more homogeneous in the high magnetization case: in the quasi-hydrodynamic model the field is weak and prone to bends and changes of direction. The projected map is the result of the integration of multiple orientations perpendicular to the line of sight which cancel each other resulting in shorter pseudovectors. In the high magnetization model, the magnetic field is stronger and therefore more coherent, resulting in the integration of pseudovectors more homogeneously oriented than in the quasi-hydrodynamic case.

\begin{figure*}
\begin{center}
    \includegraphics[angle=270,width=0.495\linewidth]{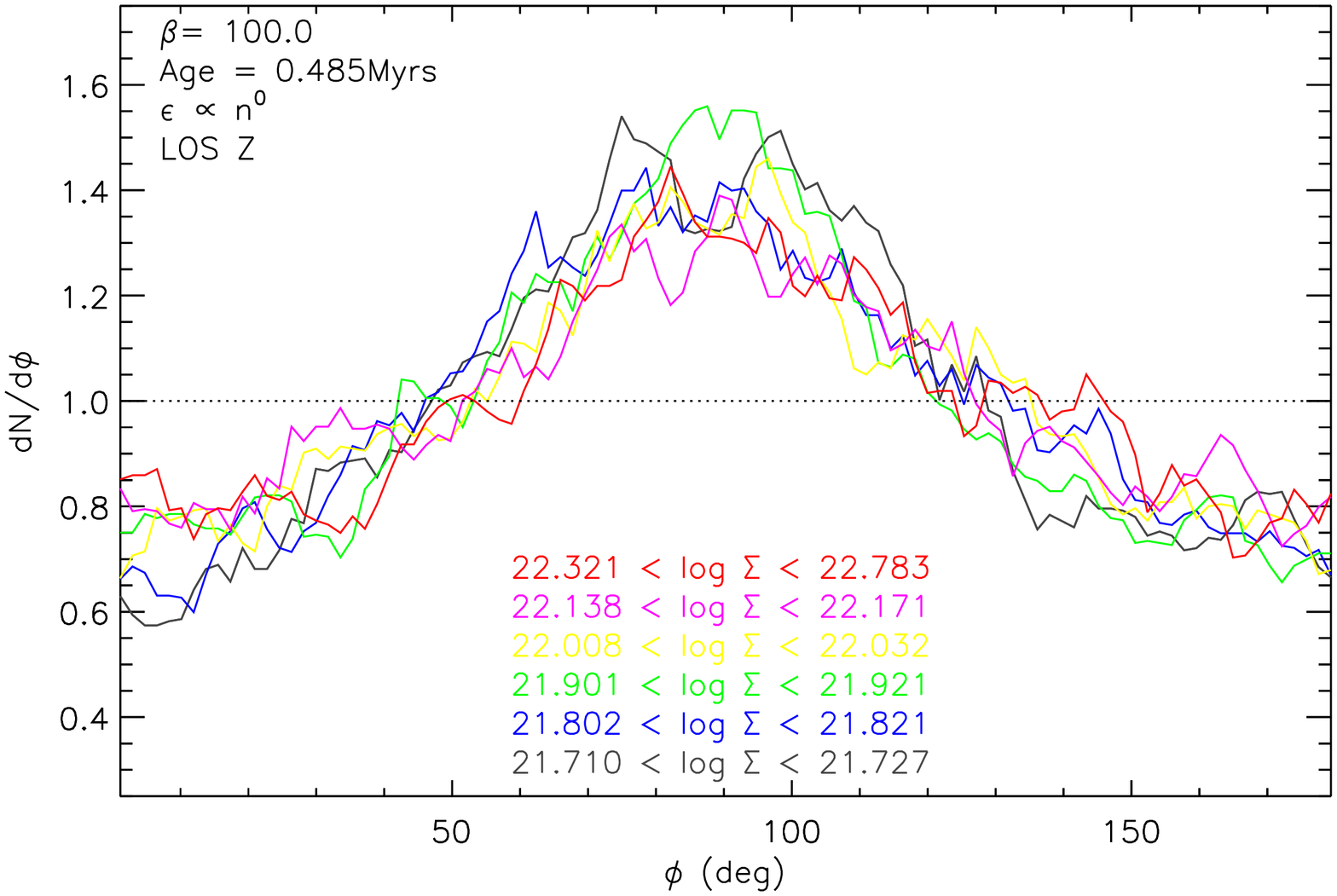}
    \includegraphics[angle=270,width=0.495\linewidth]{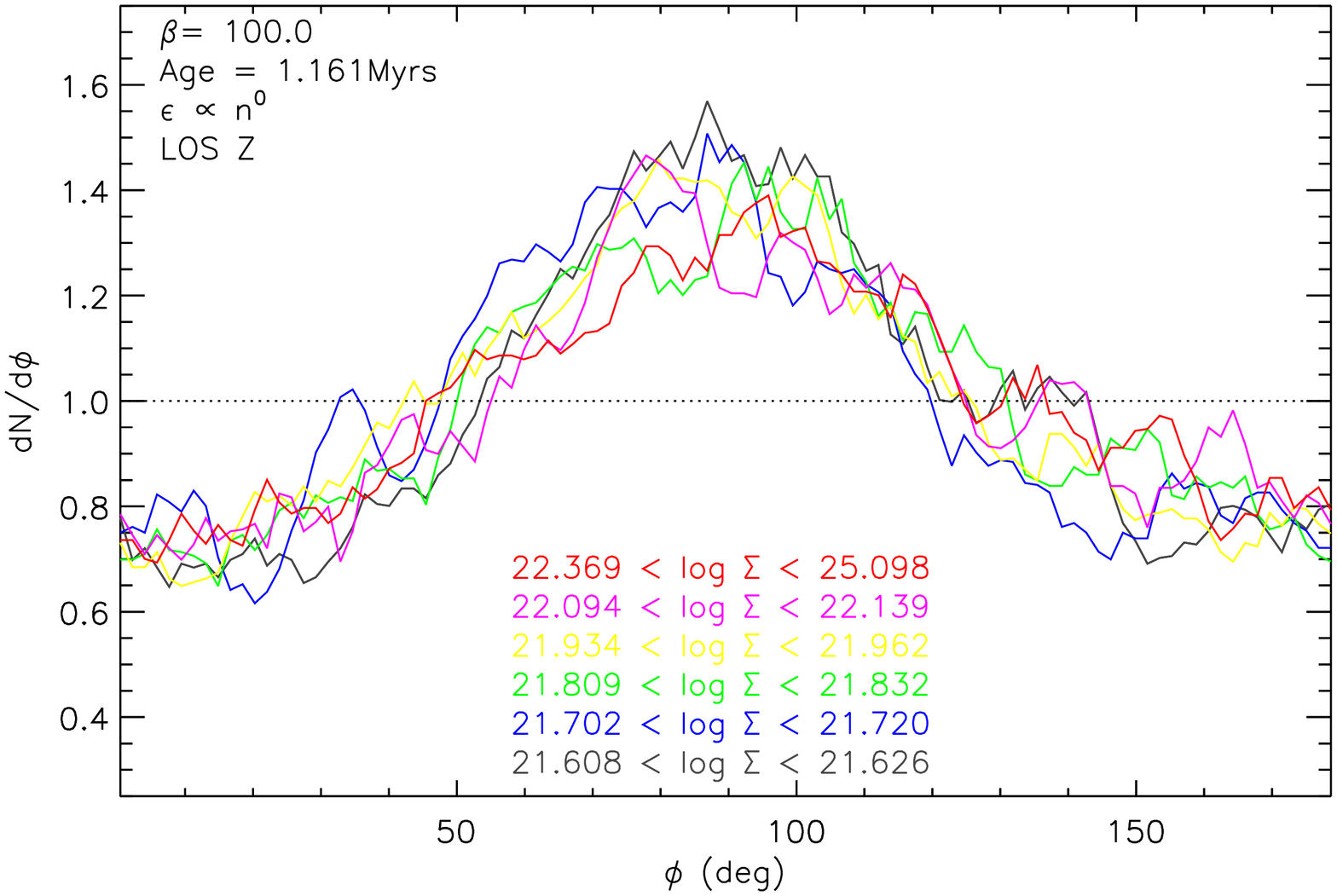}\\
    \includegraphics[angle=270,width=0.495\linewidth]{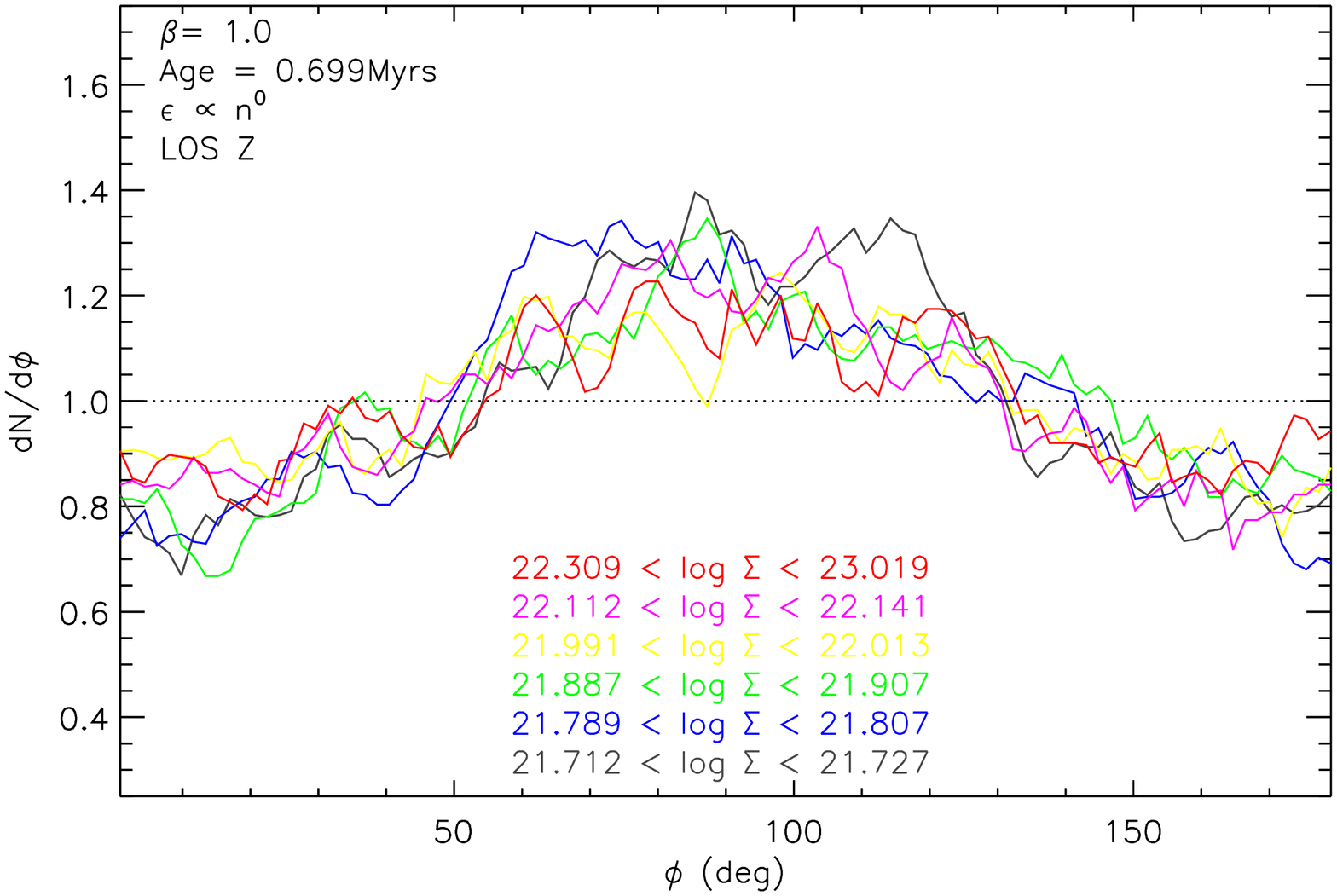}
    \includegraphics[angle=270,width=0.495\linewidth]{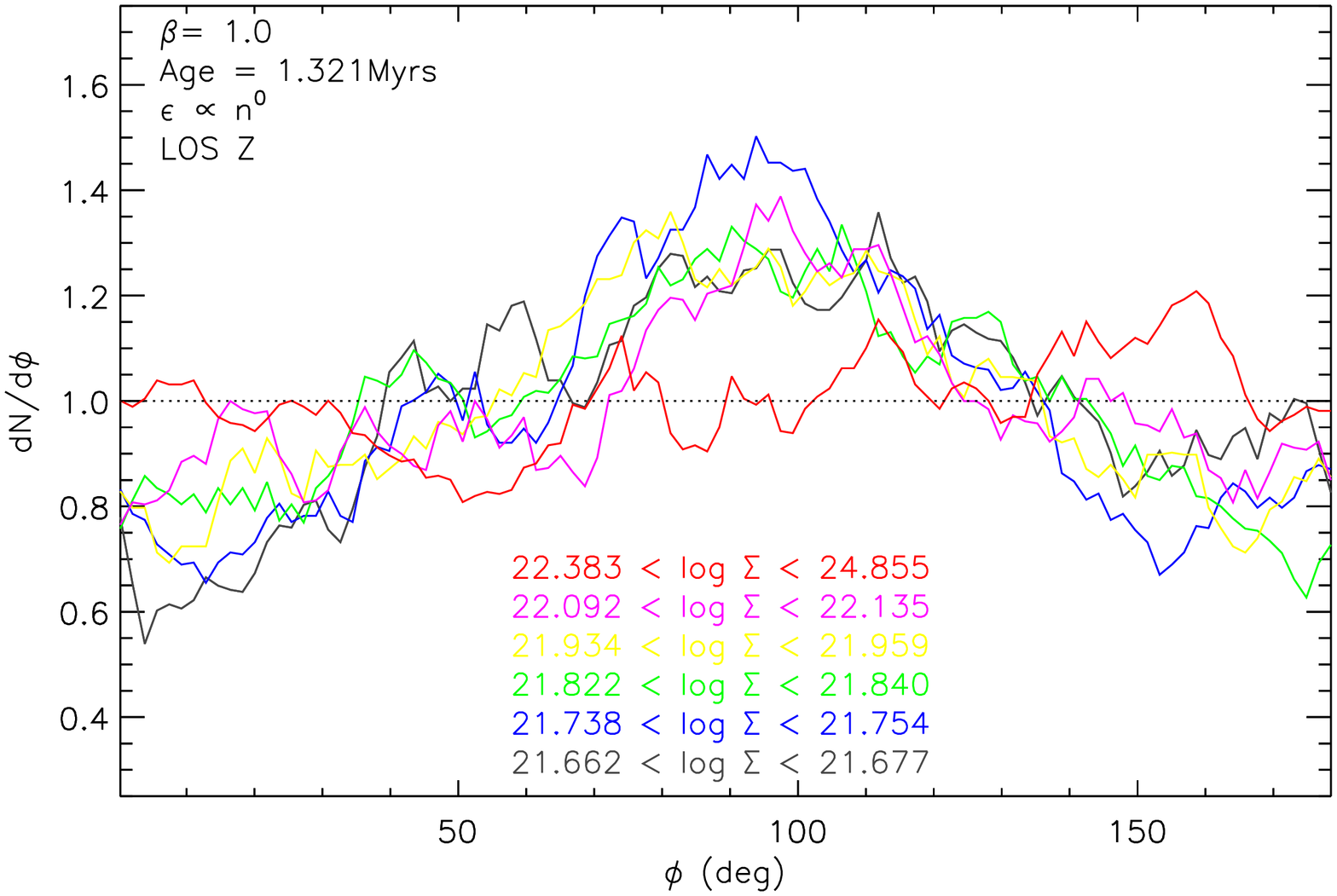}\\
    \includegraphics[angle=270,width=0.495\linewidth]{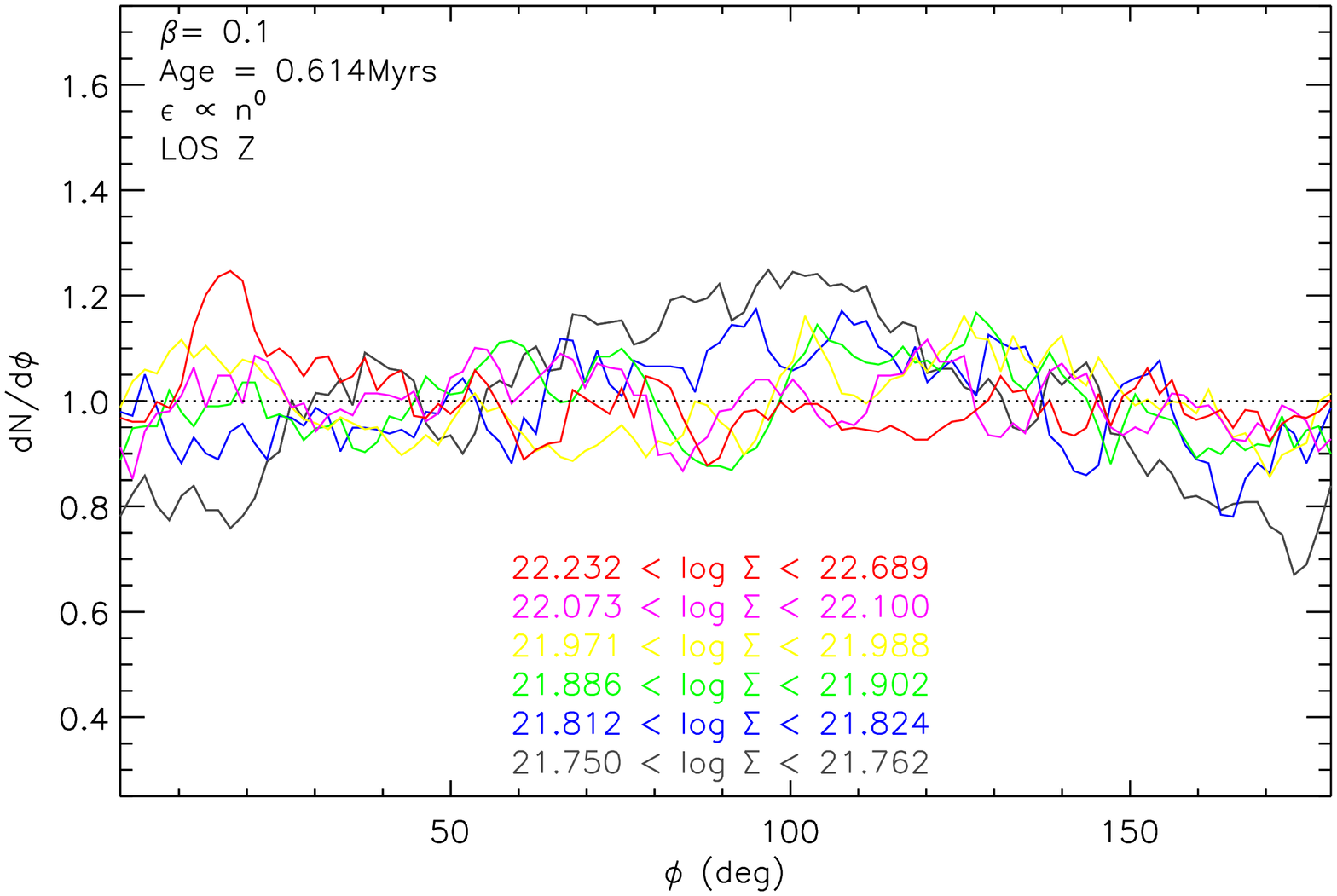}
    \includegraphics[angle=270,width=0.495\linewidth]{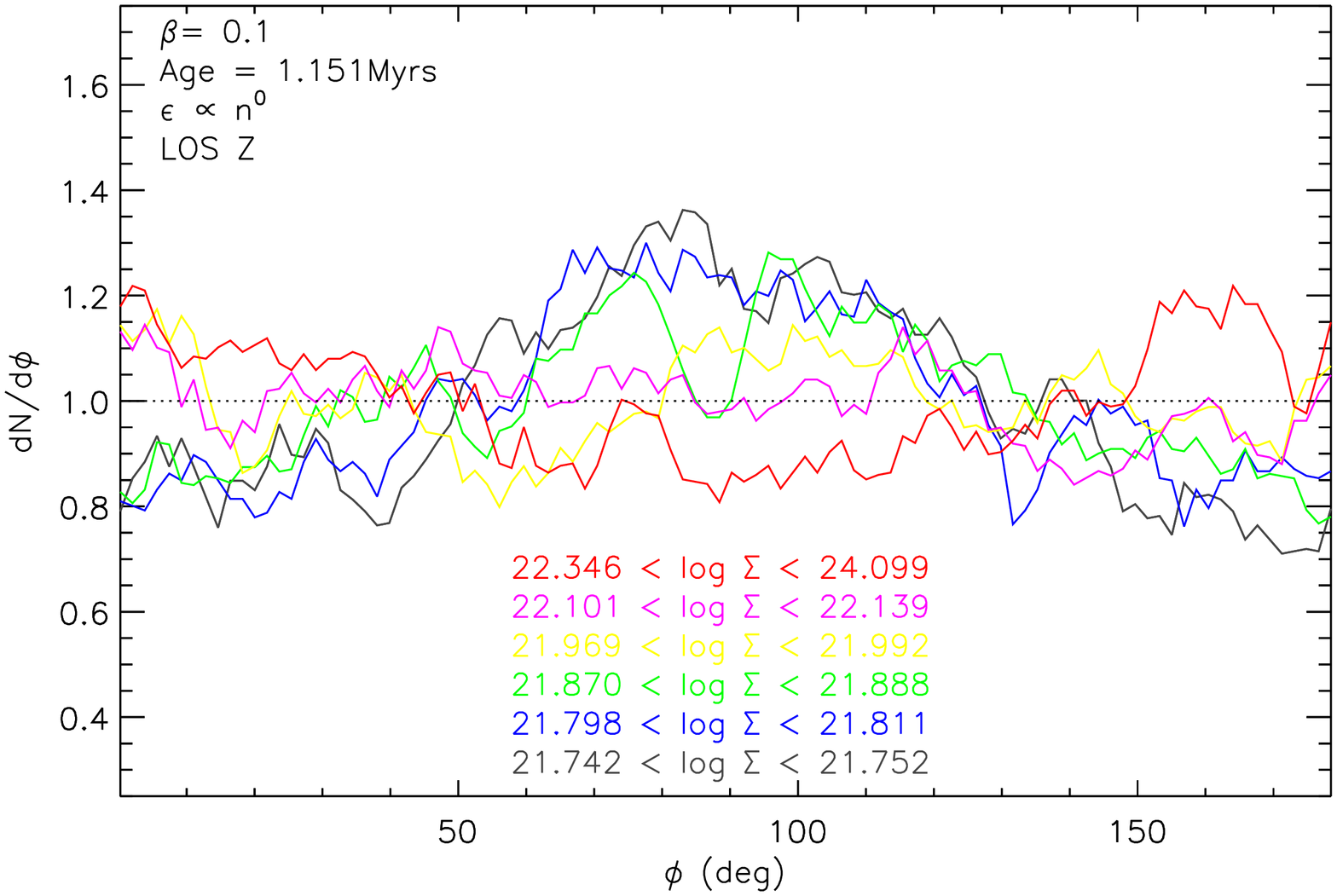}
  \caption[HROs of the Projected Simulations (Z-LOS)]{HROs corresponding to simulated cubes with $\beta$=100.0 (top), 1.0 (middle), and 0.1 (bottom) in snapshots taken at $t\sim0.03\, t_{v}$ (left) and $t\sim0.06\, t_{v}$ (right) projected along the Z-axis. The colored curves within each plot correspond to pixels in the column density ranges indicated in the figure. The HROs from the low magnetization simulation (top) peak at $\phi\sim 90$ in regions with column densities $\Sigma \geq\bar{\Sigma}$ which corresponds to the projected magnetic field ($\mathbf{B}_{POS}$) predominantly tracing the iso-$\Sigma$ contours even at the greatest column densities. The histograms from the intermediate (middle) and high magnetization (bottom) cases also peak at $\phi\sim 90$ in regions with column densities $\Sigma \sim\bar{\Sigma}$. However, the histograms flatten when considering higher $\Sigma$ regions and in the highest densities they peak at $\phi\sim 0$\degr\ or $180$\degr. This corresponds to $\mathbf{B}_{POS}$ tracing the iso-$\Sigma$ contours in regions with $\Sigma \sim\bar{\Sigma}$, then showing no particular relative orientation in intermediate $\Sigma$ regions, and predominately orienting perpendicular to the iso-$\Sigma$ contours at the highest column density regions. As in the 3D analysis, the change of the relative orientation inferred from the histogram in different column density regimes is parameterized by $\zeta$ as defined in Equation \ref{zeta}.}\label{HOG2Dsegments}
\end{center}
\end{figure*}

As in Section \ref{hro:HOG3D}, we divided the maps in column density bins to check how the relative orientation changes in the highest density regions. Figure \ref{HOG2Dsegments} shows the HRO curves corresponding to different column density regimes. The behavior of the HROs in 2D is very similar to that observed in the 3D analysis: $\mathbf{B}_{POS}$ and $\mathbf{\nabla} \Sigma$ are mainly perpendicular in all the $\Sigma$ bins of the low-magnetization map. The HROs of the intermediate and high-magnetization cases show $\mathbf{B}_{POS}$ and $\mathbf{\nabla} \Sigma$ predominantly perpendicular in regions of the map with $\Sigma\sim\bar{\Sigma}$ and parallel in the regions of the map with the greatest $\Sigma$.

In the same way as in the 3D analysis, the change in the HRO curves is parameterized by the histogram shape parameter $\zeta$ defined in Equation \ref{zeta}. In this case $A_{\mbox{c}}$ is the area under the central region of the HRO curve (67.5\degr\ $< \phi <$ 112.5\degr) and $A_{\mbox{e}}$ is the area in the extremes of the HRO curve (0\degr\ $< \phi <$ 22.5\degr\ and 157.5\degr\ $< \phi <$ 180\degr). Both in 3D and 2D, the central region and the extremes are defined based on the range of the HRO curve. The width of the central region is one quarter and the extremes are the first and last eighths of the range. For this reason, the cuts in $\phi$ in 2D are different than in the 3D HRO analysis.

\begin{figure*}
\begin{center}
    \includegraphics[angle=270,width=0.49\linewidth]{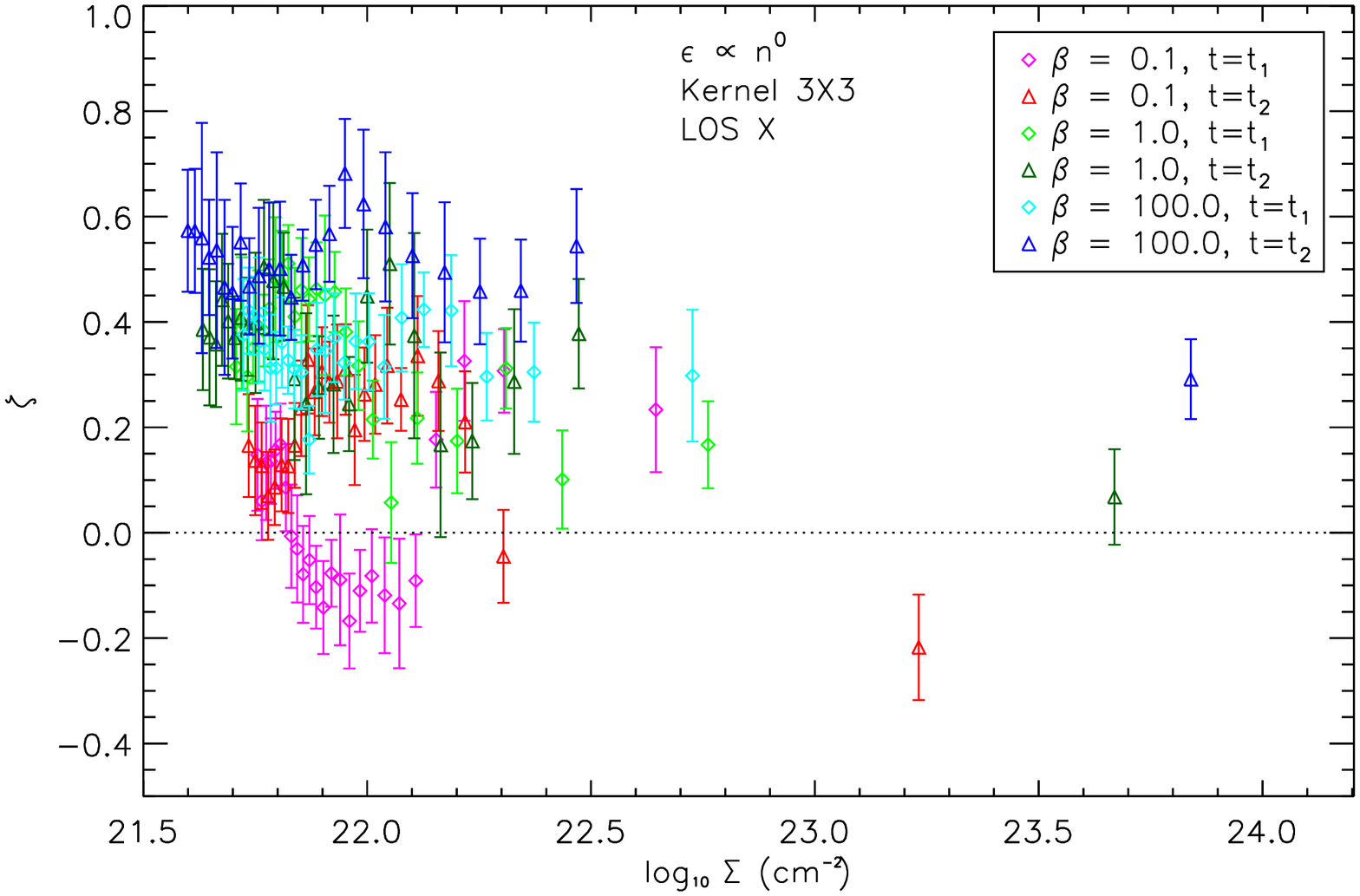}
    \includegraphics[angle=270,width=0.49\linewidth]{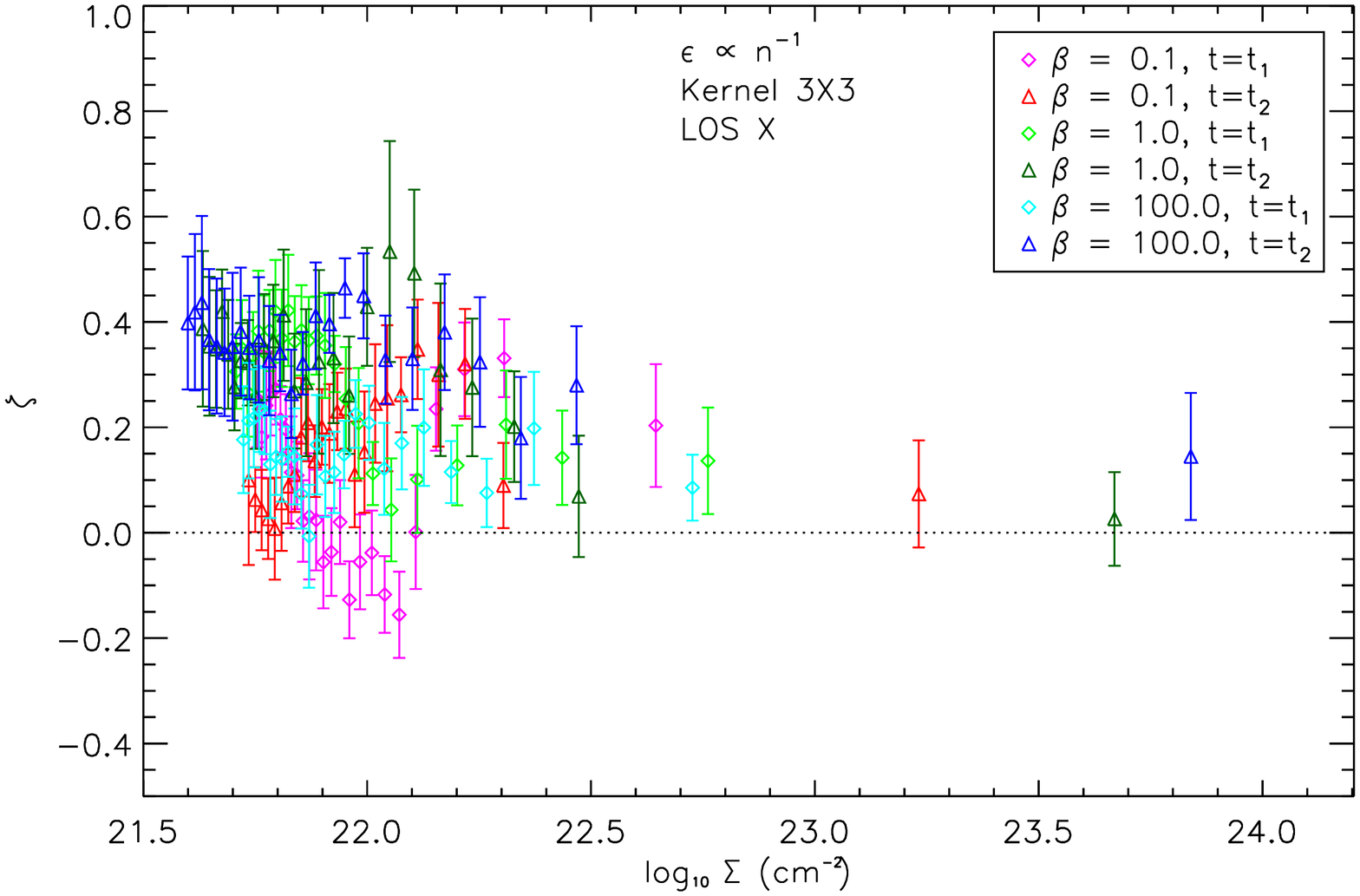}
    \includegraphics[angle=270,width=0.49\linewidth]{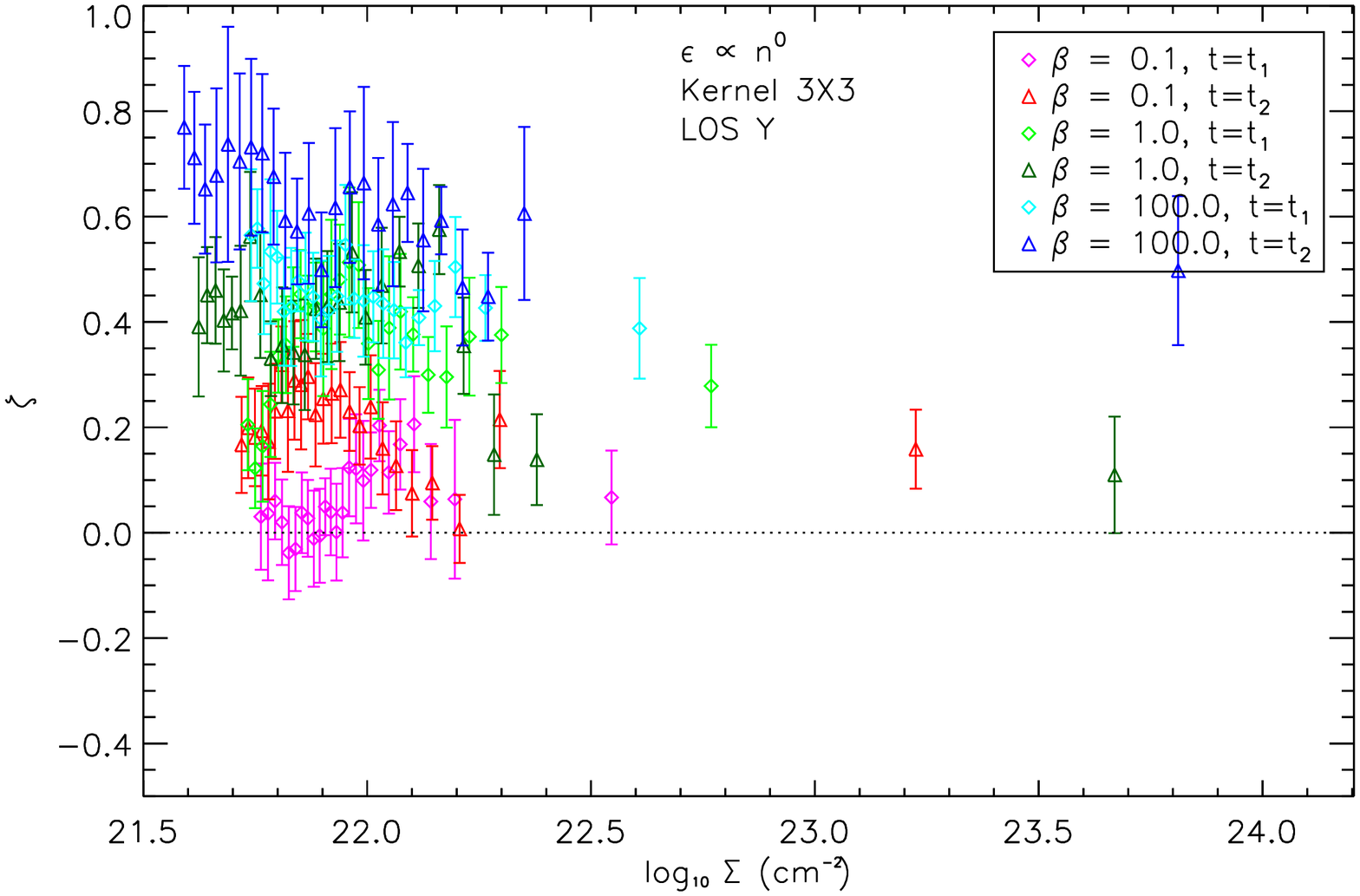}
    \includegraphics[angle=270,width=0.49\linewidth]{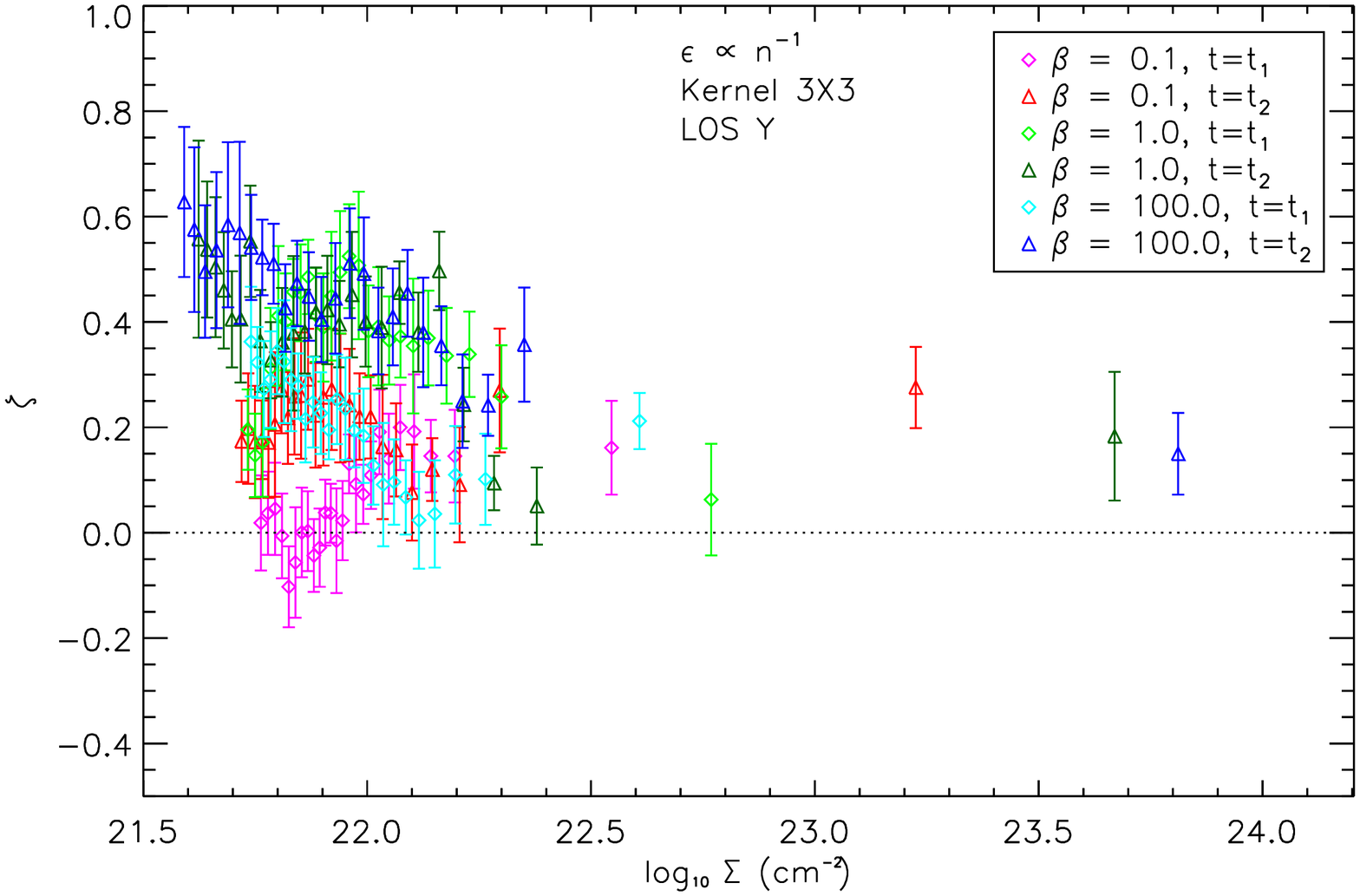}
    \includegraphics[angle=270,width=0.49\linewidth]{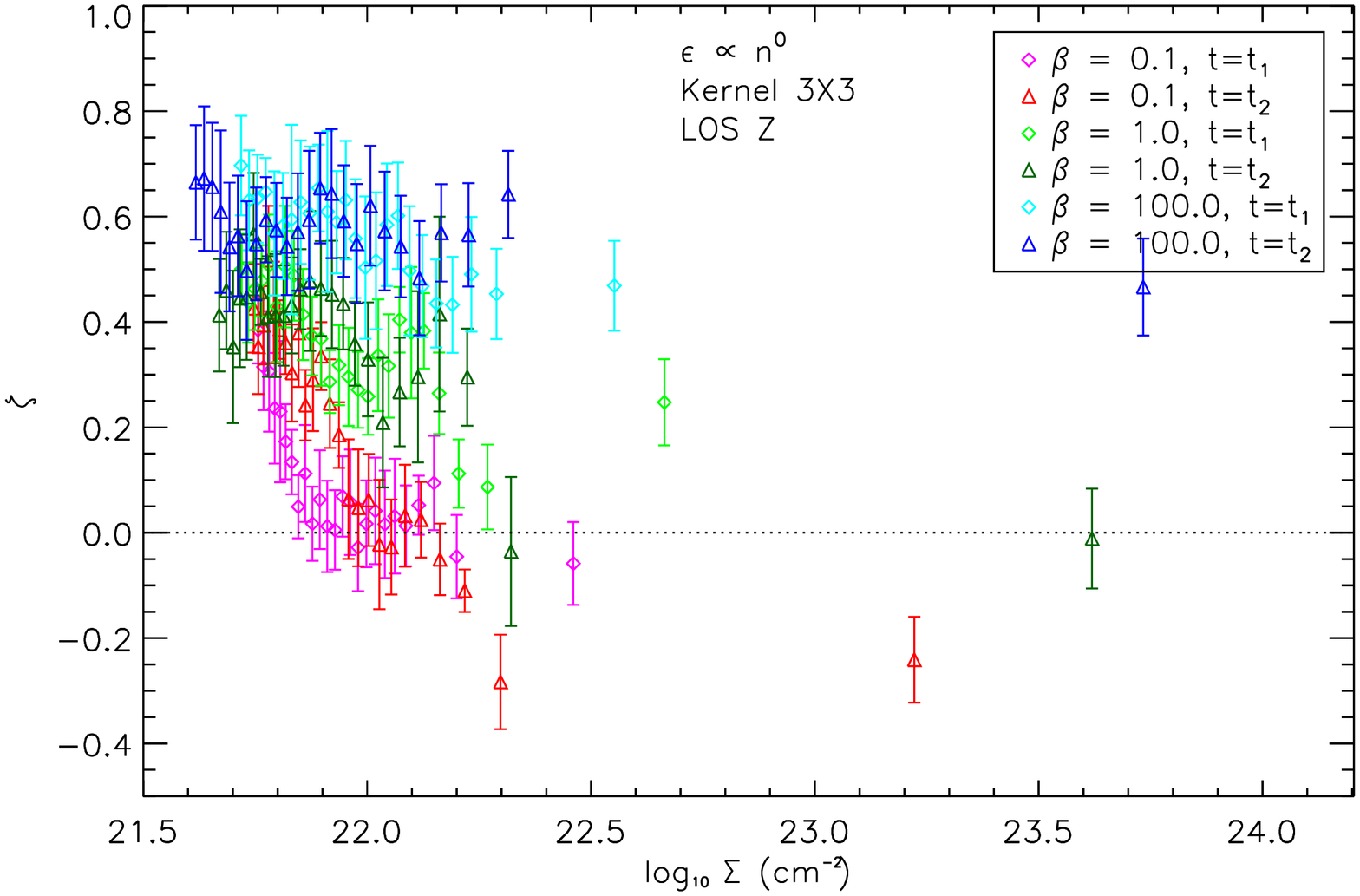}
    \includegraphics[angle=270,width=0.49\linewidth]{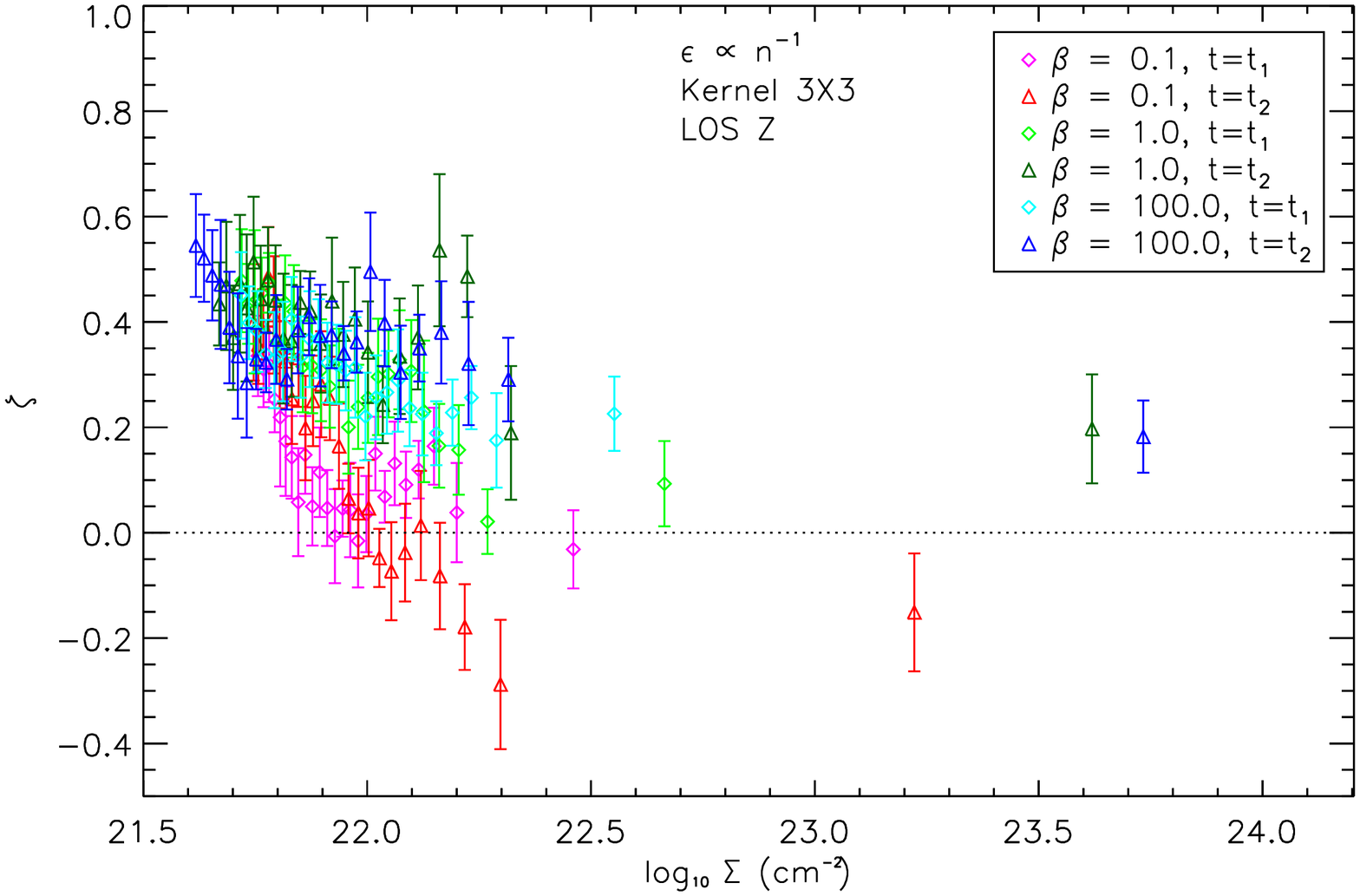}
  \caption[HRO Shape Parameter as a Function of Column Density]{HRO shape parameter $\zeta$ which parameterizes the relative orientation of the projected magnetic field ($\mathbf{B}_{POS}$) and the gradient of the column density $\mathbf{\nabla} \Sigma$. $\zeta > 0.0$ corresponds to a HRO showing $\mathbf{B}_{POS}$ predominantly perpendicular to $\mathbf{\nabla} \Sigma$ ($\mathbf{B}_{POS}$ parallel to the iso-$\Sigma$ contours). $\zeta\sim0.0$ corresponds to a flat HRO showing no predominant relative orientation between $\mathbf{B}_{POS}$ and $\mathbf{\nabla} \Sigma$. $\zeta<0.0$ corresponds to a HRO showing $\mathbf{B}_{POS}$ predominantly parallel to $\mathbf{\nabla} \Sigma$ ($\mathbf{B}_{POS}$ perpendicular to the iso-$\Sigma$ contours). As in the 3D case, the HROs of the low magnetization case show $\mathbf{B}_{POS}$ predominantly parallel to the isodensity contours with $\zeta>0.0$ even in the higher density regions. In contrast, the HROs of the intermediate and high magnetization cases show $\mathbf{B}_{POS}$ parallel to the isodensity contours with $\zeta>0.0$ at low densities and changing into $\mathbf{B}_{POS}$ perpendicular to the isodensity contours, with $\zeta<0.0$ in the highest density regions. The shape parameter $\zeta$ as a function of density decreases faster for the higher magnetization simulations.}\label{HOG2Dturnover}
\end{center}
\end{figure*}

Figure \ref{HOG2Dturnover} shows the value of $\zeta$ as a function of the central column density of each bin. The HROs of the low magnetization case show $\mathbf{B}_{POS}$ predominantly parallel to the isodensity contours with $\zeta>0.0$ even in the highest $\Sigma$ regions. In contrast, the HROs of the intermediate and high magnetization cases show $\mathbf{B}_{POS}$ parallel to the isodensity contours with $\zeta>0.0$ but changing into $\mathbf{B}_{POS}$ perpendicular to the isodensity contours with $\zeta<0.0$ predominately in the highest $\Sigma$ regions.

% CONCLUSIONS =========================================================================================================
\section{Summary and Discussion}\label{hro:conclusions}

We have introduced the Histogram of Relative Orientations (HRO), a tool to study the relative orientation of the magnetic field and the density structure in MHD simulations and polarization observations. Using HROs on a set of simulated molecular clouds with decaying supersonic turbulence and with low, intermediate, and high initial magnetization ($\beta =$ 100.0, 1.0, and 0.1) we found a significant imprint of the magnetization in the relative orientation of the magnetic field and the density structures.

\subsection{HROs in 3D}

In 3D, we found that $\mathbf{B}$ is oriented predominantly parallel to the isodensity contours in the three simulations. When dividing the the simulated regions into density bins with equal number of voxels we found a change in the relative orientation of $\mathbf{B}$ with respect to the isodensity contours in the highest density regions and this behavior is different for different initial magnetization as illustrated in Figure \ref{HOGsegments3D}. In the high-magnetization case we found that $\mathbf{B}$ changes from parallel to perpendicular to the isodensity contours in regions with density $n > 50\bar{n}$. In the intermediate magnetization case we found that $\mathbf{B}$ changes from parallel to no preferred relative orientation in regions with $n > 500\bar{n}$. In the low magnetization case $\mathbf{B}$ is parallel to the isodensity contour in all density bins.

We found two features of the relative orientation between $\mathbf{\nabla}n$ and $\mathbf{B}$ which depend on the initial magnetization: 1. The rate of change from perpendicular $\mathbf{\nabla}n$ and $\mathbf{B}$ to parallel in regions with increasing mean bin density. 2. The value of the density $n_{T}$ over which $\mathbf{\nabla}n$ and $\mathbf{B}$ are parallel. We observe that the rate of change of the relative orientation parameter $\zeta$ is greater and that $n_{T}$ is lower with higher magnetization as illustrated in Figure \ref{HOG3Dturnover}.

Both of the observed effects are related to the balance between the magnetic forces, the turbulence, and the gravitational forces in each simulation. The three simulations are initially supercritical and superalfvenic, which means that turbulence and gravitational forces are dominant over the magnetic field. Supercritical clouds can collapse gravitationally both parallel and perpendicular to the field with unlimited asymptotic density, thus gravitational collapse alone produces no preferred orientation between $\mathbf{\nabla}n$ and $\mathbf{B}$. Superalfvenic turbulence means that in scales where the kinetic energy is larger than the magnetic energy the field is dragged along with the matter resulting in density structures stretched the direction of the field, thus favoring $\mathbf{\nabla}n$ perpendicular to $\mathbf{B}$.

The supersonic perturbations resulting from the initial turbulence are amplified or suppressed depending on the strength of the initial magnetic field \citep{Draine1993}. At the time of the snapshots in this analysis, the kinetic energy has cascaded into smaller scales and the equilibrium between turbulence and magnetic field occurs at different scales for different values of the initial magnetization. In the low and intermediate magnetization cases, the kinetic energy is comparable to the magnetic field energy in scales smaller or close to the grid resolution, thus the turbulence dominates on the scales resolved by the simulation which is coincident with our observation of $\mathbf{\nabla}n$ perpendicular to $\mathbf{B}$. In the high-magnetization simulation the equilibrium between turbulence and magnetic field occurs at a scale corresponding to one-tenth of the box size. In smaller scales the magnetic field is stronger than the turbulence and the field lines are ordered in a range of scales that is well resolved by the simulation as illustrated in Figure \ref{BImap}.

The turnover in the relative orientation of $\mathbf{\nabla}n$ and $\mathbf{B}$ could be related to the ordered magnetic field routing the flow and producing collapse along the field but not perpendicular to it. Flattening of the HRO in the densest regions is consisted with isotropic gravitational collapse breaking the preferential relative orientation in the densest regions. Alternatively, the relative orientation between supersonic shock fronts and the magnetic field could produce over-densities with a particular orientation with respect to $\mathbf{B}$. Alignment between $\mathbf{B}$ and the gas velocity field has been reported in previous works \citep{ostriker2001,ossenkopf2002,banerjee2009,vazquezsemadeni2011} but the exact process which causes this relative orientation is not well understood yet. The relative orientation of $\mathbf{B}$ perpendicular to the over-densities is close to the scenario described by \citet{mouschovias1976I}. However, the interaction between supersonic turbulence and magnetic field which produces the final configuration of $\mathbf{B}$ and density structures is still the subject of active research \citep{klessen2011, schneider2011, hennebelle2012}.

\subsection{HROs in 2D}

There are two main effects to be considered in the recovery of the imprint of magnetization on the relative orientation in the projected maps. First is the effect of the integration along the line of sight. The second is the effect of the alignment efficiency in weighting the reconstruction of the projected magnetic field.

Figure \ref{HOG2Dturnover} shows that the relative orientation in the projected map is analogous to the one measured in 3D: $\mathbf{B}_{POS}$ and $\mathbf{\nabla} \Sigma$ are preferentially aligned perpendicular to each other. In 3D the highest density regions where we observed $\mathbf{B}$ perpendicular to the isodensity contours are surrounded by lower density shells where $\mathbf{B}$ is parallel to the isodensity. In 2D we observed $\mathbf{B}_{POS}$ parallel to $\mathbf{\nabla}\Sigma$ in the highest density regions of the map showing that the relative orientation in the lower density shell does not dominate in the projected map. More detailed modeling of the integration along the line of sight and its effect on the HROs and $\zeta$ will be the subject of future studies.

An interesting issue for the line of sight integration is the relative orientation of the mean magnetic field with respect to the line of sight. All of the results presented here correspond to lines of sight perpendicular to the plane of the initial magnetic field (Y or Z for initial $\mathbf{B}$ along the X axis). Small variations about this orientation of the mean magnetic field with respect to the line of sight do not affect the conclusions of the HRO study. However, the HROs corresponding to projections along the X-axis (line of sight parallel to the initial magnetic field direction) are not distinctively different for different initial magnetization as show in Figure \ref{HOG2Dturnover}. Thus, further studies are required to draw conclusions from observations of clouds where the mean magnetic field is oriented very close to the line of sight.

The effect of the alignment efficiency is illustrated in the two plots in Figure \ref{HOG2Dturnover}. The slopes of the $\zeta-\Sigma$ curves corresponding to different initial magnetization are clearly distinguishable in the case of uniform $\epsilon=1$. However, when considering a decreasing alignment efficiency, $\epsilon\propto n^{-1}$, the curves corresponding to the low and intermediate magnetization are degenerate and only the curve corresponding to the highest magnetization is distinctive. The source of this effect is the weighting of the signal coming from different regions of the cube: although the geometry of the field does not change with the alignment efficiency, the projected magnetic field is dominated by the orientation of the field in the region with better alignment efficiency.

The weighting effect is strongly dependent on the grain alignment mechanism which is yet to be understood, thus the slope of the $\zeta-\Sigma$ curve requires further characterization before it is used as a diagnostic of the magnetic field strength. In spite of that, $\zeta<0$ still corresponds to magnetic field dominance in the projected maps, therefore the column density $\Sigma_{T}$ at which the relative orientation changes can be used potentially as a comparative diagnostic of the $\mathbf{B}_{POS}$ strength.

\subsection{Relation to Existing Studies}

In the current study we focus on the characterization of the HRO using models with different magnetization. For the sake of simplicity, we choose decaying turbulence, avoiding the complexity introduced by modeling the energy injection rate and the spectrum of driven turbulence. The main purpose of turbulence driving is to maintain the energy which would continuously decay otherwise. The behavior of turbulence in the ISM is still subject of study \citep{klessen2010, schneider2011, hennebelle2012} and the continuous energy input is an approximation used to add up multiple snapshots and improve statistics in the study of the power spectrum or structure function \citep{padoan2003, padoan2006}. This study focuses in the comparison of observation for which, by definition, we get a single snapshot. Nevertheless, we expect no significant difference in the results of decaying and driven turbulence \citep{ossenkopf2002}, although the multiplicity of forcing mechanisms requires a detailed study beyond the scope of this paper \citep{federrath2010}.

Previous studies of projected maps from MHD simulations have assumed ``polarizability'' (grain alignment efficiency) proportional to the local density \citep{ostriker2001, padoan2001}. The toy-model of polarization efficiency used in this study constitutes just a zeroth-order model to take into account the grain alignment mechanism when integrating the Stokes parameters along the LOS. Further understanding of the projected magnetic field morphology requires the detailed study of the dust grain alignment mechanism and the environmental dependence of the dust chemistry.

The HRO is introduced as a statistical tool to complement the results of the CF. HRO works with polarization data only in contrast to CF which requires velocity information from spectral-line data. Given the pixel-by-pixel nature of the HRO studies, it can be used in polarization maps where the number or distribution of polarization pseudovectors is not suitable for the dispersion analysis required by CF. Although the HRO does not provide an estimate for the magnetic field magnitude yet, it can be used to quantify and extend the relative orientation studies which until now have relied on visual inspection of the maps \citep{goodman1990, tassis2009, sugitani2011, palmeirim2013}. So far, \cite{koch2012} reported a correlation between the iso-$\Sigma$ contours and the inferred magnetic field direction which is used to estimate the magnetic field strength in scales close to the size of pre-stellar cores. The HRO constitutes a tool to study similar correlations in scales which extend to entire cloud complexes.

\subsection{HROs in Observations}

The HROs of the simulations discussed in this paper evaluate the relative orientation in scales ranging from $2.34\times10^{-2}$~pc to $3.83\times10^{-1}$~pc (kernels of $3\times3$ to $49\times49$~pixels), which correspond to angular scales of $\sim$30\arcs\ to $\sim$10\arcm\ for a nearby cloud located at a distance of 150~pc, e.g. Lupus \citep{comeron2008}, or $\sim$10\arcs\ to $\sim$2\arcm\ for a nearby cloud located at a distance of 700~pc, e.g. Vela C \citep{netterfield2009}.

State-of-the-art instruments which observe the dust polarized thermal emission on large, intermediate and small scales such as \planck\ \citep{Planck2011}, ALMA \citep{peck2008}, and BLASTPol \citep{pascale2012} have angular resolutions which resolve the scales relevant for the HROs in this study: 5\arcm\ (at 857~GHz), 42\arcs\ (at 857~GHz), and 0.1\arcs\ (84 to 720~GHz) respectively. However, only instruments such as \planck, PILOT \citep{bernard2010}, and BLASTPol are designed to produce extended polarization maps resolving scales between pre-stellar cores and cloud segments which are comparable to the projected maps analyzed in this study. High angular resolution observations such as those made with CARMA \citep{hull2013}, SMA \citep{tang2013} and ALMA provide a \emph{lever arm} for HROs in the high density and small scales regime. However, the change in the relative orientation described in this study requires measurements in multiple and larger scales.

\begin{figure}
\begin{center}
    \includegraphics[angle=0,width=0.49\linewidth]{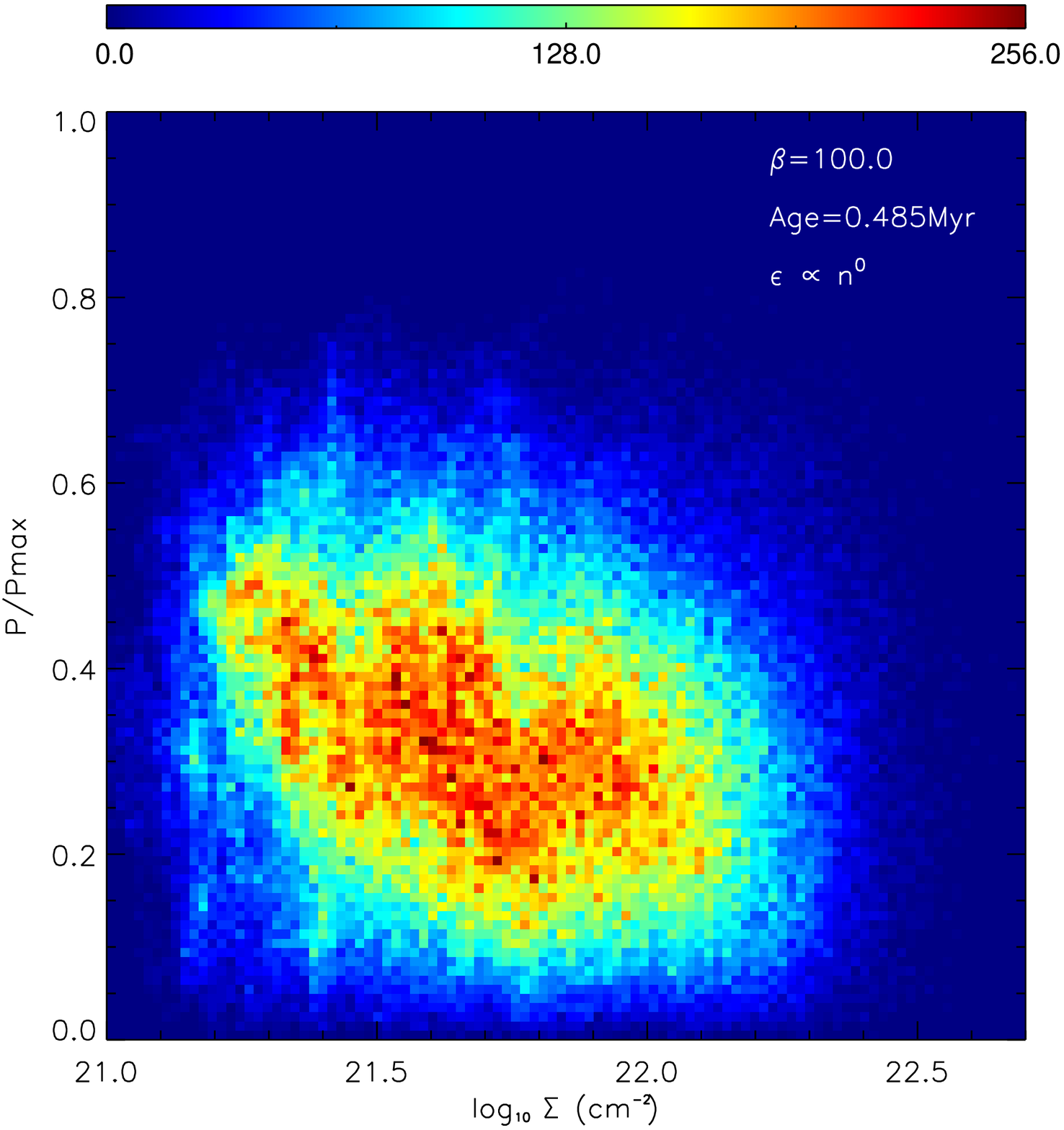}
    \includegraphics[angle=0,width=0.49\linewidth]{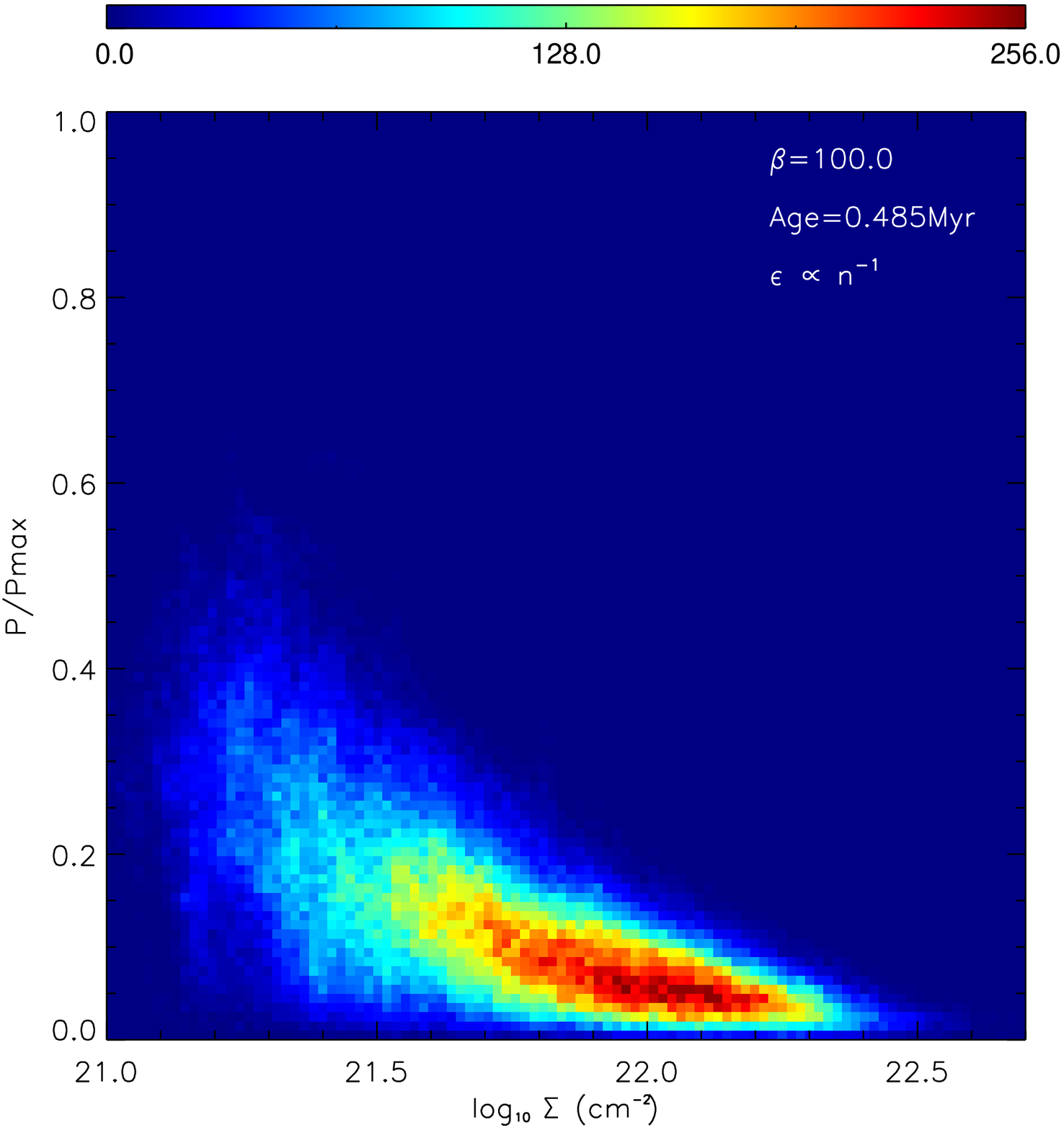}\\
    \includegraphics[angle=0,width=0.49\linewidth]{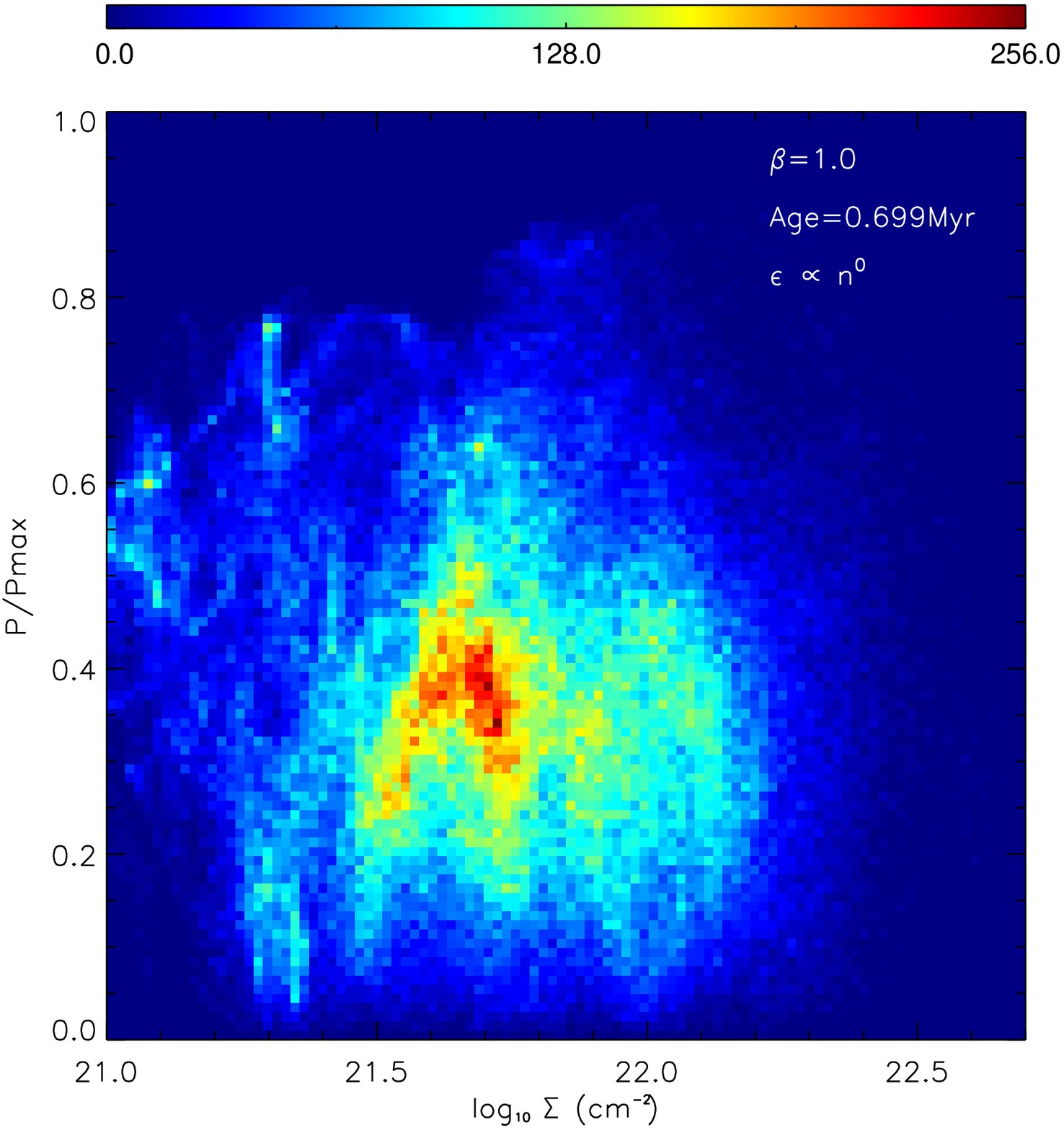}
    \includegraphics[angle=0,width=0.49\linewidth]{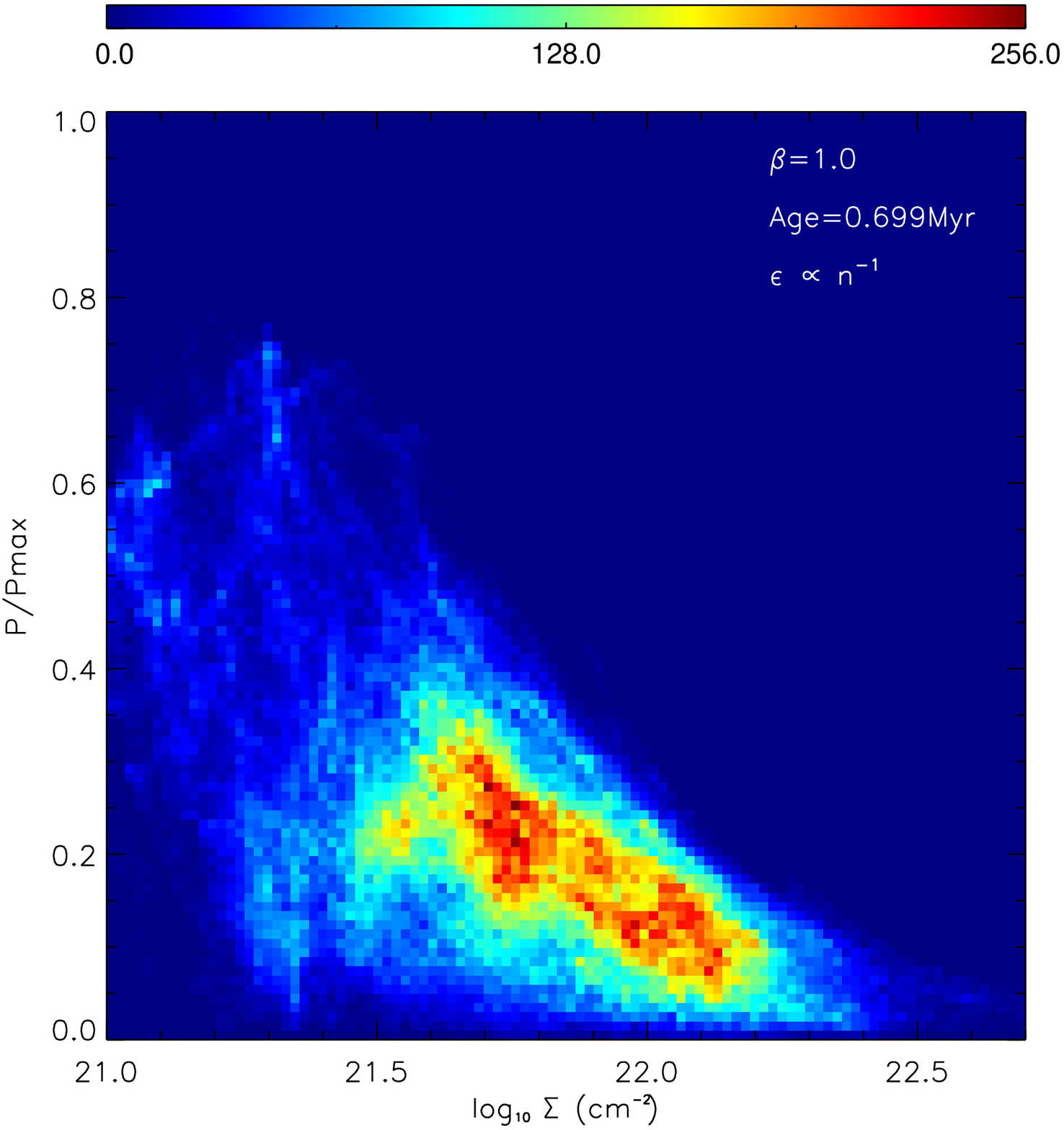}\\
    \includegraphics[angle=0,width=0.49\linewidth]{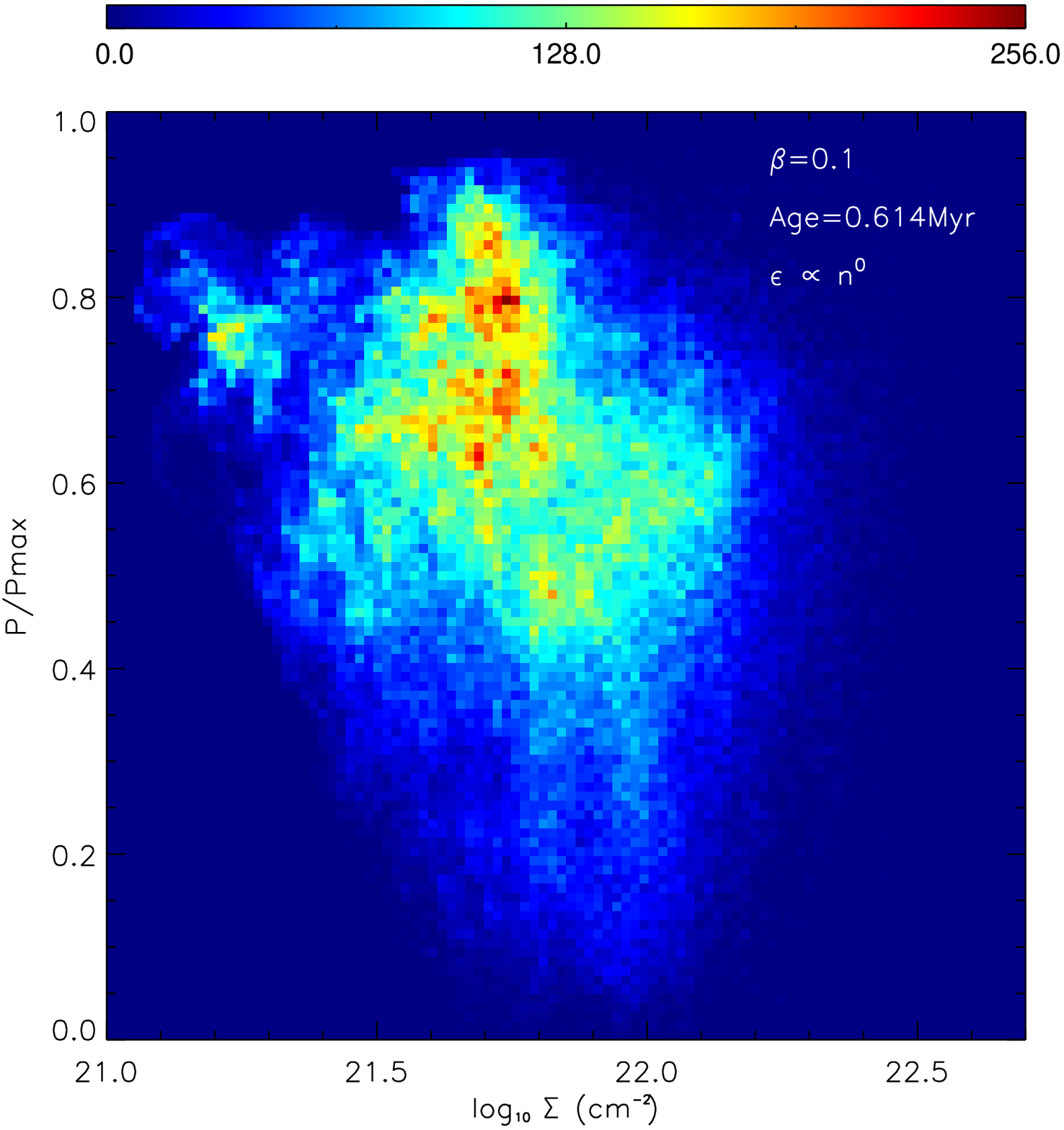}
    \includegraphics[angle=0,width=0.49\linewidth]{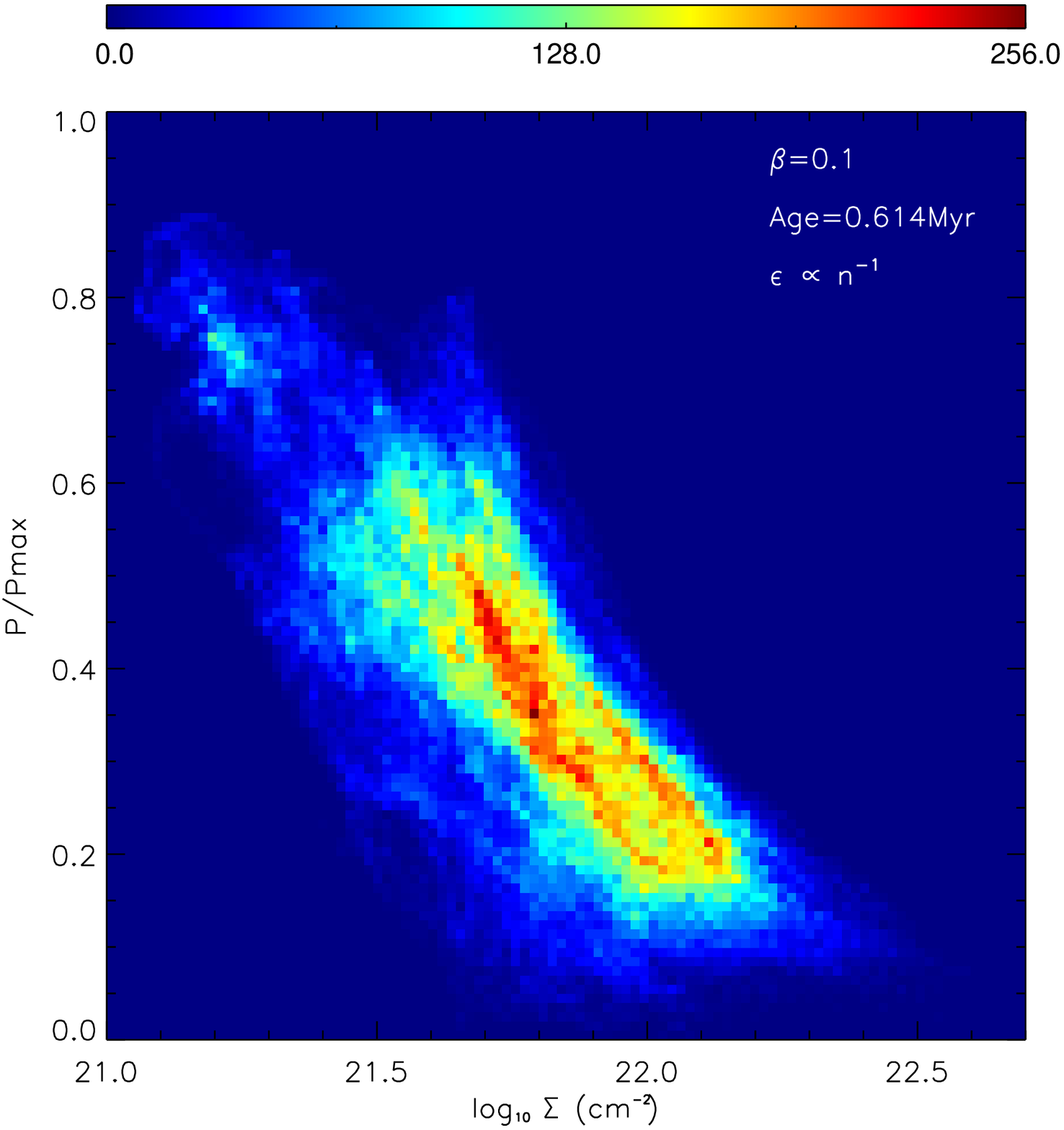}
  \caption[Polarization Percentage in Projected Simulations]{Polarization percentage $P$ in units of maximum polarization percentage $P_{max}$ as a function of column density obtained from the projections of the three simulations with uniform polarization efficiency $\epsilon=1.0$. The color scale corresponds to the density of points. $P_{max}\sim$ 20\% is determined empirically from submillimeter diffuse Galactic dust emission \citep{benoit2004}. These different distributions and the systematic behavior, especially on the right, offer diagnostics relating parameters in MHD simulations to observations.}\label{PvsSigma}
\end{center}
\end{figure}

The instrumental parameters necessary to make the polarization maps to apply HROs vary from experiment to experiment and are subject to the particular control of the systematic effects present in polarization measurements. A minimum requirement for a polarization experiment is the measurement of polarization levels of a few percent which are expected in molecular clouds and obtained in the projected maps of the three simulations as shown in Figure \ref{PvsSigma}. Estimations made using nominal values of the sensitivity of BLASTPol \citep{marsden2009} in a 50 hours observation of a 1 square degree region indicate that polarization errors of 0.5\% or less can be achieved in regions which densities $\sim2.38\times10^{22}$~N$_{H}$~cm$^{-2}$, where the HRO show the change of the relative orientation in the high-magnetization case.

The HROs can be used to characterize observations of starlight polarization, particularly those with multiple polarization pseudovectors around filamentary structures and provide enough data points to construct the histogram. However, it is important to consider that starlight polarization is limited to low density regions and the change in relative orientation described in this study is expected in regions with visual extinction $A_{\rm V}>4$.

\subsection{Future Work}

The results of the HRO analysis show that the imprint of magnetization in simulated molecular clouds is present in at least two diagnostics of the relative orientation of the magnetic field:
\begin{enumerate}
\item The change in the relative orientation of the projected magnetic field $\mathbf{B}_{POS}$ and the column density structures as a function of column density.
\item The value of the column density $\Sigma_{T}$ at which $\mathbf{B}_{POS}$ shifts from parallel to perpendicular to the iso-$\Sigma$ contours.
\end{enumerate}
These two parameters allow the systematic comparison of simulations and observations; further characterization of the magnetization conditions required detailed modeling of the grain alignment mechanism.

This study focuses on the case of an isothermal molecular cloud with a particular behavior of turbulence and magnetization with the purpose of introducing and characterizing the HRO method. Nevertheless, the HRO analysis can be extended to simulations including more complex realizations of molecular clouds \citep{nakamura2008, hennebelle2008}. The study of multiple ages, environments and dominant physics in simulations is one of the primary challenges facing the construction of a complete and coherent picture of the star formation process. Characterization using HROs is one of the multiple statistical techniques that will allow the study of multiple polarization observations and its comparison with the great diversity of relevant physical conditions which can be modeled using MHD simulations.

% ACKNOWLEDGMENTS ===================================================================================================
\acknowledgments
We thank the referee for a useful and constructive report. We gratefully acknowledge G. Novak and T. Matthews for the helpful comments on the manuscript. We are indebted to F. Flores-Mangas for providing useful insight into machine vision algorithms and to C. Matzner for his comments on the physics of shocks. This work was supported in part by the Natural Sciences and Engineering Research Council of Canada (NSERC), the Canadian Space Agency (CSA) and the Canadian Institute for Advance Research (CIFAR). Computations were performed at Sunnyvale, the beowulf cluster at the Canadian Institute for Theoretical Astrophysics. The simulations are part of the StarFormat database and are publicly available at \url{http://starformat.obspm.fr/starformat/projects}.

%More information on the AASTeX macros package is available \\ at \url{http://www.aas.org/publications/aastex}.
%For technical support, please write to \email{aastex-help@aas.org}.

%% See the AASTeX Web site at http://www.journals.uchicago.edu/AAS/AASTeX
%% for information on obtaining the facility keywords.
%{\it Facilities:} \facility{Nickel}, \facility{HST (STIS)}, \facility{CXO (ASIS)}.

%\appendix

\bibliography{jdslib}
%\bibliography{sole0305.bbl}

%\clearpage

\end{document}